\documentclass[12pt,english]{article}
\usepackage[T1]{fontenc}
\usepackage[latin9]{inputenc}
\usepackage{geometry}
\usepackage{color}
\usepackage{babel}
\usepackage{array}
\usepackage{varioref}
\usepackage{booktabs}
\usepackage{mathtools}
\usepackage{multirow}
\usepackage{amsmath}
\usepackage{amsthm}
\usepackage{amssymb}
\usepackage{graphicx}
\usepackage{setspace}
\usepackage[authoryear]{natbib}
\usepackage[unicode=true,pdfusetitle,
 bookmarks=true,bookmarksnumbered=false,bookmarksopen=false,
 breaklinks=true,pdfborder={0 0 1},backref=false,colorlinks=true]
 {hyperref}
\hypersetup{
 citecolor=darkpastelred, urlcolor=darkpastelred, linkcolor=darkpastelred}

\makeatletter

\providecommand{\tabularnewline}{\\}
\newcommand{\lyxdot}{.}

\usepackage{algorithmicx,algorithm} 
\usepackage[noend]{algpseudocode} 
\usepackage[dvipsnames,svgnames,x11names,hyperref]{xcolor}

\newcommand{\indep}{\perp \!\!\! \perp}
\usepackage{appendix}
\usepackage{chngpage}
\usepackage{lipsum}

\definecolor{harvardcrimson}{rgb}{0.79, 0.0, 0.09}
\definecolor{princetonorange}{rgb}{1.0, 0.56, 0.0}
\definecolor{brickred}{rgb}{0.8, 0.25, 0.33}
\definecolor{brightmaroon}{rgb}{0.76, 0.13, 0.28}
\definecolor{brownweb}{rgb}{0.65, 0.16, 0.16}
\definecolor{burgundy}{rgb}{0.5, 0.0, 0.13}
\definecolor{burntumber}{rgb}{0.54, 0.2, 0.14}
\definecolor{carmine}{rgb}{0.59, 0.0, 0.09}
\definecolor{carnelian}{rgb}{0.7, 0.11, 0.11}
\definecolor{airforceblue}{rgb}{0.36, 0.54, 0.66}
\definecolor{ceil}{rgb}{0.57, 0.63, 0.81}
\definecolor{cordovan}{rgb}{0.54, 0.25, 0.27}
\definecolor{cornellred}{rgb}{0.7, 0.11, 0.11}
\definecolor{crimsonglory}{rgb}{0.75, 0.0, 0.2}
\definecolor{darkpastelred}{rgb}{0.76, 0.23, 0.13}
\definecolor{oucrimsonred}{rgb}{0.6, 0.0, 0.0}
\definecolor{darkpastelred}{rgb}{0.76, 0.23, 0.13}
\definecolor{darkterracotta}{rgb}{0.8, 0.31, 0.36}
\definecolor{deepchestnut}{rgb}{0.73, 0.31, 0.28}
\definecolor{indianred}{rgb}{0.8, 0.36, 0.36}
\definecolor{maroon}{rgb}{0.5, 0.0, 0.0}
\definecolor{mediumcarmine}{rgb}{0.69, 0.25, 0.21}
\definecolor{tuftsblue}{rgb}{0.28, 0.57, 0.81}
\definecolor{yaleblue}{rgb}{0.06, 0.3, 0.57}
\definecolor{darkpastelblue}{rgb}{0.47, 0.62, 0.8}
\definecolor{bluegray}{rgb}{0.4, 0.6, 0.8}
\definecolor{airforceblue}{rgb}{0.36, 0.54, 0.66}

\@ifundefined{showcaptionsetup}{}{%
 \PassOptionsToPackage{caption=false}{subfig}}
\usepackage{subfig}
\makeatother

\begin{document}
\pagenumbering{gobble}
\title{Structural Regularization\thanks{We thank Xiaoyi Han, Thomas Sargent, Wei Song, Jiayi Wen, and seminar
audiences for many helpful discussions and suggestions. Mao acknowledges
financial support by the national natural science foundation of China.}}
\author{Jiaming Mao\thanks{Corresponding author. Xiamen University. Email: \texttt{jmao@xmu.edu.cn}}
\and Zhesheng Zheng\thanks{Xiamen University. Email: \texttt{jensenzheng99@gmail.com}}}
\maketitle
\begin{abstract}
\begin{onehalfspace}
We propose a novel method for modeling data by using structural models
based on economic theory as regularizers for statistical models. We
show that even if a structural model is misspecified, as long as it
is informative about the data-generating mechanism, our method can
outperform both the (misspecified) structural model and un-structural-regularized
statistical models. Our method permits a Bayesian interpretation of
theory as prior knowledge and can be used both for statistical prediction
and causal inference. It contributes to transfer learning by showing
how incorporating theory into statistical modeling can significantly
improve out-of-domain predictions and offers a way to synthesize reduced-form
and structural approaches for causal effect estimation. Simulation
experiments demonstrate the potential of our method in various settings,
including first-price auctions, dynamic models of entry and exit,
and demand estimation with instrumental variables. Our method has
potential applications not only in economics, but in other scientific
disciplines whose theoretical models offer important insight but are
subject to significant misspecification concerns. 
\end{onehalfspace}
\end{abstract}
\newpage{}

\pagenumbering{arabic}

\section{Introduction}

Structural models are causal models based on economic theory. A complete
structural model describes economic and social phenomena as the outcomes
of individual behavior in specific economic and social environments.
The structural approach to data analysis takes a structural model
as a truthful representation of the data-generating mechanism and
estimates the model parameters from observed data. The estimated model
can then be used to make predictions, evaluate causal effects, and
conduct welfare analyses\footnote{See \citet{reiss_structural_2007}, \citet{heckman_econometric_2007}
and \citet{low_use_2017} for surveys on structural estimation.}. 

One of the main strengths of structural estimation lies in its ability
to make claims of \emph{generalizability} or \emph{external validity}.
Because a structural model is based on economic theory, its parameters
-- such as those governing preferences and technology -- can be
``deep,'' or policy-invariant, so that the estimated model can be
used to generate predictions in different environments. A key assumption
involved, however, is that the model is correctly specified. In practice,
there is no such guarantee and structural models are often criticized
for relying on strong, unrealistic assumptions and identification
by functional form. This has limited the usefulness of the structural
approach and its empirical success\footnote{\citet{heckman_causal_2000}: ``The empirical track record of the
structural approach is, at best, mixed. Economic data, both micro
and macro, have not yielded many stable structural parameters. Parameter
estimates from the structural research program are widely held not
to be credible.'' \citet{rust_limits_2014}: ``Looking back nearly
four decades after the Lucas critique paper, it is fair to ask whether
structural models really have succeeded and resulted in significantly
more accurate and reliable policy forecasting and evaluation.''}. 

In this paper, we propose a new methodology for modeling data that
both inherits the desirable property of structural estimation --
the ability to make claims of external validity -- and incorporates
a robustness against model misspecification. The method, which we
call the \emph{structural regularization estimator} (\emph{SRE}),
treats a given structural model as the benchmark model and estimates
a flexible statistical model with a penalty on deviance from the structural
benchmark. Equivalently, we select the best statistical model to describe
the data within a neighborhood of the structural model. We show that
even if the structural model is misspecified, as long as it is \emph{informative}
about the true data-generating mechanism, our method can outperform
both the (misspecified) structural model and un-structural-regularized
statistical models. 

Our method belongs to a class of regularized regression models. In
contrast to popular methods such as ridge regression and the lasso,
which shrink the parameters of a regression model toward zero to achieve
a balance between bias and variance, the SRE shrinks the parameters
of a statistical model toward those values \emph{implied }by the structural
model so as to achieve a balance between maximizing statistical fit
and minimizing deviance from theory. 

The SRE permits a Bayesian interpretation of using theory as prior
knowledge. From a Bayesian perspective, regularization amounts to
the use of informative priors that introduce our beliefs about the
observed data \citep{li_regularized_2006}. Classic priors used for
regularization in statistics and machine learning include sparsity
and smoothness priors. In this paper, we argue that since theoretical
models are formulated based on the results of previously observed
information and conducted studies, they should naturally serve as
priors for analyzing new evidence. 

Our method can be used both for statistical prediction and causal
inference. When used for statistical prediction, it contributes to
the literature on transfer learning by showing how incorporating theory
into statistical modeling can significantly improve out-of-domain
prediction. Given a predictive task involving inputs $x$ and outcome
$y$, a key limitation with most statistical methods is that they
require the distributions governing the training and the test data
to be the same in order to guarantee performance\footnote{This remains true for state-of-the-art deep learning models. See \citet{donahue_decaf_2014}
and \citet{yosinski_how_2014} for discussions on how features extracted
from deep convolutional neural networks trained on large image datasets
are susceptible to various domain shifts.}. In the machine learning literature, the problem of applying a model
trained on a \emph{source} domain with distribution $\mathbb{P}_{xy}^{S}$
to a \emph{target }domain with distribution $\mathbb{P}_{xy}^{T}\ne\mathbb{P}_{xy}^{S}$
is known as \emph{transfer learning}\footnote{Several definitions of \emph{domain }exist in the transfer learning
literature. In this paper, given $\left(x,y\right)\in\mathcal{O}$
and a joint distribution $\mathbb{P}_{xy}$ on $\mathcal{O}$, we
define \emph{domain} as a pair $\left\langle \mathcal{O},\mathbb{P}_{xy}\right\rangle $.
Note that this notion of domain is different from that of the domain
of a function.}$^{,}$\footnote{See \citet{pan_survey_2010} for a survey on transfer learning. \citet{ben-david_theory_2010}
provides a theoretical treatment on learning from different domains.}. A majority of research on transfer learning so far has focused on
\emph{domain adaptation}\footnote{Also known as \emph{covariate shift} or \emph{transductive transfer
learning} \citep{pan_survey_2010}.}, where the marginal distributions of the inputs differ across domains,
i.e. $\mathbb{P}_{x}^{S}\ne\mathbb{P}_{x}^{T}$, but the conditional
outcome distributions remain the same, i.e. $\mathbb{P}_{y|x}^{S}=\mathbb{P}_{y|x}^{T}$.
Methods that have been proposed aim to reduce the difference in input
distributions either by sample-reweighting \citep{zadrozny_learning_2004,huang_correcting_2007,jiang_instance_2007,sugiyama_direct_2008}
or by finding a domain-invariant transformation \citep{pan_domain_2010,gopalan_domain_2011}\footnote{This includes the more recent \emph{deep domain adaptation} literature
that employs deep neural networks for domain adaptation. See \citet{glorot_domain_2011,chopra_dlid_2013,ganin_unsupervised_2014,tzeng_deep_2014,long_learning_2015}.
\citet{wang_deep_2018} provides an overview of this literature in
the context of computer vision.}. Few studies, however, have dealt with the more difficult problem
of when both $\mathbb{P}_{x}$ and $\mathbb{P}_{y|x}$ change across
domains\footnote{The problem is known as \emph{inductive transfer learning} \citep{pan_survey_2010}.
While a number of methods have been proposed to deal with this problem,
they all require target domain data in training -- we need to observe
some $\left\{ x_{i},y_{i}\right\} $ in the target domain. See \citet{schwaighofer_learning_2005,dai_boosting_2007,gao_knowledge_2008,wang_flexible_2014}.
These methods mostly adapt multi-task learning algorithms and are
\emph{not} solutions to the problem of generalizing model predictions
to different domains in a strict sense.}.

In this paper, we note that transfer learning can be viewed as a counterfactual
prediction problem. If the source and the target domain are governed
by the same data-generating mechanism, then a structural model that
\emph{correctly} describes this mechanism, when estimated on the source
domain, will generalize naturally to the target domain, even if both
the marginal and the conditional distributions have changed. In the
context of transfer learning, the external validity of a structural
model translates into \emph{domain-invariance}. Fundamentally, this
is because causal relationships are more stable than statistical relationships
\citep{pearl_causality_2009}\footnote{Motivated by the idea that causal relationships are more stable, \citet{rojas-carulla_invariant_2018}
propose\emph{ }``causal transfer learning.'' \citet{kuang_stable_2020}
propose ``stable prediction''. Both studies rely on the assumption
that a subset of the input variables $v\subseteq x$ have a causal
relation with the outcome $y$ and the conditional probability $\mathbb{P}\left(y|v\right)$
is invariant across domains. However, it is \emph{not} true that having
a causal relationship implies $\mathbb{P}\left(y|v\right)$ is domain-invariant.
Let $w=x\backslash v$. The assumption only holds under very limited
and untestable conditions, namely that $y\perp w|v$ and that the
causal effect of $v$ on $y$ is homogeneous.}. On the other hand, if a structural model is misspecified yet informative
about the data-generating mechanism, then it may not compete with
the best statistical models \emph{in-domain}, but can still provide
useful guidance for extrapolating \emph{out-of-domain}\footnote{In this paper, we distinguish between the notion of \emph{out-of-domain}
and \emph{out-of-sample}. Out-of-sample data are test data drawn from
the same distribution as the training data.}. This intuition motivates our estimator. Indeed, we show that the
SRE can significantly outperform un-structural-regularized statistical
models in \emph{out-of-domain} prediction whether we are given a correctly
specified or a misspecified but informative structural model\footnote{Note that we do not claim superiority over un-structural-regularized
statistical models \emph{in-domain}, since one can \emph{always} pick
a statistical model flexible enough to generate good in-domain (out-of-sample)
performance -- performance on test data drawn from the same distribution
on which the model is estimated. Hence the main contribution of the
SRE to statistical prediction is in terms of its out-of-domain performance,
i.e. the ability to extrapolate.\label{fn:superior}}.

Our method also contributes to the literature on causal effect estimation
by offering a way to combine the nonstructural statistical approach
to causal inference with the structural approach. The nonstructural
approach, also known as the \emph{reduced-form} approach\footnote{As \citet{chetty_sufficient_2009} points out, the term ``reduced-form''
is largely a misnomer, whose meaning in the econometrics literature
today has departed from its historical root. Historically, a reduced-form
model is an alternative representation of a structural model. Given
a structural model $\mathcal{M}(x,y,\epsilon)=0$, where $x$ is exogenous,
$y$ is endogenous, and $\epsilon$ is unobserved, if we write $y$
as a function of $x$ and $\epsilon$, $y=f(x,\epsilon)$, then $f$
is the\emph{ reduced-form} of $\mathcal{M}$ \citep{reiss_structural_2007}.
Today, however, applied economists typically refer to nonstructural,
statistical treatment effect models as ``reduced-form'' models.
Perhaps reflecting the informal nature of the terminology today, \citet{rust_limits_2014}
gives the following definitions of the two approaches: ``At the risk
of oversimplifying, empirical work that takes theory \textquotedblleft seriously\textquotedblright{}
is referred to as \emph{structural econometrics} whereas empirical
work that avoids a tight integration of theory and empirical work
is referred to as \emph{reduced form econometrics}.''}, estimates causal effects from observational data using statistical
models. Knowledge of the data-generating mechanism is used not to
specify a complete causal model, but to inform research designs that
can identify the causal effects of interest by exploiting exogenous
variations in the data. Reduced-form methods, including selection
on observables regression, instrumental variables regression, difference-in-differences
estimation and so on, are widely used in applied economic analyses.
At their best, these methods take advantage of credible sources of
identifying information to deliver estimates that have high \emph{internal
validity}\footnote{\citet{angrist_credibility_2010} offer an account of what they call
``the credibility revolution'' -- the increasing popularity of
quasi-experimental methods that seek natural experiments as sources
of identifying information. Our definition of reduced-form methods
include both quasi-experimental and more traditional, non-quasi-experimental
methods that use prior information to locate exogenous sources of
variation.}. On the other hand, they have also been criticized for learning effects
that are \emph{local} and lack justifications for \emph{external validity}.
Which approach should be preferred -- the structural or the reduced-form
-- has been the subject of a long-standing debate within the economics
profession\footnote{See \citet{rosenzweig_natural_2000,angrist_credibility_2010,keane_structural_2010,keane_structural_2010-1,nevo_taking_2010,deaton_instruments_2010,heckman_building_2010}
for different perspectives on the structural vs. reduced-form debate.}. This debate has at times been framed as a disagreement over the
role of theory in data analysis, with some authors emphasizing the
limits to inference \emph{without} theory \citep{wolpin_limits_2013}
and others emphasizing the limits\emph{ with} theory \citep{rust_limits_2014}. 

We show that the SRE offers a way to reconcile and synthesize these
two competing approaches and philosophies. Theory, in our approach,
informs but not dictates data analysis. Technically, by using structural
models to regularize the functional form of reduced-form models, we
can effectively select models that sit ``in the interior of the continuum
between reduced-form and structural estimation'' \citep{chetty_sufficient_2009}.
The resulting estimator has the ability to leverage the strengths
of both approaches -- the internal validity of reduced-form methods
and the external validity of structural estimation -- while defending
against their weaknesses\footnote{As a price to pay, the SRE largely loses its structural interpretation
and cannot be used to conduct welfare analyses. We discuss this limitation
in section \vpageref{subsec:Discussion}.}. 

We demonstrate the effectiveness of our approach using a set of simulation
experiments designed to showcase its power under a variety of realistic
settings in applied economic analyses, including first-price auctions,
dynamic models of entry and exit, and demand estimation with instrumental
variables. For each experiment, we compare the in-domain and out-of-domain
performance of our estimator with that of structural and (reduced-form)
statistical estimation. We consider a number of scenarios in which
the benchmark structural model is misspecified. In particular, we
consider cases in which individual agents deviate from perfect rationality
and display various degrees of non-optimizing behavior or boundedly-rational
expectations. These cases pose significant challenges to structural
estimation due to a lack of identifiable, consensus models for non-rational
behavior. Dynamic models in both macro- and microeconomics, for example,
have long relied on the rational expectations assumption despite its
well-known limitations. In all of these cases, we show that based
on benchmark models that assume perfect rationality, the SRE is nevertheless
able to obtain results that are much closer to the true non-rational
models and, as a consequence, generates much more accurate out-of-domain
predictions than (reduced-form) statistical models. 

Several authors have proposed combining structural and reduced-form
estimation \citep{chetty_sufficient_2009,heckman_building_2010}.
Their solution is to use structural models to derive sufficient statistics
for the intended analysis and then use reduced-form methods to estimate
them. In comparison, we offer a general algorithm rather than relying
on ad hoc derivations\footnote{However, our method cannot be used to conduct welfare analysis, which
is the focus of \citet{chetty_sufficient_2009}.}. In a paper concurrent with ours, \citet{mao_ensemble_2020} propose
two novel ways for combining structural and reduced-form models, one
with a doubly robust construction and the other a weighted ensemble.
Their methods can be viewed as complementary to ours.

Our method is most closely related to \citet{fessler_how_2019} (FK)
who also propose the idea of using theory to regularize statistical
models. In their framework, theory is represented as a set of constraints
on the parameters of a statistical model. They propose an empirical
Bayes approach that first estimates the statistical model without
constraints and then project the estimated parameters, $\widehat{\beta}$,
onto a subspace defined by theoretical restrictions. These projected
values, $\widehat{\beta}_{0}$ are then used as priors to obtain the
parameters' posterior means which shrink $\widehat{\beta}$ towards
$\widehat{\beta}_{0}$. Compared with their approach, our method is
different in its construction and has arguably a number of key advantages.
First, FK assumes a statistical model whose parameters are identified
and are consistently and unbiasedly estimated in the absence of theoretical
constraints. We do not impose such assumptions. The statistical model
that we shrink toward our structural benchmark can be complex and
high-dimensional. While FK also assumes their statistical model to
be correctly specified, we regard ours as an approximation to an unknown
target function, allowing the potential use of adaptive methods such
as random forests and neural nets. Second, a key requirement for FK
is that theory has to be expressed as a set of constraints on the
statistical parameters. This puts significant limitations on the type
of theoretical models that can be considered as well as requires an
ad hoc approach to find a statistical model that nests the theoretical
model for each application. In contrast, the structural model that
we use as our benchmark can be highly complicated, whose assumptions
have no obvious ways of being expressed as a set of constraints on
a statistical model, and whose parameter space can have a higher dimension.
This include models such as dynamic discrete choice models and dynamic
games that are widely used in empirical applications. Our method is
general and does not require ad hoc constructions. Third and perhaps
most importantly, while FK focuses on improving the in-domain performance
of statistical estimators with theory, our goal is to achieve both
good in-domain and out-of-domain predictive performance and obtain
estimates with both internal and external validity. Moreover, we note
that when it comes to in-domain performance, compared to a purely
statistical approach, FK's method is mainly useful in a limited setting
in which the sample size $N$ is larger but not significantly larger
than the number of parameters $p$. This is because the empirical
Bayes estimator can improve the precision of estimates when the sample
size is small, but the improvement vanishes as the sample size grows
large. In contrast, we show that the advantage of our SRE relative
to a purely statistical approach is in its out-of-domain performance,
i.e. its ability to extrapolate\footnote{\citet{fessler_how_2019} prove that their estimator dominates (in
terms of MSE) the unconstrained statistical model that they estimate
in the first step. But this result holds for all James-Stein type
shrinkage estimators. Conceptually, shrinking to \emph{anything} has
the effect of trampling down the variability of a statistical model
when $p$ is large relative to $N$ and thereby helping to lower the
MSE. In our simulations, therefore, we compare the performance of
the SRE not against the statistical model that we regularize, but
against the best statistical model we obtain using model selection.
We argue that this is the more meaningful comparison.}. Such advantage does not disappear no matter how much data we observe
in-domain.

Finally, this paper is related to the robustness literature in economics
and statistics. Motivated by \citet{hansen_robust_2001,hansen_wanting_2010,hansen_structured_2020}'s
work on robust decision making under model misspecification\footnote{See \citet{watson_approximate_2016} and \citet{hansen_ambiguity_2016}
for surveys of recent developments in statistical decision theory
and robust estimation in the presence of model misspecification.}, \citet{bonhomme_minimizing_2018} develop a locally robust minimax
estimator that minimizes maximum expected loss over a statistical
neighborhood of a benchmark model using local linearization techniques.
In a Bayesian setting, \citet{giacomini_estimation_2019} analyzes
partially identified models by constructing a class of priors in a
neighborhood of a benchmark prior and obtaining the optimal posterior
minimax decision over this class. Working on structural models, \citet{christensen_counterfactual_2019}
consider a class of models defined by equilibrium conditions and characterize
the sensitivity of their counterfactuals to deviations from benchmark
specifications of the distribution of unobservables. Like these studies,
we are motivated by concerns over model misspecifications. However,
our goal is not to achieve robustness in the sense of minimizing the
worst-case impact of misspecification in a given neighborhood of the
benchmark model or quantifying its local or global sensitivity. The
shrinkage method we employ allows arbitrary deviation from the structural
benchmark, so that when it is uninformative, the SRE is ``reduced''
to a (reduced-form) statistical model\footnote{Pun intended.}. Our
method allows arbitrary misspecification of the structural model,
unlike \citet{christensen_counterfactual_2019} whose misspecification
concerns are limited to the distribution of unobservables. 

The rest of this paper is organized as follows. Section \ref{sec:Motivating-Example}
provides a motivating example of how our method works in the context
of a simple demand estimation problem. Section \ref{sec:Methodology}
lays out the details of our algorithm. In section \ref{sec:Applications}
we apply our method to three sets of simulation experiments in the
settings of first-price auctions, dynamic models of entry and exit,
and demand estimation with instrumental variables and report their
results. Section \ref{sec:Conclusion} concludes. 

\section{Motivating Example\label{sec:Motivating-Example}}

\begin{figure}[t]
\subfloat[\label{fig:consumption_a}]{\includegraphics[width=0.5\columnwidth]{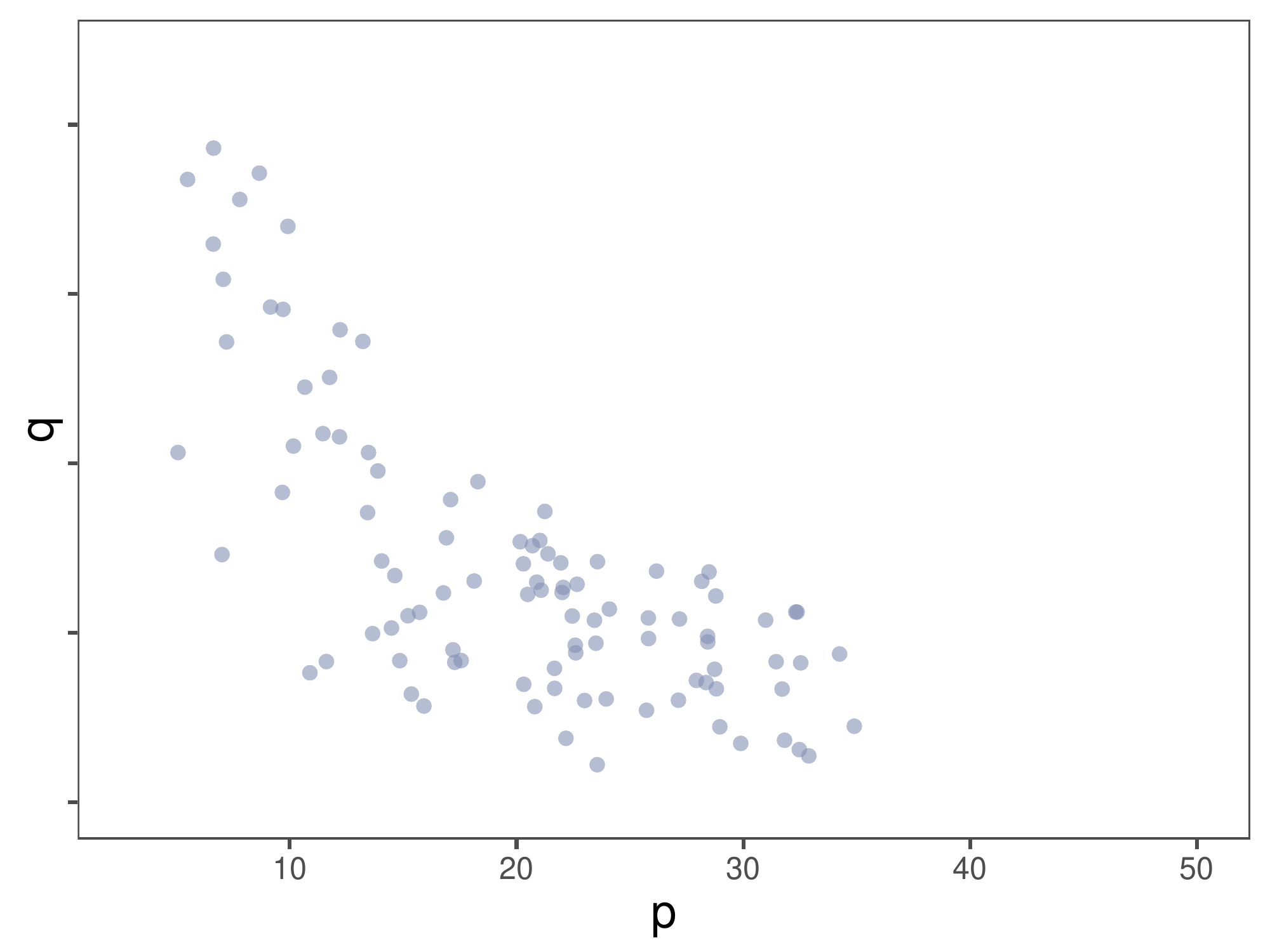}

}\subfloat[\label{fig:consumption_b}]{\includegraphics[width=0.5\columnwidth]{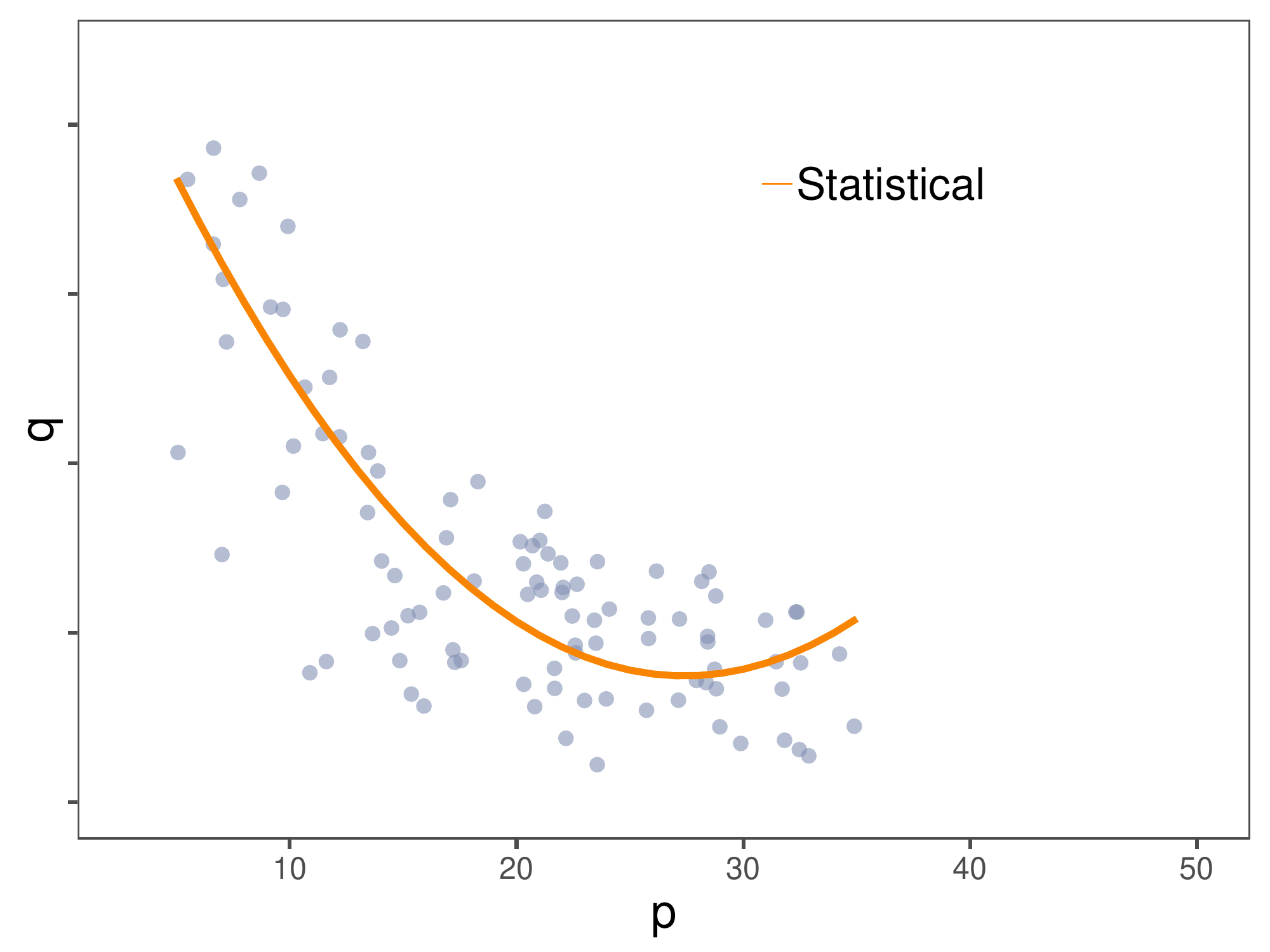}

}

\subfloat[\label{fig:consumption_c}]{\includegraphics[width=0.5\columnwidth]{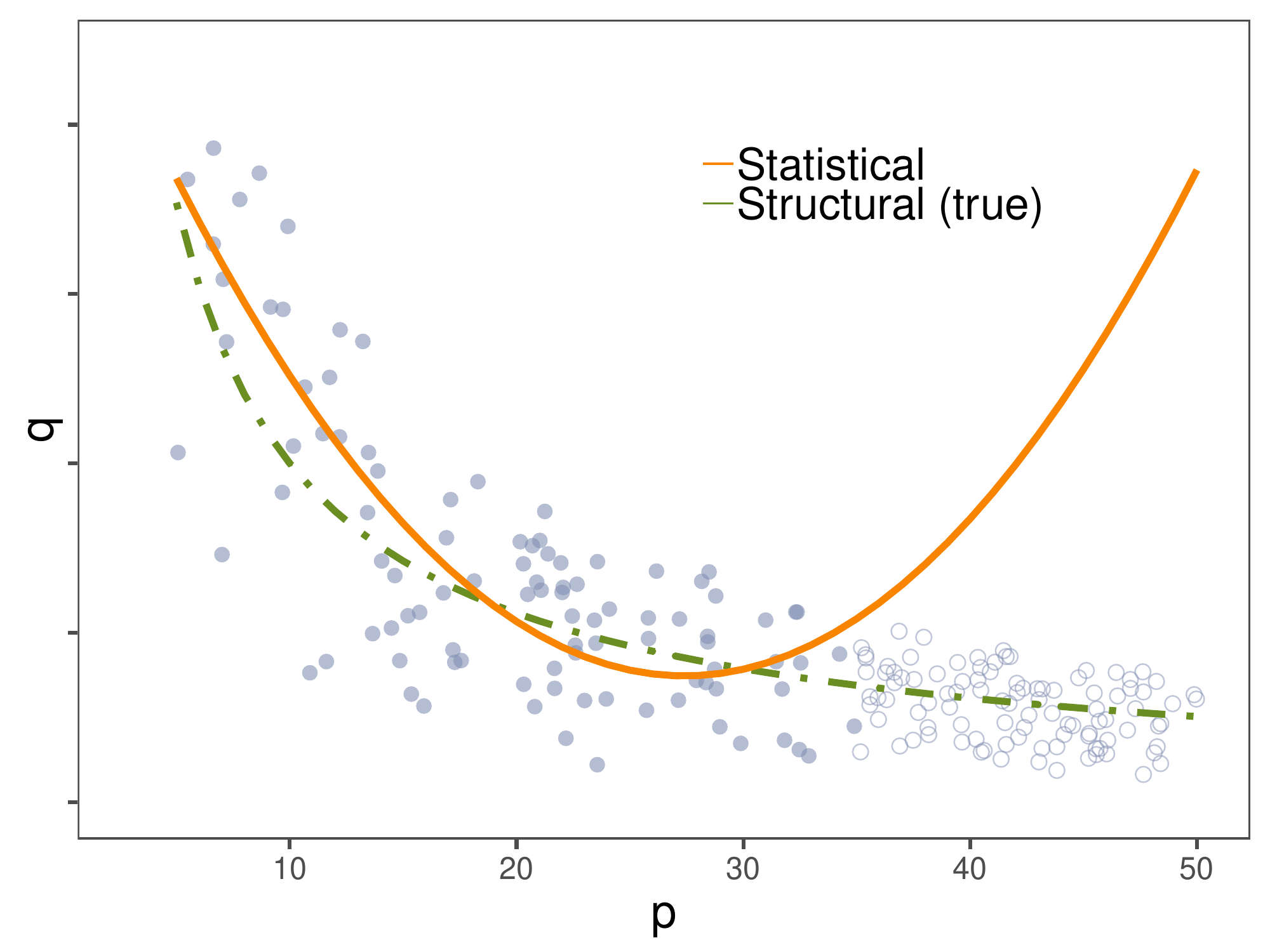}

}\subfloat[\label{fig:consumption_d}]{\includegraphics[width=0.5\columnwidth]{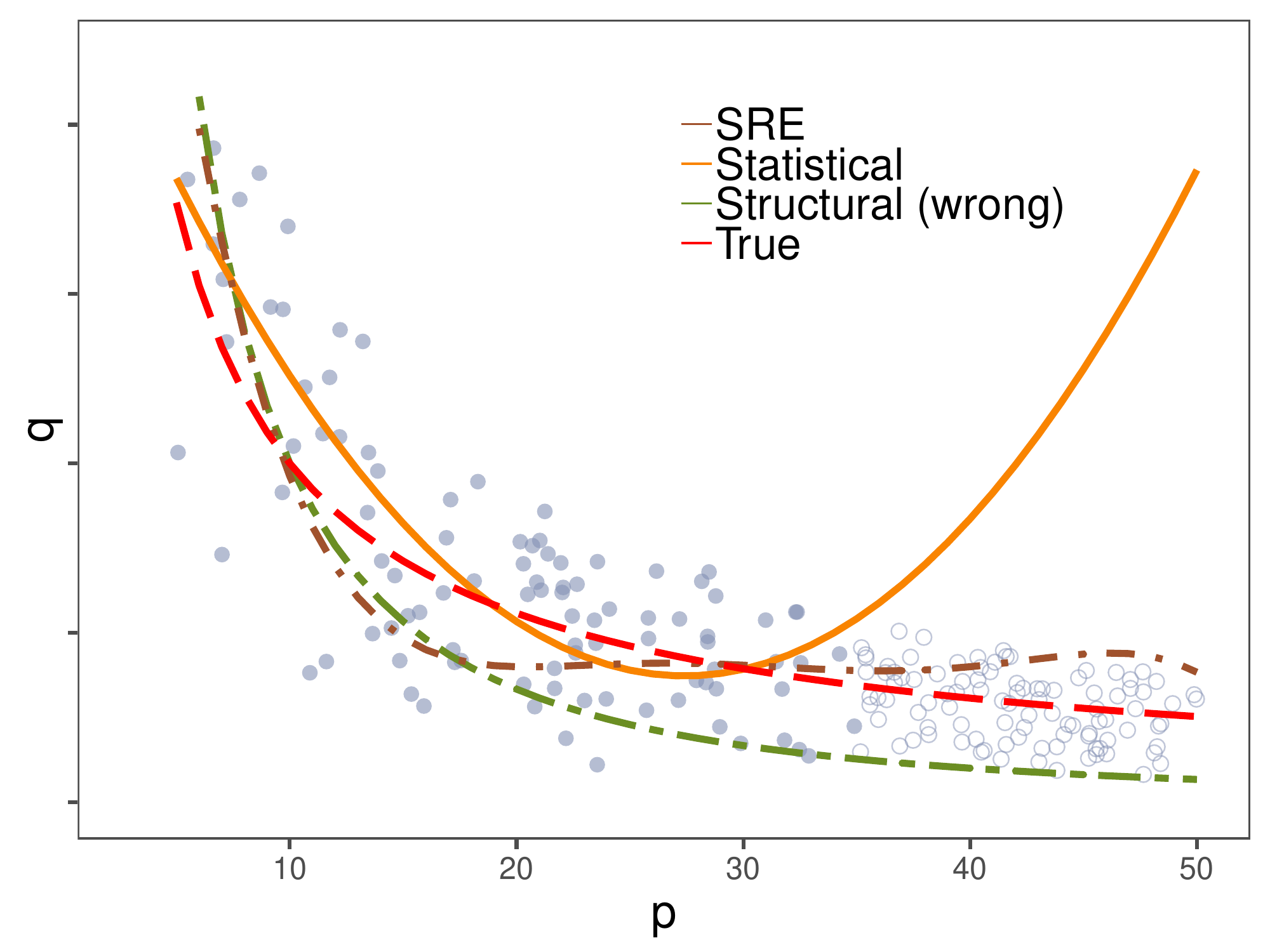}

}

\caption{{\small{}Demand Estimation. Dots represent training data. Circles
represent out-of-domain test data.}}
\end{figure}

As a motivating example, consider a simple demand estimation problem.
We observe the prices and quantities sold of a good $x$, as plotted
in Figure \ref{fig:consumption_a}. Suppose the data are generated
by the consumption decisions of $N$ consumers who purchased $x$
at different prices. Each consumer had fixed income $I$ and decided
how much to purchase $x$ by solving the problem
\begin{equation}
\max_{q,q^{o}}u_{i}\left(q,q^{o}\right)\quad\text{subject to }pq+p^{o}q^{o}\le I\label{eq:motivate}
\end{equation}
, where $\left(p,q,p^{o},q^{o}\right)$ denote respectively the price
and quantity of $x$ and of an outside good $o$, and 
\[
u_{i}\left(q,q^{o}\right)=\left[\alpha_{i}q^{\rho}+\left(1-\alpha_{i}\right)\left(q^{o}\right)^{\rho}\right]^{\frac{1}{\rho}}
\]
, with $\rho=-\frac{1}{2}$, implying an elasticity of substitution
equal to $0.67$\footnote{$\left\{ \alpha_{i}\right\} $ are generated as follows: 
\[
\alpha_{i}=\frac{\exp\left(\xi_{i}\right)}{1+\exp\left(\xi_{i}\right)},\quad\xi_{i}\sim\mathcal{N}\left(0,0.5\right)
\]
}. 

We can fit the following statistical model to the data:
\begin{equation}
q_{i}=\beta_{0}+\beta_{1}p_{i}+\beta_{2}p_{i}^{2}+\epsilon_{i}\label{eq:appetizier02}
\end{equation}

The result is plotted in Figure \ref{fig:consumption_b}. If we further
make the causal assumption that changes in $p$ are exogenous, then
\eqref{eq:appetizier02} represents a reduced-form estimate of the
individual demand curve. 

The model fits the data well\footnote{In practice, \eqref{eq:appetizier02} is selected from a set of nested
polynomial models based on AIC.} and would suffice if our goal is to make sales predictions \emph{in-domain}
or obtain an \emph{internally valid} demand curve estimate. However,
the fit becomes bad once we extrapolate outside the observed range
of prices, as shown in Figure \ref{fig:consumption_c}. More sophisticated
statistical and machine learning models wouldn't help. In particular,
domain adaptation methods do not apply since both the marginal distribution
of $p$ and the conditional distribution of $q$ change once we extrapolate
outside the observed domain. 

On the other hand, structurally estimating model \eqref{eq:motivate}
would yield an estimated curve that has both internal and external
validity (Figure \ref{fig:consumption_c}). This is not surprising
since \eqref{eq:motivate} describes the true data-generating mechanism.
What happens if we estimate a structural model that is incorrectly
specified? Figure \ref{fig:consumption_d} shows the result of estimating
\eqref{eq:motivate} but assuming $\rho=0.5$\footnote{That is, instead of estimating both $\left(\alpha_{i},\rho\right)$
from the data, we estimate $\alpha_{i}$ only while treating $\rho=0.5$
as an assumption of the model. The assumption, of course, is incorrect
in this case.}. The structural fit is now poor both in-domain and out-of-domain,
highlighting the fact that the validity of the structural approach
hinges crucially on the model being correct. 

Our structural regularization approach offers a way to combine statistical
and structural models to address the shortcomings of each. Figure
\ref{fig:consumption_d} also shows the result of structural regularization
using the misspecified structural model as the benchmark. Compared
to the structural fit, the SRE fit is closer to the true model both
in-domain and out-of-domain. Compared to the statistical fit, the
SRE performs slightly worse in-domain but significantly better out-of-domain\footnote{To generate the results of this example, we shrink a $5-$ degree
polynomial toward the structural benchmark. Note that if we compare
the resulting SRE fit with this $5-$degree polynomial \emph{in-domain},
as in \citet{fessler_how_2018}, the SRE fit will always perform no
worse.}. While we formally present the structural regularization method in
the next section, this example helps illustrate why a misspecified
structural model can be useful: although we misspecify the utility
function, the assumption of consumer utility maximization subject
to budget constraints provides important information on the relationship
between price and demand that can be used to regulate the behavior
of statistical models. The SRE is therefore able to achieve a balance
between producing accurate descriptions of the data and incorporating
theoretical (economic/behavioral) insight that allows it to better
extrapolate beyond the observed domain. 

\section{Methodology\label{sec:Methodology}}

In this section, we first lay out our method in the context of conditional
mean estimation. We then present it in the general framework of penalized
extremum estimation and show how it can used to fit quantities identified
via moment conditions. In each case, we discuss how our method can
be used both for statistical prediction and causal inference. 

\subsection{Overview\label{subsec:Overview}}

We begin by considering the following statistical prediction problem:
given variables $\left(x,y\right)\in\mathcal{O}$, let $\mathbb{P}_{xy}$
be a joint distribution defined on $\mathcal{O}$ that governs $\left(x,y\right)$.
Our goal is to learn a target function $f\left(x\right)$ that minimizes
the expected $\ell_{2}$ loss $\mathbb{E}_{\mathbb{P}_{xy}}\left[\left(y-f\left(x\right)\right)^{2}\right]$.
Equivalently, we are interested in estimating the conditional expectation
function $f\left(x\right)=\mathbb{E}_{\mathbb{P}_{y|x}}\left[y|x\right]$.
We may not have access, however, to a random sample from $\mathbb{P}_{xy}$.
Instead, we observe data $\mathcal{D}=\left\{ \left(x_{i},y_{i}\right)\right\} _{i=1}^{N},\ \left(x_{i},y_{i}\right)\in\mathcal{O}'\subseteq\mathcal{O}$,
with data-generating probability distribution $\mathbb{P}'_{xy}=\left.\mathbb{P}_{xy}\right|_{\left(x,y\right)\in\mathcal{O}'}$.
Classic statistical and machine learning algorithms built on the assumption
that the training data is a random sample of the distribution of interest
will thus have difficulty learning $f$ from $\mathcal{D}$. 

We assume that we have at our disposal an identifiable structural
model $\mathcal{M}$ that we believe \emph{may} describe the causal
mechanism that generates $\mathbb{P}_{xy}$. However, the model may
also be misspecified. With this setup, our structural regularization
estimator proceeds in two stages. In the first stage, we estimate
the structural model $\mathcal{M}$ on the data $\mathcal{D}$ to
obtain $\widehat{\mathcal{M}}$. We then use $\widehat{\mathcal{M}}$
to generate \emph{synthetic }data $\mathcal{D}^{\mathcal{M}}=\left\{ \left(x_{i}^{\mathcal{M}},y_{i}^{\mathcal{M}}\right)\right\} _{i=1}^{M},\ \left(x_{i}^{\mathcal{M}},y_{i}^{\mathcal{M}}\right)\in\mathcal{O}$.
This is generally feasible since structural models are \emph{generative}
models capable of simulating new data and since $\mathcal{M}$ is
a causal model for $\mathbb{P}_{xy}$, it can be used to simulate
data on the entire domain $\left\langle \mathcal{O},\mathbb{P}_{xy}\right\rangle $
rather than on $\left\langle \mathcal{O}',\mathbb{P}'_{xy}\right\rangle $
only\footnote{In practice, this means that if we know, at the time of estimation,
where we want to apply our model, i.e. the target domain input space,
then we can use the estimated structural model to generate synthetic
data on the target domain in addition to the source domain in this
first stage.}. Based on the estimated model $\mathcal{\widehat{M}}$, we can also
compute $f^{\mathcal{M}}\left(x\right)=\mathbb{E}^{\mathcal{M}}\left[y|x\right]$
-- the \emph{implied} conditional expectation of $y$ according to
$\widehat{\mathcal{M}}$.

In the second stage, we estimate a flexible statistical model $g\left(x;\theta\right)$
by seeking solution to the following problem:
\begin{equation}
\underset{\theta\in\Theta}{\min}\left\{ \sum_{i=1}^{N}\left(y_{i}-g\left(x_{i};\theta\right)\right)^{2}+\lambda\cdot\Omega\left(\theta,\widehat{\theta}^{\mathcal{M}}\right)\right\} \label{eq:main}
\end{equation}
, where $\Omega\left(.,.\right)$ is a distance function, $\lambda\ge0$
is a penalty parameter, and $\widehat{\theta}^{\mathcal{M}}$ is obtained
by fitting $g$ to $\mathcal{D}^{\mathcal{M}}$, i.e.\footnote{In practice, to avoid overfitting, one can either generate a very
large synthetic set $\left(x_{i}^{\mathcal{M}},y_{i}^{\mathcal{M}}\right)$
or fit $g$ directly to $\left(x_{i}^{\mathcal{M}},f^{\mathcal{M}}\left(x_{i}^{\mathcal{M}}\right)\right)$,
where $x_{i}^{\mathcal{M}}$ belongs to a grid of possible values
of $x$.},
\begin{align}
\widehat{\theta}^{\mathcal{M}} & =\underset{\theta\in\Theta}{\arg\min\ }\sum_{i=1}^{M}\left(y_{i}^{\mathcal{M}}-g\left(x_{i}^{\mathcal{M}};\theta\right)\right)^{2}\label{eq:thetafit}\\
 & =\underset{\theta\in\Theta}{\arg\min\ }\sum_{i=1}^{M}\left(f^{\mathcal{M}}\left(x_{i}^{\mathcal{M}}\right)-g\left(x_{i}^{\mathcal{M}};\theta\right)\right)^{2}\label{eq:thetafit2}
\end{align}

$g\left(x;\widehat{\theta}^{\mathcal{M}}\right)$ represents a statistical
approximation to the model derived conditional mean, $f^{\mathcal{M}}\left(x\right)$\footnote{The method of indirect inference \citep{gourieroux_indirect_1993},
widely used for fitting structural models whose complexity makes direct
likelihood evaluation difficult, also relies on the use of approximating
statistical models generated by fitting to synthetic data.}$^{,}$\footnote{Thus, in contrast to \citet{fessler_how_2019}, our method does not
require the structural model $\mathcal{M}$ to be expressed as a set
of constraints on the parameters of a statistical model $g\left(x;\theta\right)$.}. $\widehat{\theta}^{\mathcal{M}}$ can therefore be viewed as the
``structural'' or ``theoretical'' value of $\theta$. The term
$\Omega\left(\theta,\widehat{\theta}^{\mathcal{M}}\right)$ is a regularizer
that penalizes the distance between $\theta$ and $\widehat{\theta}^{\mathcal{M}}$.
Typical choices for $\Omega\left(.,.\right)$ include the $\ell_{1}$
norm, $\Omega\left(\theta,\widehat{\theta}^{\mathcal{M}}\right)=\left\Vert \theta-\widehat{\theta}^{\mathcal{M}}\right\Vert _{1}$,
or the squared $\ell_{2}$ norm, $\Omega\left(\theta,\widehat{\theta}^{\mathcal{M}}\right)=\left\Vert \theta-\widehat{\theta}^{\mathcal{M}}\right\Vert _{2}^{2}$. 

The penalty, or \emph{tuning}, parameter $\lambda$ controls the tradeoff
between goodness of in-sample fit and deviance of the statistical
model from its structural counterpart. Let $\widehat{\theta}$ be
the solution to problem \eqref{eq:main}. Then $g\left(x;\widehat{\theta}\right)$
is our SRE estimate of $f\left(x\right)$. At one extreme, when $\lambda=0$,
$g\left(x;\widehat{\theta}\right)$ is a completely unregularized
statistical fit. At the other extreme, as $\lambda\rightarrow\infty$,
$g\left(x;\widehat{\theta}\right)$ approaches its structural counterpart.
It is in this sense that the SRE sits in the interior of the continuum
between statistical and structural estimation. 

As in standard penalized regression models, \eqref{eq:main} is known
as the \emph{Tikhonov} form. Equivalently, it can be expressed in
the \emph{Ivanov} form:
\begin{align}
 & \underset{\theta\in\Theta}{\min}\ \sum_{i=1}^{N}\left(y_{i}-g\left(x_{i};\theta\right)\right)^{2}\label{eq:main-ivanov}\\
 & \text{\ \ \ \ \ \ \ subject to }\Omega\left(\theta,\widehat{\theta}^{\mathcal{M}}\right)\le C\nonumber 
\end{align}
, which makes it transparent that we can likewise think of the SRE
as selecting the best statistical model to fit the data within a neighborhood
$C$ of the structural benchmark.

In practice, the choice of $\lambda$ or $C$ are determined via cross-validation.
To improve the out-of-domain performance of our estimator, in addition
to standard cross-validation procedures for \emph{i.i.d.} data, we
propose a \emph{forward cross-validation} procedure that can be useful
in situations in which we know the target domain input space at the
time of training. In addition, to avoid overfitting due to first stage
structural estimation, we adopt a \emph{sample-splitting} strategy
that splits the available training data into independent sets for
first and second stage estimation. These and other details of implementation
are given in section \ref{subsec:Implementation}.

\subsection{Bayesian Interpretation}

\begin{figure}
\begin{centering}
\includegraphics[width=0.55\columnwidth]{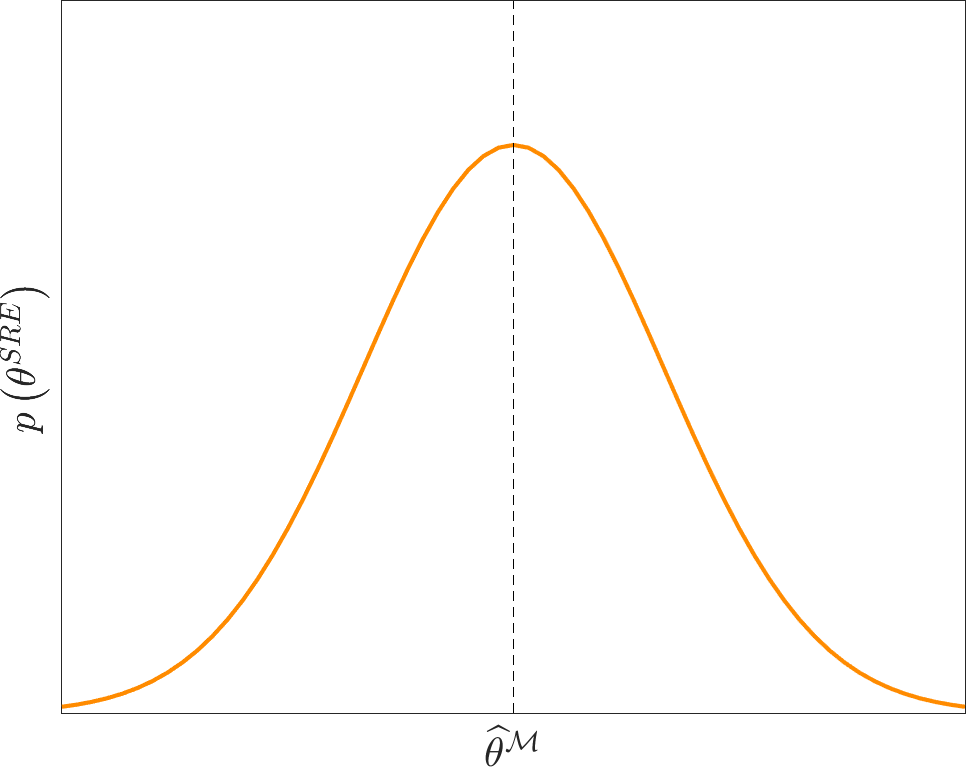}
\par\end{centering}
\caption{{\small{}Gaussian Prior for SRE ($\ell_{2}$ regularizer)\label{fig:prior}}}
\end{figure}

The SRE permits a Bayesian interpretation\footnote{Our method has a proper Bayesian interpretation due to our sample-splitting
strategy (section \ref{subsec:Implementation}) that separates the
training data used for first and second stage estimation. Thus, from
the perspective of second stage estimation, $\widehat{\theta}^{\mathcal{M}}$
is exogenously given, so that the prior distribution centered around
it does not depend on the data.}. Specifically, in the case of $\ell_{2}$ regularization, $\widehat{\theta}$
is the posterior mode of $\theta$ under a Gaussian prior centered
around $\widehat{\theta}^{\mathcal{M}}$\footnote{See, e.g. \citet{james_introduction_2013}. For $\ell_{1}$ regularization,
the corresponding prior is a double-exponential (Laplace) distribution
\citep{tibshirani_regression_1996}. \citet{murphy_machine_2012}
provides more general discussions on the connection between regularization
and MAP (\emph{maximum à posteriori}) Bayesian inference.}. This is illustrated in Figure \ref{fig:prior} for a one-dimensional
parameter. The standard deviation of the prior distribution is inversely
proportional to regularization strength -- the smaller the standard
deviation, the larger the corresponding $\lambda$ is and the more
confidence is placed on $\widehat{\theta}^{\mathcal{M}}$ being the
``true value.'' As in \citet{fessler_how_2019}, the use of informative
priors centered around theoretically derived values gives the resulting
estimator an appealing understanding of using theory as prior knowledge
for analyzing new evidence\footnote{Note that the estimator obtained by \citet{fessler_how_2019} is the
posterior mean rather than the posterior mode.}. 

\subsection{Causal Inference}

\paragraph{Causal Effect Estimation under Unconfoundedness}

In this section, we adapt the estimator introduced in section \ref{subsec:Overview}
to the problem of causal effect estimation under unconfoundedness.
Let the observed variables be $\left(y,d,w\right)\in\mathbb{R}\times\mathbb{R}\times\mathcal{W}$,
where $y$ is the outcome variable, $d$ is the treatment variable,
and $w$ is a set of control variables. We are interested in the causal
effect of $d$ on $y$. Specifically, let our target be the average
treatment effect (ATE) denoted by $\tau$. We allow $\tau$ to be
fully nonlinear and heterogeneous, i.e. $\tau=\tau\left(d,w\right)$.
Under the unconfoundedness assumption of \citet{rosenbaum_central_1983}\footnote{Using the notations of the Rubin causal model \citep{rubin_estimating_1974},
suppose the treatment variable $d$ takes on a discrete set of values,
$d\in\left\{ 1,\ldots,D\right\} $, then the \emph{unconfoundedness}
-- or \emph{conditional exchangeability} -- assumption can be stated
as \[\left.d\indep\left(y\left(1\right),\ldots,y\left(D\right)\right)\right|w\],
where $y\left(d\right)$ is the \emph{potential outcome} of $y$ associated
with treatment $d$. This assumption is satisfied if $d$ is not associated
with any other causes of $y$ conditional on $w$. A more precise
statement on the sufficient conditions for satisfying this assumption,
made in the language of causal graphical models based on directed
acyclic graphs (DAGs), is that $w$ satisfies the \emph{back-door
criterion} \citep{pearl_causality_2009}.},
\begin{equation}
\tau\left(d,w\right)=\frac{\partial}{\partial d}\mathbb{E}\left[y|d,w\right]\label{eq:ATE}
\end{equation}

Let $x=\left(d,w\right)$. The task of estimating $\tau\left(d,w\right)$
is thus equivalent to the task of estimating $\mathbb{E}\left[y|x\right]$.
Suppose now that we have a structural model $\mathcal{M}$ that describes
the causal mechanism that generates these variables\footnote{Importantly, $\mathcal{M}$ does not have to support the unconfoundedness
assumption, i.e. in the causal structure assumed by $\mathcal{M}$
, $w$ does not have to satisfy the back-door criterion. This is because
the structural model is used to aid the estimation of $\mathbb{E}\left[y|d,w\right]$.
The identifying assumption required for interpreting $\left.\partial\mathbb{E}\left[y|d,w\right]\right/\partial d$
as the (conditional) ATE remains that of unconfoundedness.}, then we can use the SRE to produce $g\left(x;\widehat{\theta}\right)=\widehat{\mathbb{E}}\left[y|x\right]$,
from which we can derive $\widehat{\tau}\left(d,w\right)$\footnote{Technically, $\tau\left(d,w\right)$ is the conditional ATE. With
a slight abuse of notation, the population ATE $\tau\left(d\right)=\mathbb{E}_{w}\left[\tau\left(d,w\right)\right]$.}.

As the preceding discussion shows, when the goal is to estimate the
ATE under unconfoundedness, the difference between the reduced-form
statistical approach and the structural approach boils down to a difference
in the choice of the functional form of $\mathbb{E}\left[y|x\right]$,
with the former traditionally relying on simple linear models --
although recent studies increasingly adopt more complex nonlinear
and adaptive machine learning models, while the latter derive the
functional form from theory. In a sense, one can argue that critics
on either side of the methodological debate are motivated by a shared
concern over model misspecification. Advocates for the reduced-form
approach are concerned about the misspecification of $\mathbb{E}\left[y|x\right]$
due to the often strong and unrealistic assumptions -- causal as
well as parametric -- made in structural models, while those advocating
for the structural approach are concerned about misspecifications
due to not incorporating theoretical insight -- functional forms
such as constant elasticity of substitution (CES) aggregation and
the gravity equation of trade often encode important prior economic
knowledge that sophisticated statistical and machine learning methods
would not be able to capture based on training data alone\footnote{\citet{rust_limits_2014} makes a similar point: \textquotedblleft Notice
the huge difference in world views. The primary concern of Leamer,
Manski, Pischke, and Angrist is that we rely too much on assumptions
that could be wrong, and which could result in incorrect empirical
conclusions and policy decisions. Wolpin argues that assumptions and
models could be right, or at least they may provide reasonable first
approximations to reality.\textquotedblright{}}. The SRE addresses both of these concerns: our two-stage procedure
effectively searches through a combined statistical and structural
model space to arrive at an optimal functional form of $\mathbb{E}\left[y|x\right]$
that defends against both types of misspecifications. 

\paragraph{Instrumental Variables}

When the unconfoundedness condition does not hold -- when there is
\emph{unmeasured} confounding -- one of the most widely used strategies
in reduced-form inference is to rely on the use of instrumental variables,
which are auxiliary sources of randomness that can be used to identify
causal effects. Let our reduced-form statistical model be $y=g\left(x;\theta\right)+\epsilon,\ x=\left(d,w\right)$,
where $\tau\left(d,w\right)=\left.\partial g\left(x;\theta\right)\right/\partial d$
and $\epsilon$ is a noise term that may be correlated with $d$\footnote{In this reduced-form model, $g\left(x;\theta\right)$ is a statistical
model for $\mathbb{E}\left[\left.y\left(d\right)\right|w\right]$
-- the conditional expectation of the potential outcome of $y$ under
treatment $d$ and $\epsilon$ is \emph{defined} as $y-g\left(x;\theta\right)$.
Thus by definition, the conditional ATE $\tau\left(d,w\right)=\left.\partial g\left(x;\theta\right)\right/\partial d$.
When $\mathbb{E}\left[d\epsilon\right]\ne0$, the received treatment
$d$ is related to unobserved factors that also affect $y$, thus
violating the unconfoundedness condition.}. If we have access to a variable $z$ that is correlated with treatment
$d$ and is related to outcome $y$ only through its association with
$d$, then $z$ can serve as an instrument for $d$\footnote{More precisely, the requirement is that $\mathbb{E}\left[z\epsilon\right]=0$
and $\rho_{dz\cdot w}\ne0$, where $\rho_{zd\cdot w}$ is the partial
correlation of $z$ and $d$ given $w$. On a causal graph, this translates
into the requirement that $z$ is correlated with $d$ and that every
open path connecting $z$ and $y$ has an arrow pointing into $d$.}. In general, given $\theta\in\mathbb{R}^{k}$ and instrument $z\in\mathbb{R}^{l},\ l\ge k$,
$\theta$ can be identified via the following moment conditions:
\begin{equation}
\mathbb{E}\left[z\left(y-g\left(x;\theta\right)\right)\right]=0\label{eq:ivmoment}
\end{equation}

Assume again that we have a structural model $\mathcal{M}$ that describes
the causal mechanism governing these variables\footnote{$\mathcal{M}$ does not have to contain $z$. Once we have an estimated
model $\widehat{\mathcal{M}}$, we can use it to generate and fit
$g$ directly to a synthetic data set $\left(x_{i}^{\mathcal{M}},g^{\mathcal{M}}\left(x_{i}^{\mathcal{M}}\right)\right)$
to obtain $\widehat{\theta}^{\mathcal{M}}$, where $g^{\mathcal{M}}\left(x_{i}^{\mathcal{M}}\right)=\mathbb{E}^{\mathcal{M}}\left[\left.y\left(d_{i}\right)\right|w_{i}\right]$
is the model derived conditional expectation of the potential outcome
under treatment $d$.}. Our SRE would proceed as before in the first stage and solve the
following problem in the second stage:
\begin{equation}
\underset{\theta\in\Theta}{\min}\left\{ \overline{m}\left(\theta\right)'W\overline{m}\left(\theta\right)+\lambda\cdot\Omega\left(\theta,\widehat{\theta}^{\mathcal{M}}\right)\right\} \label{eq:GMM-main}
\end{equation}
, where $\overline{m}\left(\theta\right)=\frac{1}{N}\sum_{i=1}^{N}m_{i}\left(\theta\right)$,
$m_{i}\left(\theta\right)=z_{i}\left(y_{i}-g\left(x_{i};\theta\right)\right)$
are the moment functions and $W$ is a $l\times l$ weight matrix.
Once we obtain $\widehat{\theta}$ as a solution to \eqref{eq:GMM-main},
we can derive the conditional ATE as $\widehat{\tau}\left(d,w\right)=\left.\partial g\left(x;\widehat{\theta}\right)\right/\partial d$. 

\subsection{Implementation\label{subsec:Implementation}}

\begin{algorithm}[t] 
{\fontsize{11}{13}\selectfont  
\renewcommand{\algorithmicloop}{\textbf{procedure:}}
\caption{Structural Regularization with Sample-Splitting}
\label{alg:sp}
\ 
\begin{algorithmic}[1]
\Require Observed data $\mathcal{D}$ \medskip{}
\State samples $\left\{\mathcal{D}_{1},\mathcal{D}_{2}\right\} \leftarrow$ \textsc{Partition}$\left(\mathcal{D},K=2\right)$
\State \textbf{output} $\widehat{\theta} \leftarrow$ \textsc{StructuralRegularization}$\left(\mathcal{D}_{1},\mathcal{D}_{2}\right)$
\end{algorithmic} 
\medskip{}
The function \textsc{Partition} randomly partitions a sample into $K$ equal sized parts. The function \textsc{StructuralRegularization} takes in two  data samples and uses them to produce the SRE estimates as follows:
\medskip{}
\begin{algorithmic}[1]
\Loop \ \textsc{StructuralRegularization}$\left(\text{sample }\mathcal{I},\text{ sample }\mathcal{J}\right)$
\State fit the structural model $\mathcal{M}$ on sample $\mathcal{I}$ to obtain $\widehat{\mathcal{M}}$
\State use $\widehat{\mathcal{M}}$ to generate a synthetic data set $\mathcal{D}^{\mathcal{M}}$
\State solve problem \eqref{eq:thetafit} on $\mathcal{D}^{\mathcal{M}}$ to obtain $\widehat{\theta}^{\mathcal{M}}$
\State substitute $\widehat{\theta}^{\mathcal{M}}$ into problem \eqref{eq:main}
\State solve problem \eqref{eq:main} on sample $\mathcal{J}$ for a grid of $\lambda$ values and find the optimal $\lambda^{*}$ by cross-validation \label{alg:cv}
\State \Return $\widehat{\theta}$ as the solution to problem \eqref{eq:main} on sample $\mathcal{J}$ at $\lambda=\lambda^{*}$
\EndLoop
\end{algorithmic} 
}
\end{algorithm}

In this section, we detail the implementation of our algorithm. We
begin by showing that under the setup of section \ref{subsec:Overview},
our estimator has a closed form solution at any given $\lambda$ in
the special case of $\Omega\left(.,.\right)$ being an $\ell_{2}$
regularizer and $g\left(x;\theta\right)$ being linear in $\theta$. 

Consider $g\left(x;\theta\right)=\alpha+x'\beta$, where $x\in\mathbb{R}^{p}$,
$\theta=\left(\alpha,\beta\right)$. In practice, the constant term
$\alpha$ should not be penalized. Let $x$ be  standardized into
$\widetilde{x}$ with mean zero. Then we can write our model as $g\left(\widetilde{x};\widetilde{\theta}\right)=\widetilde{\alpha}+\widetilde{x}'\widetilde{\beta},\ \widetilde{\theta}=\left(\widetilde{\alpha},\widetilde{\beta}\right)$.
We estimate $g\left(\widetilde{x};\widetilde{\theta}\right)$ as follows:
in the first stage, after generating synthetic data $\mathcal{D}^{\mathcal{M}}$
based on the estimated structural model $\widehat{\mathcal{M}}$,
fitting $g\left(\widetilde{x};\widetilde{\theta}\right)$ to $\mathcal{D}^{\mathcal{M}}$
gives $\left(\widehat{\widetilde{\alpha}}^{\mathcal{M}},\widehat{\widetilde{\beta}}^{\mathcal{M}}\right)$.
In the second stage, because $\widetilde{x}$ is centered, we have
$\widehat{\widetilde{\alpha}}=\overline{y}$. Let $\widetilde{y}=y-\overline{y}$.
Let $\Omega\left(\widetilde{\beta},\widehat{\widetilde{\beta}}^{\mathcal{M}}\right)=\left\Vert \widetilde{\beta}-\widehat{\widetilde{\beta}}^{\mathcal{M}}\right\Vert _{2}^{2}$.
Then
\begin{align}
\widehat{\widetilde{\beta}} & =\text{argmin}_{\widetilde{\beta}}\left\{ \left\Vert \widetilde{y}-\widetilde{X}\widetilde{\beta}\right\Vert _{2}^{2}+\lambda\left\Vert \widetilde{\beta}-\widehat{\widetilde{\beta}}^{\mathcal{M}}\right\Vert _{2}^{2}\right\} \label{eq:sol}\\
 & =\left(\widetilde{X}'\widetilde{X}+\lambda I\right)^{-1}\left(\widetilde{X}'\widetilde{y}+\lambda\widehat{\widetilde{\beta}}^{\mathcal{M}}\right)\label{eq:closesol}
\end{align}
, where $\widetilde{X}=\left[\widetilde{x}_{1},\ldots,\widetilde{x}_{N}\right]'$
and $I$ is the $p\times p$ identity matrix. In the case that $\widetilde{X}$
is orthonormal, \eqref{eq:closesol} can be expressed as
\begin{equation}
\widehat{\widetilde{\beta}}=\frac{1}{1+\lambda}\widehat{\widetilde{\beta}}^{OLS}+\frac{\lambda}{1+\lambda}\widehat{\widetilde{\beta}}^{\mathcal{M}}\label{eq:interpol}
\end{equation}
, where $\widehat{\widetilde{\beta}}^{OLS}$ is the least squares
estimate. In this case, the SRE can be viewed as a weighted average
of statistical and structural estimation.

\paragraph{Sample Splitting}

\begin{algorithm}[t] 
{\fontsize{11}{13}\selectfont  
\renewcommand{\algorithmicloop}{\textbf{procedure:}}
\caption{Structural Regularization with Cross-Fitting}
\label{alg:cf}
\ 
\begin{algorithmic}[1]
\State samples $\left\{\mathcal{D}_{1},\mathcal{D}_{2}\right\} \leftarrow$ \textsc{Partition}$\left(\mathcal{D},K=2\right)$
\State $\widehat{\theta}_{1} \leftarrow$ \textsc{StructuralRegularization}$\left(\mathcal{D}_{1},\mathcal{D}_{2}\right)$
\State $\widehat{\theta}_{2} \leftarrow$ \textsc{StructuralRegularization}$\left(\mathcal{D}_{2},\mathcal{D}_{1}\right)$
\State \textbf{output} $\widehat{\theta}\leftarrow\frac{1}{2}\left(\widehat{\theta}_{1}+\widehat{\theta}_{2}\right)$
\end{algorithmic} 
}
\end{algorithm}

We use the technique of \emph{sample-splitting} \citep{angrist_split-sample_1995}
to avoid overfitting and ensure good statistical behavior especially
when complex structural models are employed in the first stage\footnote{\citet{angrist_split-sample_1995} propose the use of sample-splitting
in the context of instrumental variable estimation. Related ideas
in the statistical literature goes back at least to \citet{bickel_adaptive_1982}.}. The idea of sample-splitting is to split the training data into
two parts to be used respectively for the two stages of estimation,
so that $\widehat{\theta}^{\mathcal{M}}$ can be treated as exogenously
given when $g\left(x;\theta\right)$ is fit in the second stage. The
details of our algorithm with sample-splitting are given in Algorithm~\ref{alg:sp}. 

The sample-splitting procedure reduces overfitting at a cost of wasting
half of the data in each stage of estimation. To improve efficiency,
we can use the \emph{cross-fitting} procedure of \citet{chernozhukov_double_2016,chernozhukov_doubledebiasedneyman_2017}.
The idea is to similarly split the original sample into two parts,
but alternately use each part for first and second stage estimation,
so that each data point will participate in both stages albeit not
at the same time. The details of our algorithm with cross-fitting
are given in Algorithm~\ref{alg:cf}. 

\paragraph{Forward Cross-Validation}

\begin{algorithm}[t] 
{\fontsize{11}{13}\selectfont  
\renewcommand{\algorithmicloop}{\textbf{procedure:}}
\caption{K-Fold Forward Cross-Validation}
\label{alg:fcv}
\medskip{}
\textsc{FowardCV} is a subroutine for performing cross-validation in \textsc{StructuralRegularization}
\begin{algorithmic}[1]
\Loop \ \textsc{FowardCV}$\left(\text{sample }\mathcal{S},K\right)$
\State samples $\ensuremath{\left\{ \mathcal{S}_{1},\mathcal{S}_{2}\right\} \leftarrow}\textsc{FowardSplit}\ensuremath{\left(\mathcal{S}\right)}$
\State folds $\left\{\mathcal{I}_{1},\ldots,\mathcal{I}_{K}\right\} \leftarrow$ \textsc{Partition}$\left(\mathcal{S}_{1},K\right)$  
\ForAll{$\lambda \in \Lambda$}
\For{$k = 1:K$} 
\State validation set $\mathcal{V}_{k} \leftarrow \mathcal{I}_{k}\cup\mathcal{S}_{2}$
\State training set $\mathcal{T}_{k} \leftarrow \left.\mathcal{S}_{1}\right\backslash \mathcal{I}_{k}$
\State solve problem \eqref{eq:main} on $\mathcal{T}_{k}$ to obtain $\widehat{\theta}_{k}\left(\lambda\right)$ 
\State compute the predicton error of $g\left(x;\widehat{\theta}_{k}\left(\lambda\right)\right)$ on $\mathcal{V}_{k}$ to obtain validation error $e_{k}\left(\lambda\right)$
\EndFor
\EndFor
\State \Return optimal tuning parameter $\lambda^{*} \leftarrow \underset{\lambda\in\Lambda}{\arg\min\ }\frac{1}{K}\sum_{k=1}^{K}e_{k}\left(\lambda\right)$
\EndLoop
\end{algorithmic} 
$\Lambda$ is a grid of $\lambda$ values. The function \textsc{FowardSplit} randomly partitions $\mathcal{S}$ into $\left\{ \mathcal{S}_{1},\mathcal{S}_{2}\right\}$, satisfying the following condition: let $\mathcal{X}_{1}$, $\mathcal{X}_{2}$, and $\mathcal{X}_{T}$ be compact input spaces associated respectively with $\mathcal{S}_{1}$, $\mathcal{S}_{2}$ and the target domain, then $d_{H}\left(\mathcal{X}_{2},\mathcal{X}_{T}\right)<d_{H}\left(\mathcal{X}_{1},\mathcal{X}_{T}\right)$, where $d_{H}\left(.,.\right)$ denotes the Hausdorff distance.
\medskip{}
}
\end{algorithm}

We use cross-validation to choose the optimal penalty $\lambda$ in
\eqref{eq:main}. If we know, at the time of estimation, the target
domain on which we want to apply our model, then there are two ways
to further improve the out-of-domain performance of our estimator.
One is to use the estimated structural model to generate synthetic
data on both the source and the target domain in the first stage,
as discussed in section \ref{subsec:Overview}. In this section, we
introduce a \emph{forward cross-validation} procedure as another way
to improve out-of-domain performance. The idea is to validate on subsets
of the data that are ``closer'' to the target domain than the subsets
on which the model is trained. More specifically, given a sample $\mathcal{S}$\footnote{In practice, $\mathcal{S}$ would be the subsample of $\mathcal{D}$
on which the second stage estimation is conducted.}, we partition $\mathcal{S}$ into two parts, $\mathcal{S}_{1}$ and
$\mathcal{S}_{2}$ with associated input spaces $\mathcal{X}_{1}$
and $\mathcal{X}_{2}$, such that $d_{H}\left(\mathcal{X}_{2},\mathcal{X}_{T}\right)<d_{H}\left(\mathcal{X}_{1},\mathcal{X}_{T}\right)$,
where $\mathcal{X}_{T}$ is the target domain input space and $d_{H}\left(.,.\right)$
is the Hausdorff distance. We then further partition $\mathcal{S}_{1}$
randomly into $K-1$ equal sized subsets and perform $K-$ fold cross
validation, each time using $K-2$ subsets of $\mathcal{S}_{1}$ for
training and validating on a validation set that contains $\mathcal{S}_{2}$
and the remaining subset of $\mathcal{S}_{1}$. See Algorithm~\ref{alg:fcv}
for more details.

\begin{figure}
\begin{centering}
\includegraphics[width=0.9\columnwidth]{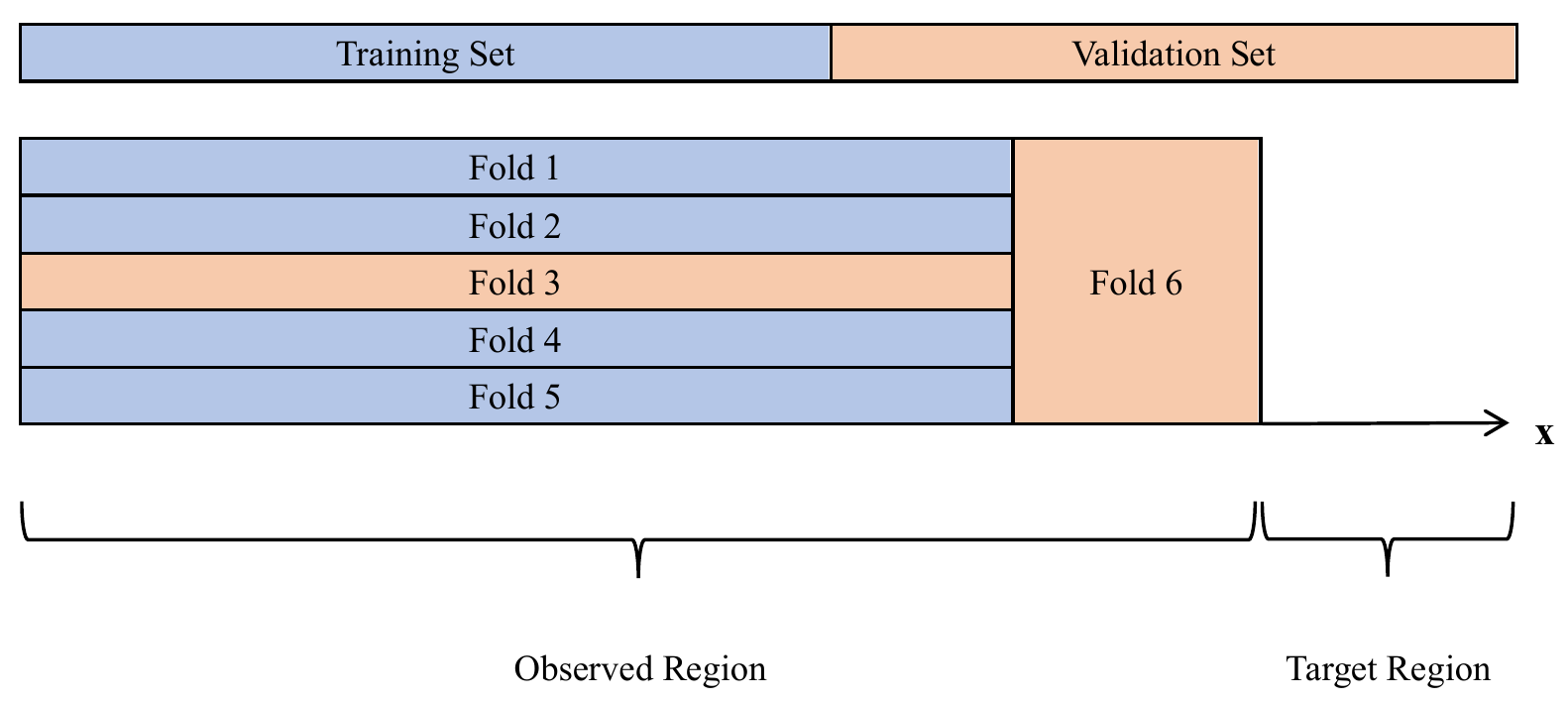}
\par\end{centering}
\caption{{\small{}Forward Cross-Validation. Illustrated is one iteration of
the procedure, in which the folds in blue are used for training and
the folds in brown are used for validation. While fold $1-5$ are
iteratively used for training and validation, fold $6$ is }\emph{\small{}always}{\small{}
used for validation.\label{fig:fcv}}}
\end{figure}

The idea of forward CV is perhaps best illustrated in the one-dimensional
setting (Figure \ref{fig:fcv}). Here we want to extrapolate the estimated
model in the direction of increasing $x$. To this end, we perform
cross validation by creating a six-fold partition of the sample data,
where the sixth fold lies in the direction of increasing $x$ compared
to the remaining five and is \emph{always} in the validation set.
Doing so helps produce tuning parameters whose corresponding models
have superior extrapolation performance in the intended direction\footnote{When the data has a time series structure, a rolling-window design
can be used for cross-validation, as is commonly used for model selection
in time series forecasting. See section \ref{subsec:DDCM} for an
application.}. 

\paragraph*{Adaptive Models }

The statistical model that we shrink toward the structural benchmark
can be adaptive itself, allowing the potential use of machine learning
methods like random forests and neural nets with structural regularization.
Let $h\left(x;\theta,\gamma\right)$ be such a model with hyperparameter
$\gamma$. To incorporate $h$ into our estimator, we can modify Algorithm~\ref{alg:sp}
as follows. Split the initial training data into three parts: $\left\{ \mathcal{D}_{1},\mathcal{D}_{2},\mathcal{D}_{3}\right\} $.
We first fit $h$ to $\mathcal{D}_{1}$ to obtain the optimal $\lambda=\lambda^{*}$.
Let $g\left(x;\theta\right)=h\left(x;\theta,\lambda^{*}\right)$.
$g\left(x;\theta\right)$ then enters into the standard SRE algorithm,
with $\mathcal{D}_{2}$ and $\mathcal{D}_{3}$ used respectively for
structural estimation and regularization. The modified algorithm thus
becomes a three-stage procedure. Algorithm~\ref{alg:cf} can be adapted
similarly with the three parts of data used alternately for the three
stages of estimation.

\subsection{Extension}

In general, the SRE can be formulated as a penalized extremum estimator
that seeks solution to the following problem in the second stage:
\begin{equation}
\underset{\theta\in\Theta}{\min}\left\{ \mathcal{L}_{g}\left(\mathcal{S};\theta\right)+\lambda\cdot\Omega\left(\theta,\widehat{\theta}^{\mathcal{M}}\right)\right\} \label{eq:extremum}
\end{equation}
, where $\mathcal{L}_{g}\left(\mathcal{S};\theta\right)$ is an objective
function associated with $g\left(x;\theta\right)$ and evaluated on
sample $\mathcal{S}$. 

This setup encompasses many possibilities. The statistical model $g$
can be discriminative or generative. The objective function can be
based on any appropriate loss functions such as the quadratic loss
and the cross-entropy (negative likelihood) loss. When $\theta$ is
identified via moment functions $\mathbb{E}\left[m\left(\theta\right)\right]=0$,
we obtain \eqref{eq:GMM-main} as a special case of \eqref{eq:extremum}.
In addition, in the special case that $g\left(x;\theta\right)=x'\theta,\ x\in\mathbb{R}^{p}$,
$\Omega\left(\theta,\widehat{\theta}^{\mathcal{M}}\right)=\left\Vert \theta-\widehat{\theta}^{\mathcal{M}}\right\Vert _{2}^{2}$,
and $m\left(\theta\right)=z\left(y-g\left(x;\theta\right)\right)$,
where $z\in\mathbb{R}^{l},\ l\ge p$ is an instrument for $x$, we
have the following analytical solution to \eqref{eq:GMM-main} for
a given $\lambda$:
\begin{equation}
\widehat{\theta}=\left(X'ZWZ'X+\lambda I\right)^{-1}\left(X'ZWZ'Y+\lambda\widehat{\theta}^{\mathcal{M}}\right)\label{eq:solGMM}
\end{equation}
, where $X=\left[x_{1},\ldots,x_{N}\right]'$, $Z=\left[z_{1},\ldots,z_{N}\right]'$,
and $I$ is the $p\times p$ identity matrix\footnote{In practice, it is often desirable as in \eqref{eq:sol} not to penalize
the constant term in $g$.}$^{,}$\footnote{One can use $W=\mathbb{E}\left[m\left(\theta_{0}\right)m\left(\theta_{0}\right)'\right]^{-1}$,
the efficient weight of \citet{hansen_large_1982}, and obtain $\widehat{\theta}$
via a two-step procedure. This however may not be the optimal weight
for our estimator. We leave the characterization of the asymptotic
properties of the SRE as well as the optimal weighting matrix for
\eqref{eq:GMM-main} to future work. \label{fn:OptimalW}}.

\section{Applications\label{sec:Applications}}

\renewcommand\thesubsection{\Alph{subsection}}

In this section, we demonstrate the effectiveness of our method and
compare its finite-sample performance with that of statistical and
structural estimation in three economic applications using Monte Carlo
simulations. Taken together, these exercises cover prediction and
causal inference (both under unconfoundedness and confounding) problems,
static and dynamic settings, and individual behavior that deviates
in various ways from perfect rationality. 

\subsection{First-Price Auction}

\begin{table}
\caption{First-price Auction - Setup\label{tab:AuctionSetup}}

\medskip{}

\begin{centering}
\begin{tabular}{clc}
\toprule 
\toprule Experiment & \multicolumn{1}{c}{True Mechanism} & Structural Model\tabularnewline
\midrule 
1 & $v_{i}\overset{\text{i.i.d.}}{\sim}U(0,1)$, $b_{i}=b\left(v_{i}\right)$ & \multirow{3}{*}{$v_{i}\overset{\text{i.i.d.}}{\sim}U(0,1)$, $b_{i}=b\left(v_{i}\right)$}\tabularnewline
2 & $v_{i}\overset{\text{i.i.d.}}{\sim}\text{Beta}(2,5)$, $b_{i}=b\left(v_{i}\right)$ & \tabularnewline
3 & $v_{i}\overset{\text{i.i.d.}}{\sim}U(0,1)$, $b_{i}=\eta_{i}\cdot b\left(v_{i}\right)$ & \tabularnewline
\bottomrule
\end{tabular}
\par\end{centering}
\medskip{}

\centering{}\emph{\small{}Notes:}{\small{} $b\left(v_{i}\right)$
is the equilibrium bid function \eqref{eq:auctionBidding}. $\eta_{i}\overset{\text{i.i.d.}}{\sim}\text{TN}\left(0,0.25,0,\infty\right)$.}{\small\par}
\end{table}

In our first application, we consider first-price sealed-bid auctions.
Auctions are one of the most important market allocation mechanisms.
Over the past twenty years, empirical analysis of auction data has
been transformed by structural estimation of auction models based
on games of incomplete information\footnote{See \citet{paarsch_introduction_2006,athey_nonparametric_2007,hickman_structural_2012,perrigne_econometrics_2019}
for surveys on econometric analysis of auction data}. Structural analysis of auction data views the observed bids as equilibrium
outcomes and attempts to recover the distribution of bidders' private
values by estimating relationships derived directly from equilibrium
bid functions. This approach, while offering a tight integration of
theory and observations, relies on a set of strong assumptions on
the information structure and rationality of bidders \citep{bajari_are_2005}. 

In this exercise, we conduct three experiments by simulating auction
data with varying number of participants under three scenarios. The
first scenario features rational bidders with independent private
values drawn from a uniform distribution. The second scenario features
rational bidders whose values are drawn from a beta distribution.
The third scenario features boundedly-rational bidders whose bids
deviate from optimal bidding strategies. Assume that we are interested
in the relationship between the number of bidders $n$ and the winning
bid $b^{*}$, $\mathbb{E}\left[\left.b^{*}\right|n\right]$. In each
experiment, we estimate $\mathbb{E}\left[\left.b^{*}\right|n\right]$
using (a) a statistical model, (b) a structural model, and (c) the
SRE. The structural model we use assumes rational bidders with uniform
private value distribution and is thus correctly specified for Experiment
1 but misspecified in Experiment 2 and 3. Table \ref{tab:AuctionSetup}
summarizes this setup. Below we detail the data-generating models
of the three experiments.

\paragraph{Setup}

Consider a first-price sealed-bid auction with $n$ risk-neutral bidders
with independent private value $v_{i}\sim^{i.i.d.}F(v)$. Each bidder
submits a bid $b_{i}$ to maximize her expected return 
\begin{equation}
\pi_{i}=\left(v_{i}-b_{i}\right)\times\Pr\left(b_{i}>\max\left\{ b_{-i}\right\} \right)\label{eq:auctionProfit}
\end{equation}
, where $b_{-i}$ denotes the other submitted bids. In Bayesian-Nash
equilibrium, each bidder's bidding strategy is given by 
\begin{equation}
b\left(v\right)=v-\frac{1}{F\left(v\right)^{n-1}}\int_{0}^{v_{i}}F(x)^{n-1}dx\label{eq:auctionBidding}
\end{equation}

For Experiment 1 and 3, we let $F$ be $U\left(0,1\right)$. In this
case, the equilibrium bid function simplifies to $b\left(v\right)=\frac{n-1}{n}v$.
For Experiment 2, we let $F$ be $\text{Beta}\left(2,5\right)$. In
each experiment, we simulate repeated auctions with varying number
of bidders\footnote{Assuming the same object is being repeatedly auctioned.}.
For Experiment 1 and 2, the observed bids $b_{i}$ are the equilibrium
outcomes, i.e. $b_{i}=b\left(v_{i}\right)$. For experiment 3, we
let $b_{i}=\eta_{i}\cdot b\left(v_{i}\right)$, where $\eta_{i}$
follows a normal distribution left-truncated at $0$, $\eta_{i}\overset{\text{i.i.d.}}{\sim}\text{TN}\left(0,0.25,0,\infty\right)$.
Bidders in Experiment 3 thus ``overbid'' relative to the Bayesian-Nash
equilibrium. 

\paragraph*{Simulation}

For each experiment, we simulate $M=100$ auctions with number of
bidders $n_{m}$ varying between $5$ and $30$. The observed data
thus consist of $\text{\ensuremath{\mathcal{D}}}=\left\{ \left\{ b_{i}^{m}\right\} _{i=1}^{n_{m}}\right\} _{m=1}^{M}$.
In this exercise, our goal is to learn $\mathbb{E}\left[\left.b^{*}\right|n\right]$,
the relationship between the number of bidders and the winning bid.
To this end, three different types of estimators are used to estimate
$\mathbb{E}\left[\left.b^{*}\right|n\right]$ from the training data.
To assess their performance, we use the true data-generating models
to compute $\mathbb{E}\left[\left.b^{*}\right|n\right]$ for $n\in\left[5,50\right]$,
so that we can compare the predictions of the estimators with the
true value both in-domain and out-of-domain. 

\paragraph*{Statistical Estimation }

To estimate $\mathbb{E}\left[\left.b^{*}\right|n\right]$ using a
statistical model\footnote{Since $n$ is exogenous, $\mathbb{E}\left[\left.b^{*}\right|n\right]$
is also a causal relationship and \eqref{eq:AuctionStatModel} can
also be thought of a reduced-form model of the effect of the number
of bidders on the winning bid.}, the data we need are $\left\{ \left(n_{m},b_{m}^{*}\right)\right\} _{m=1}^{M}$,
where $b_{m}^{*}$ is the winning bid of auction $m$. We fit the
following $p-$degree polynomial to the data:
\begin{equation}
b_{m}^{*}=\beta_{0}+\sum_{j=1}^{p}\beta_{j}n_{m}^{j}+e_{m}\label{eq:AuctionStatModel}
\end{equation}
, where the optimal degree $p$ is determined based on information
criteria. 

\paragraph*{Structural Estimation}

We structurally estimate the data from each experiment using a model
$\mathcal{M}$ that assumes bidders are rational, risk-neutral, and
have independent private values drawn from a $U\left(0,1\right)$
distribution. Under these assumptions, the bidders' private values
can be easily identified from the observed bids in each auction by
$v_{i}=\frac{n}{n-1}b_{i}$\footnote{In general, if we do not impose the assumption that $v_{i}\overset{\text{i.i.d.}}{\sim}U(0,1)$
and assume instead that $v_{i}\overset{\text{i.i.d.}}{\sim}F\left(v\right)$,
with $F$ unknown, then we can identify and estimate $v_{i}$ using
the following strategy based on \citet{guerre_optimal_2000}: let
$G\left(b\right)$ and $g\left(b\right)$ be the distribution and
density of the bids. \eqref{eq:auctionBidding} implies 
\[
v_{i}=b_{i}+\frac{1}{n-1}\frac{G\left(b_{i}\right)}{g\left(b_{i}\right)}
\]
Thus, by nonparametrically estimating $G\left(b\right)$ and $g\left(b\right)$
from the observed bids, we can obtain an estimate of $v_{i}$.}. The structural model makes it even easier to make predictions on
the winning bid: the model implies that $\mathbb{E}\left[\left.b^{*}\right|n\right]=\frac{n}{n+1}$.
No estimation is necessary. 

\paragraph*{Structural Regularization}

We use $\mathcal{M}$ as the benchmark model for the SRE and specify
a $5-$degree polynomial for the statistical model $g\left(n;\theta\right)$
that we shrink toward the structural benchmark. Let $g\left(\widetilde{n};\widetilde{\theta}\right)$
be the model after $n$ is standardized, as described in section \ref{subsec:Implementation}.
For regularizer, we use
\begin{equation}
\Omega\left(\widetilde{\beta},\widehat{\widetilde{\beta}}^{\mathcal{M}}\right)=\sum_{j=1}^{5}j\cdot\left(\widetilde{\beta}_{j}-\widehat{\widetilde{\beta}}_{j}^{\mathcal{M}}\right)^{2}\label{eq:AuctionRegularizer}
\end{equation}
, where $\widetilde{\beta}$ are the non-intercept coefficients of
$\widetilde{\theta}$. \eqref{eq:AuctionRegularizer} is commonly
used for regularizing polynomial models. It puts more penalty on higher
degrees of a polynomial and has the effect of making the resulting
fit more stable\footnote{Assuming that the model parametrized by $\widehat{\widetilde{\beta}}^{\mathcal{M}}$
is more stable, which is typically the case, since it is obtained
by fitting to a very large synthetic data set generated by the structural
model.}. The regularization procedure follows Algorithm~\ref{alg:sp} with
sample-splitting and forward CV.

\paragraph*{Results}

\begin{figure}
\subfloat[\label{fig:auction1_a}]{\includegraphics[width=0.5\columnwidth]{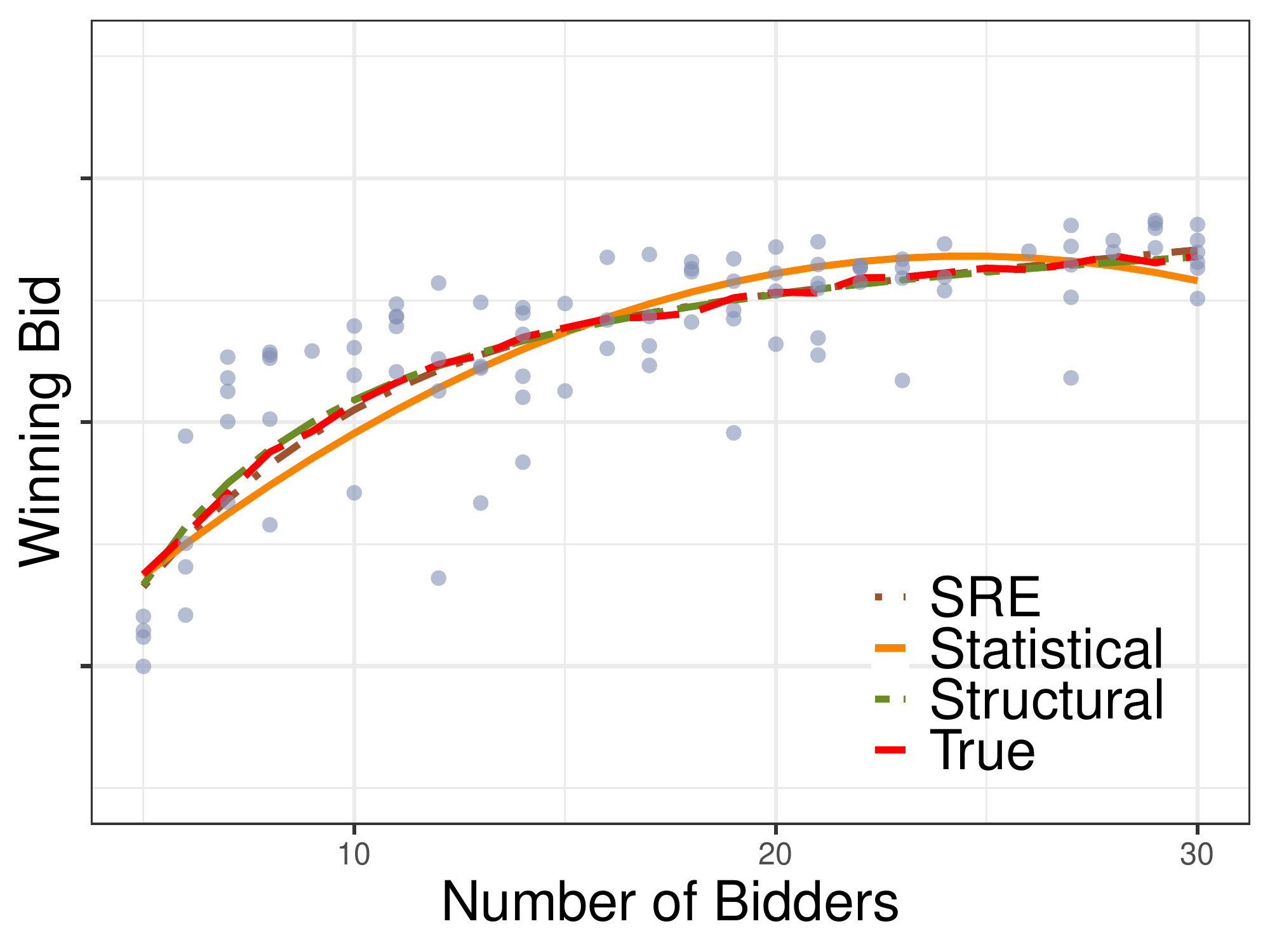}

}\subfloat[\label{fig:auction1_b}]{\includegraphics[width=0.5\columnwidth]{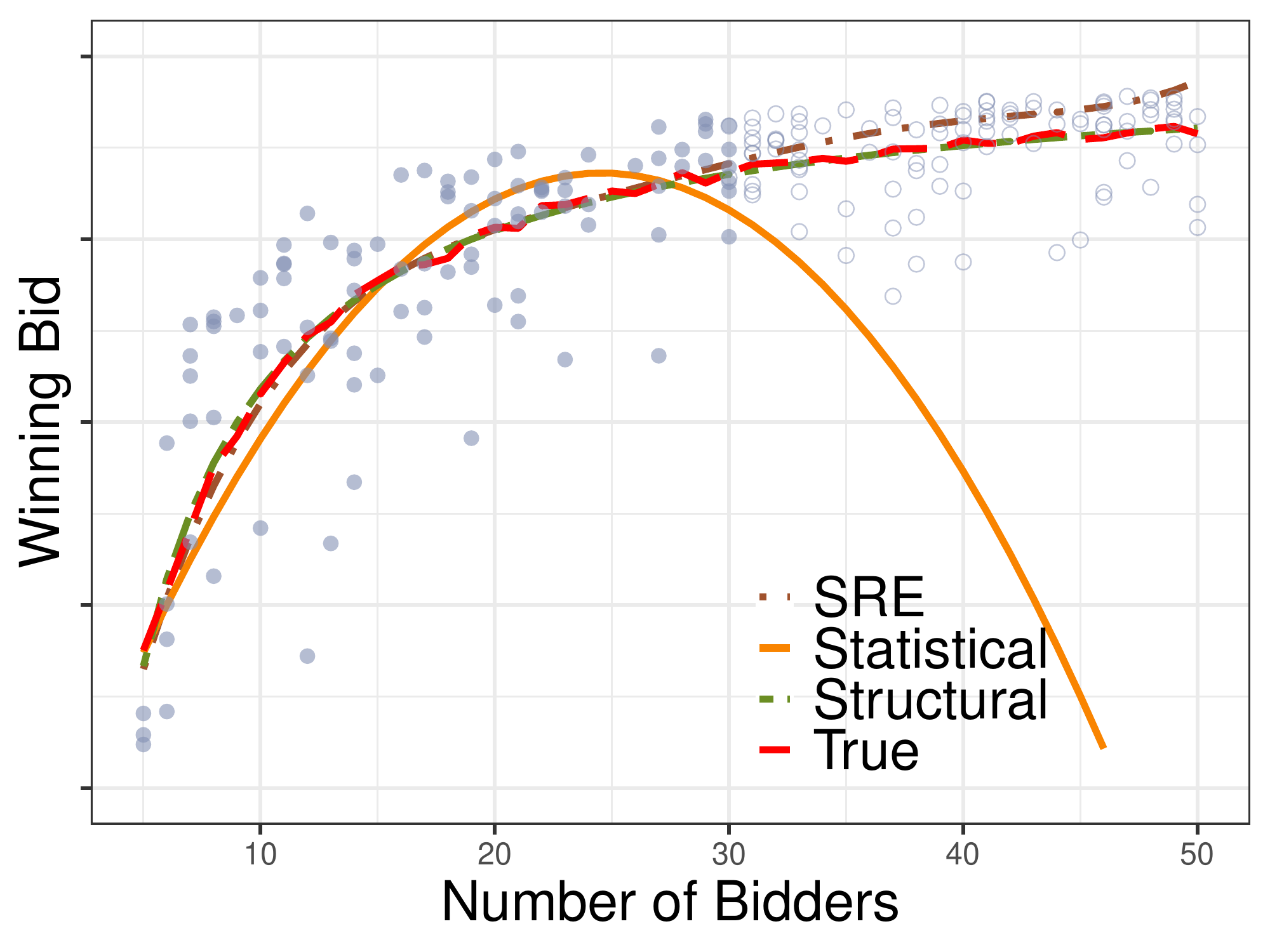}

}

\subfloat[\label{fig:auction2_a}]{\includegraphics[width=0.5\columnwidth]{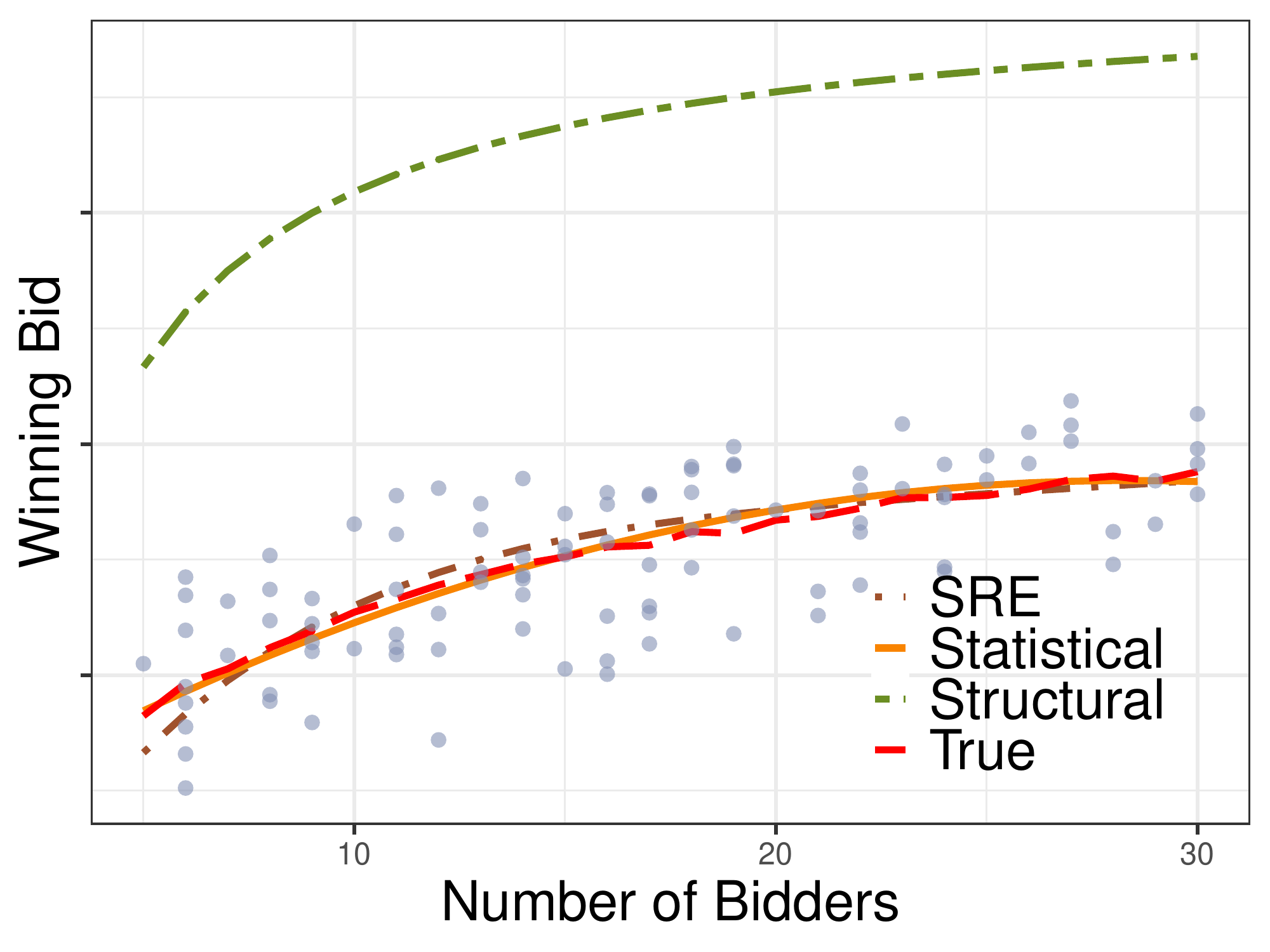}

}\subfloat[\label{fig:auction2_b}]{\includegraphics[width=0.5\columnwidth]{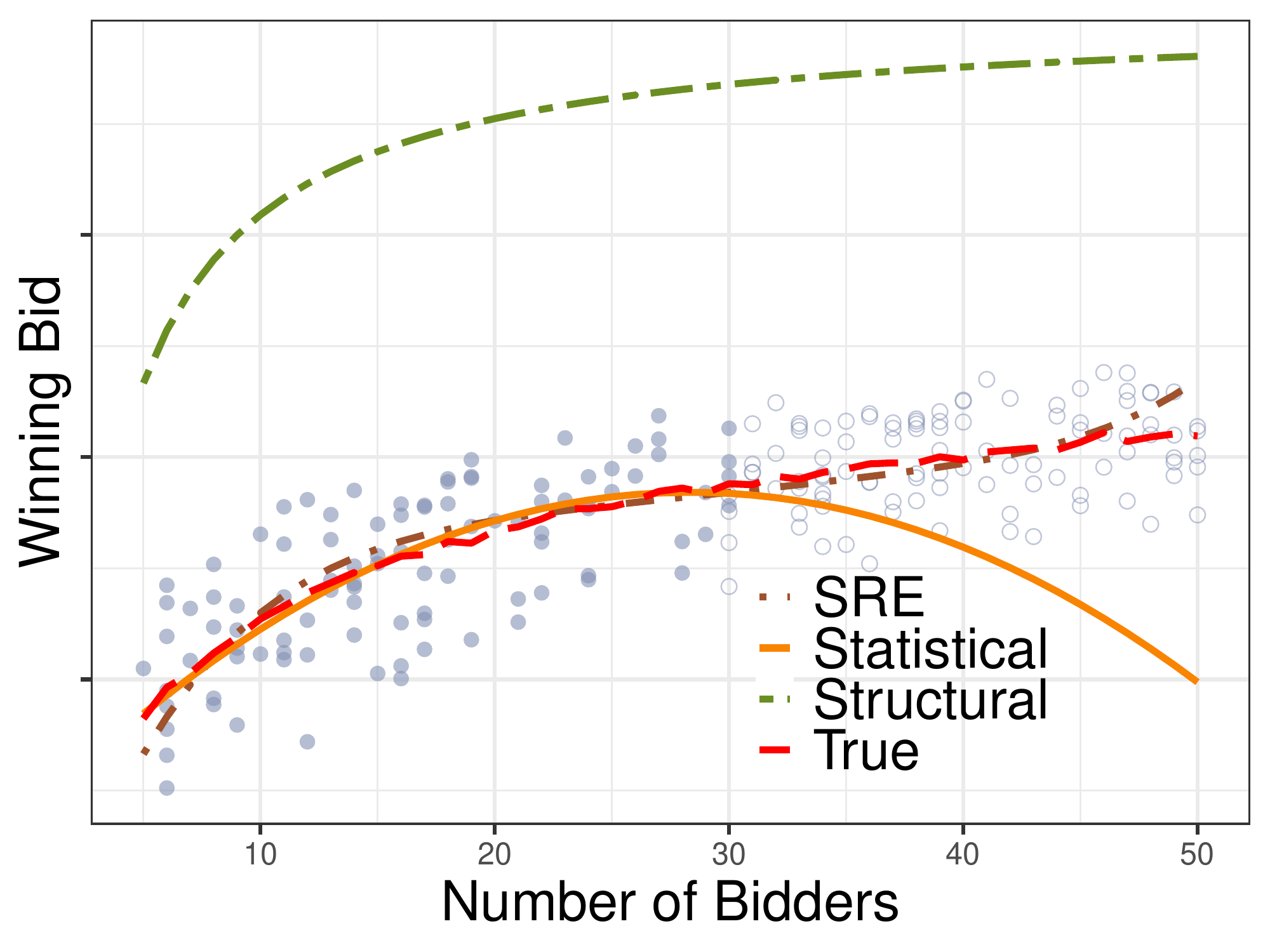}

}

\subfloat[\label{fig:auction3_a}]{\includegraphics[width=0.5\columnwidth]{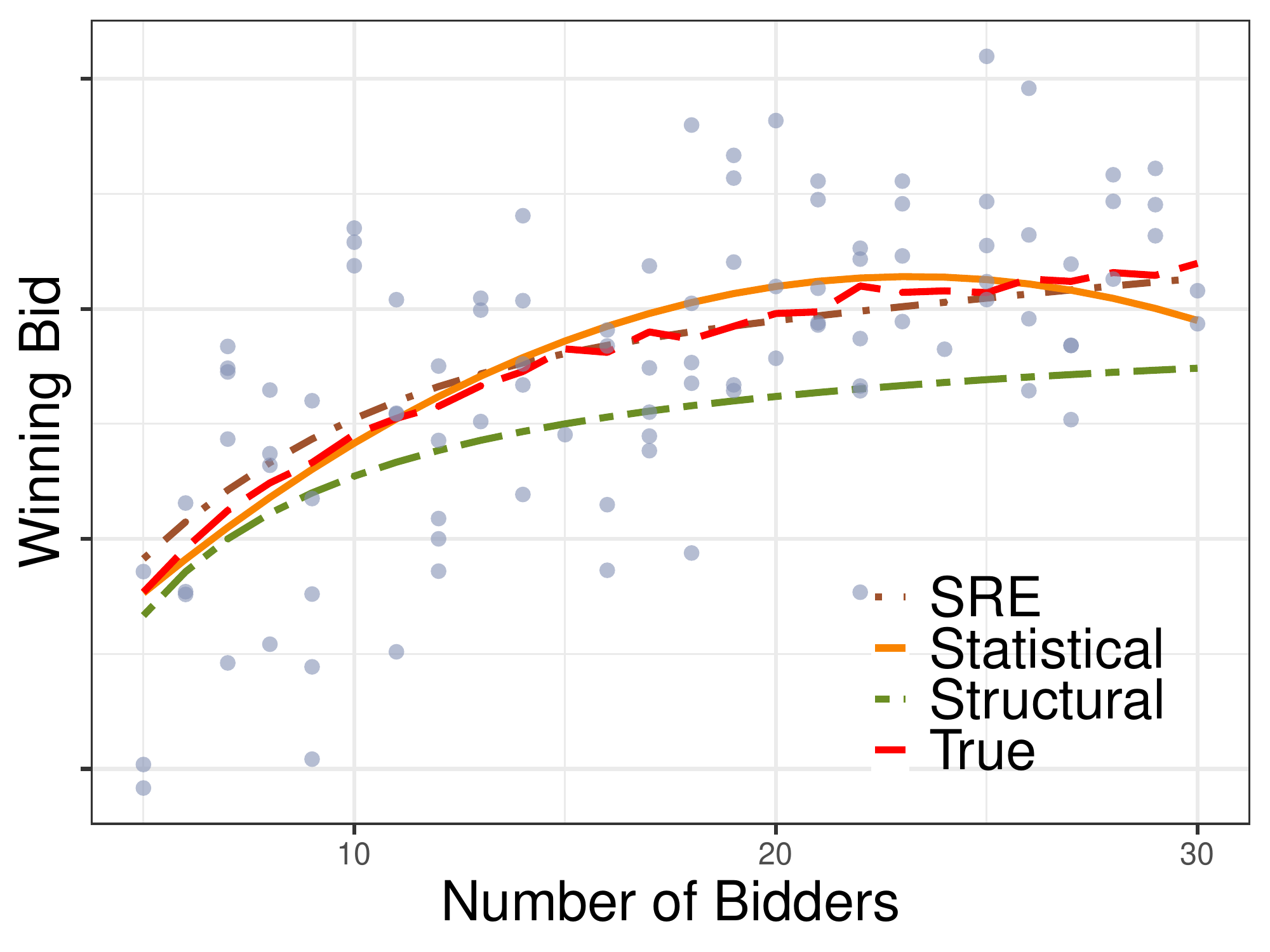}

}\subfloat[\label{fig:auction3_b}]{\includegraphics[width=0.5\columnwidth]{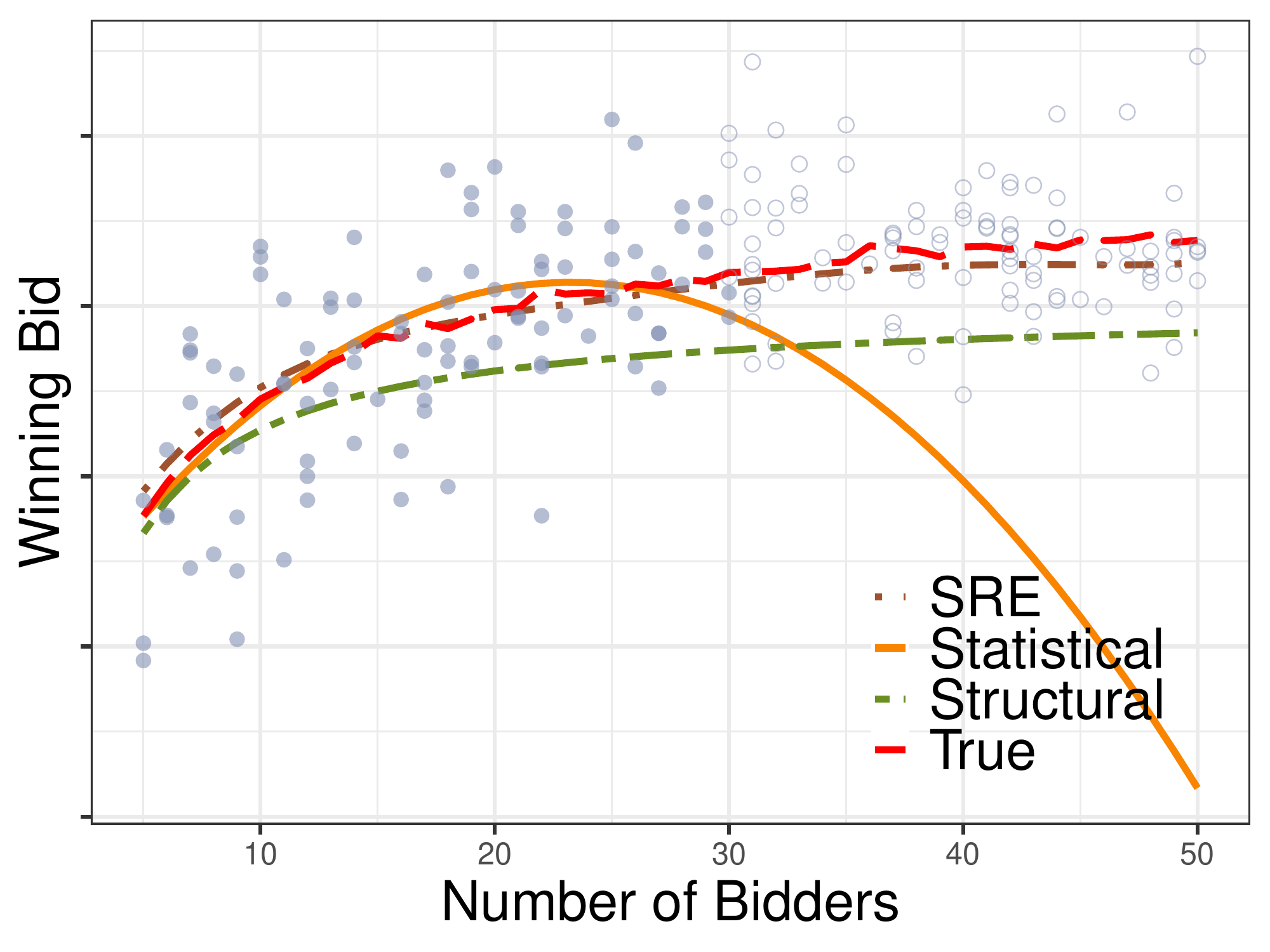}

}

\caption{{\small{}First-price Auction - The relationship between the number
of bidders and the winning bid. Dots represent training data. Circles
represent out-of-domain test data. (a) - (b): Experiment 1; (c) -
(d): Experiment 2; (e) - (f): Experiment 3.}}
\end{figure}

\begin{table}
\caption{First-price Auction - Results\label{tab:AuctionTable}}

\medskip{}

\begin{centering}
\begin{tabular}{clccccccc}
\toprule 
\toprule &  & \multicolumn{3}{c}{In-Domain} & \  & \multicolumn{3}{c}{Out-of-Domain}\tabularnewline
\cmidrule{3-5} \cmidrule{4-5} \cmidrule{5-5} \cmidrule{7-9} \cmidrule{8-9} \cmidrule{9-9} 
Experiment & Estimator & Bias & Var & MSE &  & Bias & Var & MSE\tabularnewline
\midrule 
\multirow{5}{*}{1} &  &  &  &  &  &  &  & \tabularnewline
 & Statistical & 0.0037 & 0.0003 & 0.0003 &  & 0.1256 & 118.8697 & 118.9157\tabularnewline
 & Structural & 0.0000 & 0.0000 & 0.0000 &  & 0.0000 & 0.0000 & 0.0000\tabularnewline
 & SRE & 0.0035 & 0.001 & 0.0001 &  & 0.0431 & 0.2690 & 0.2734\tabularnewline
 &  &  &  &  &  &  &  & \tabularnewline
\midrule
\multirow{5}{*}{2} &  &  &  &  &  &  &  & \tabularnewline
 & Statistical & 0.1251 & 0.1129 & 0.1426 &  & 3.5506 & 572.1151 & 595.3736\tabularnewline
 & Structural & 9.8255 & 0.0000 & 98.5233 &  & 12.6631 & 0.0000 & 160.6319\tabularnewline
 & SRE & 0.2772 & 0.0883 & 0.2399 &  & 0.3359 & 12.7092 & 12.8710\tabularnewline
 &  &  &  &  &  &  &  & \tabularnewline
\midrule
\multirow{5}{*}{3} &  &  &  &  &  &  &  & \tabularnewline
 & Statistical & 0.1455 & 0.1987 & 0.2307 &  & 0.5285 & 18.2734 & 18.5692\tabularnewline
 & Structural & 7.7095 & 0.0000 & 60.4870 &  & 9.8956 & 0.0000 & 98.0982\tabularnewline
 & SRE & 0.2787 & 0.1991 & 0.3215 &  & 0.4517 & 817.5363 & 817.8250\tabularnewline
 &  &  &  &  &  &  &  & \tabularnewline
\bottomrule
\end{tabular}
\par\end{centering}
\bigskip{}

\emph{\small{}\hspace{10bp}Notes:}{\small{} results are based on
100 simulation trials. Reported are the mean bias, variance, and }{\small\par}

\emph{\small{}\hspace{10bp}}{\small{}MSE, averaged over the number
of bidders $n$. Since the structural model predicts $\mathbb{E}\left[\left.b^{*}\right|n\right]=$}{\small\par}

\emph{\small{}\hspace{10bp}}{\small{}$\left.\left(n-1\right)\right/\left(n+1\right)$,
its predictions have zero variance.}{\small\par}
\end{table}

Figure \ref{fig:auction1_a} and \ref{fig:auction1_b} show the results
of the first experiment. Figure \ref{fig:auction1_a} plots the number
of participants $n$ against the winning bid $b^{*}$, the true $\mathbb{E}\left[\left.b^{*}\right|n\right]$,
as well as predictions by the estimated statistical, structural, and
SRE model. All three models fit very well \emph{in-domain}. Since
the structural model is the true model, it predicts the true expected
winning bids. The statistical model -- here a 2nd degree polynomial
-- also closely approximates the target function and could suffice
if our goal is to obtain a good in-domain fit. Figure \ref{fig:auction1_b}
plots the results of extrapolating the model predictions from $n\in\left[5,30\right]$
to $n\in\left[30,50\right]$. While the structural predictions still
hold true, the statistical fit becomes very bad. On the other hand,
the SRE fit remains close to the true relationship both in-domain
and out-of-domain and can accurately predict winning bids well beyond
the observed range of $n$.

Figure \ref{fig:auction2_a} $-$ \ref{fig:auction3_b} show the results
of Experiment 2 and 3. In both experiments, the structural model is
misspecified. In Experiment 2, it misspecifies the private value distribution.
In Experiment 3, it assumes that bidders are rational and the observed
bids are Bayesian-Nash equilibrium outcomes when they are not. As
a consequence, in both cases, the structural fit deviates from the
true model significantly. The statistical model, like in Experiment
1, is able to fit well in-domain but poorly out-of-domain. Remarkably,
the SRE continues to perform well despite relying on a misspecified
benchmark model. Its predictions are close to the true expected winning
bids both in-domain and out-of-domain. Intuitively, the misspecified
structural models still provide useful guidance on the functional
form of $\mathbb{E}\left[\left.b^{*}\right|n\right]$ when we extrapolate
beyond the observed domain, as evidenced in Figure \ref{fig:auction2_b}
and \ref{fig:auction3_b}.

In Table \ref{tab:AuctionTable}, we report the mean bias, variance,
and mean squared error of the three estimators for 100 simulation
runs\footnote{Reported are the \emph{mean} bias, variance, and mean squared error,
averaged over the number of bidder $n$. Given an estimator $f$,
let $f^{\left(r\right)}\left(n\right)$ denote the estimator's prediction
of the winning bid in simulation $r$, then the empirical pointwise
bias of $f$ at $n=i$ is $\text{pwbias\ensuremath{\left(f,i\right)}}=\frac{1}{R}\sum_{r=1}^{R}\left|f^{\left(r\right)}\left(n=i\right)-\mathbb{E}\left[\left.b^{*}\right|n=i\right]\right|$,
where $R$ is the total number of simulations. The empirical overall
bias, or mean bias, of $f$ is $\text{bias\ensuremath{\left(f\right)}}=\frac{1}{\left|\mathcal{X}\right|}\sum_{n\in\mathcal{X}}\text{pwbias\ensuremath{\left(f,n\right)}}$,
where $\mathcal{X}$ is the space associated with $n$. The mean variance
and the mean MSE are likewise defined.}. For all three experiments, the SRE has a low MSE comparable to those
of the statistical model and of the true structural model in-domain,
while achieving a significantly lower out-of-domain MSE than both
the statistical model and the structural model when the latter is
misspecified. 

\subsection{Dynamic Entry and Exit\label{subsec:DDCM}}

\begin{table}
\caption{Dynamic Entry and Exit - Setup\label{tab:DDCSetup}}

\medskip{}

\centering{}%
\begin{tabular}{clc}
\toprule 
\toprule Experiment & \multicolumn{1}{c}{True Mechanism} & Structural Model\tabularnewline
\midrule 
1 & Rational Expectations & \multirow{3}{*}{Rational Expectations}\tabularnewline
2 & Adaptive Expectations & \tabularnewline
3 & Myopic & \tabularnewline
\bottomrule
\end{tabular}
\end{table}

Our second application concerns the modeling and estimation of firm
entry and exit dynamics. Structural analysis of dynamic firm behavior
based on dynamic discrete choice (DDC) and dynamic game models has
been an important part of empirical industrial organization\footnote{See \citet{aguirregabiria_dynamic_2010,bajari_game_2013} for surveys
on structural estimation of dynamic discrete choice and dynamic game
models.}. These dynamic structural models capture the path dependence and
forward-looking behavior of agents, but pays the price of imposing
strong behavioral and parametric assumptions for tractability and
computational convenience. 

In this exercise, we focus our attention on the rational expectations
assumption that has been a key building block of dynamic structural
models in macro- and microeconomic analyses. The assumption and its
variants state that agents have expectations that do not systematically
differ from the realized outcomes\footnote{More precisely, rational expectations are mathematical expectations
based on information and probabilities that are model-consistent \citep{muth_rational_1961}.}. Despite having long been criticized as unrealistic, the rational
expectations paradigm has remained dominant due to a lack of tractable
alternatives and the fact that economists still know preciously little
about belief formation.

We conduct three experiments in the context of the dynamic entry and
exit of firms in competitive markets in non-stationary environments.
Our data-generating models are DDC models of entry and exit with entry
costs and exogenously evolving economic conditions. In our first experiment,
agents have rational expectations about future economic conditions.
In the second experiment, agents have a simple form of adaptive expectations
that assume the future is always like the past. The third experiment
features myopic agents who optimize only their current period returns.
In all experiments, we are interested in predicting the number of
firms that are in each market each period. To this end, we estimate
(a) a statistical model, (b) a structural model, and (c) the SRE.
The structural model we estimate assumes rational expectations and
is thus correctly specified only in Experiment 1. Table \ref{tab:DDCSetup}
summarizes this setup.

\paragraph{Setup}

Consider a market with $N$ firms. In each period, the market structure
consists of $n_{t}$ incumbent firms and $N-n_{t}$ potential entrants.
The profit to operating in the market at time $t$ is $R_{t}$, which
we assume to be exogenous and time-varying. At the beginning of each
period, both incumbents and potential entrants observe the current
period payoff $R_{t}$ and each draws an idiosyncratic utility shock
$\epsilon_{it}$. Incumbent firms then decide whether to remain or
exit the market by weighing the expected present values of each option,
while potential incumbents decide whether or not to enter the market,
which will incur a one-time entry cost $c$. Specifically, let the
entry status of a firm be represented by $\left(0,1\right)$. The
time-$t$ flow utility of a firm, who is in state $j\in\left\{ 0,1\right\} $
in time $t-1$ and state $k\in\left\{ 0,1\right\} $ in time $t$,
is given by
\begin{equation}
u_{it}^{jk}=\pi_{t}^{jk}+\epsilon_{it}^{k}\label{eq:DDCUtility}
\end{equation}
, where 
\begin{equation}
\pi_{t}^{jk}=\left(\mu+\alpha\cdot R_{t}-c\cdot\mathcal{I}\left(j=0\right)\right)\cdot\mathcal{I}\left(k=1\right)\label{eq:DDCPayoff}
\end{equation}
is the deterministic payoff function and $\epsilon_{it}=\left(\epsilon_{it}^{0},\epsilon_{it}^{1}\right)$
are idiosyncratic shocks, which we assume are \emph{i.i.d.} type-I
extreme value distributed. The parameter $\alpha$ measures the importance
of operating profits to entry-exit decisions relative to the idiosyncratic
utility shocks. 

The \emph{ex-ante} value function of a firm at the beginning of a
period is given by
\begin{align}
V_{t}^{j}\left(\epsilon_{it}\right) & =\max_{k\in\left\{ 0,1\right\} }\left\{ \pi_{t}^{jk}+\epsilon_{it}^{k}+\beta\cdot\mathbb{E}_{t}\left[\overline{V}_{t+1}^{k}\right]\right\} \label{eq:DDCMValueFcn}\\
 & =\max_{k\in\left\{ 0,1\right\} }\left\{ \mathcal{V}_{t}^{jk}+\epsilon_{it}^{k}\right\} 
\end{align}
, where $j$ is the firm's state in $t-1$, $\beta$ is the discount
factor, $\overline{V}_{t}^{j}\coloneqq\mathbb{E}_{\epsilon}\left[V_{t}^{j}\left(\epsilon_{it}\right)\right]$
is the expected value integrated over idiosyncratic shocks, and $\mathcal{V}_{t}^{jk}\coloneqq\pi_{t}^{jk}+\beta\cdot\mathbb{E}_{t}\left[\overline{V}_{t+1}^{k}\right]$
is the choice-specific conditional value function. 

At the beginning of each period, after idiosyncratic shocks are realized,
each firm thus chooses its action, $a_{it}\in\left\{ 0,1\right\} $,
by solving the following problem:
\begin{equation}
a_{it}=\underset{k\in\left\{ 0,1\right\} }{\arg\max}\left\{ \mathcal{V}_{t}^{jk}+\epsilon_{it}^{k}\right\} \label{eq:DDCAction}
\end{equation}
, which gives rise to the conditional choice probability (CCP) function:
\begin{equation}
p_{t}\left(k|j\right)\coloneqq\Pr\left(\left.a_{it}=k\right|a_{i,t-1}=j\right)=\frac{e^{\mathcal{V}_{t}^{jk}}}{\sum_{\ell=0}^{1}e^{\mathcal{V}_{t}^{j\ell}}}\label{eq:DDCCCP}
\end{equation}
, which follows from the extreme value distribution assumption. 

Since the value function involves the continuation values $\mathbb{E}_{t}\left[\overline{V}_{t+1}^{k}\right]$,
which requires expectations of the future profits $\left(R_{t+1},R_{t+2},\ldots\right)$,
its solution requires us to specify how such expectations are formed.
In Experiment 1, we assume firms have \emph{perfect foresight} on
$R_{t}$. This is a stronger form of rational expectations that assumes
individuals \emph{knows} the future realized values. Firms can then
compute $\overline{V}_{t}^{j}=\mathbb{E}_{\epsilon}\left[V_{t}^{j}\left(\epsilon_{it}\right)\right],\ j\in\left\{ 0,1\right\} $
in a model-consistent way, i.e. based on the distributional assumption
of $\epsilon_{it}$. In Experiment 2, we assume firms have a form
of \emph{adaptive expectations}, according to which beliefs about
the future are formed based on past values. Here for simplicity, we
assume that firms expect future profits to be always the same as in
current period, i.e. $R_{t}=R_{t+1}=R_{t+2}=\cdots$. Finally, in
Experiment 3, we allow firms to be \emph{myopic}, so that they do
not care about the future and only maximize current payoffs.

\paragraph*{Simulation}

\begin{figure}
\begin{centering}
\includegraphics[width=0.8\columnwidth]{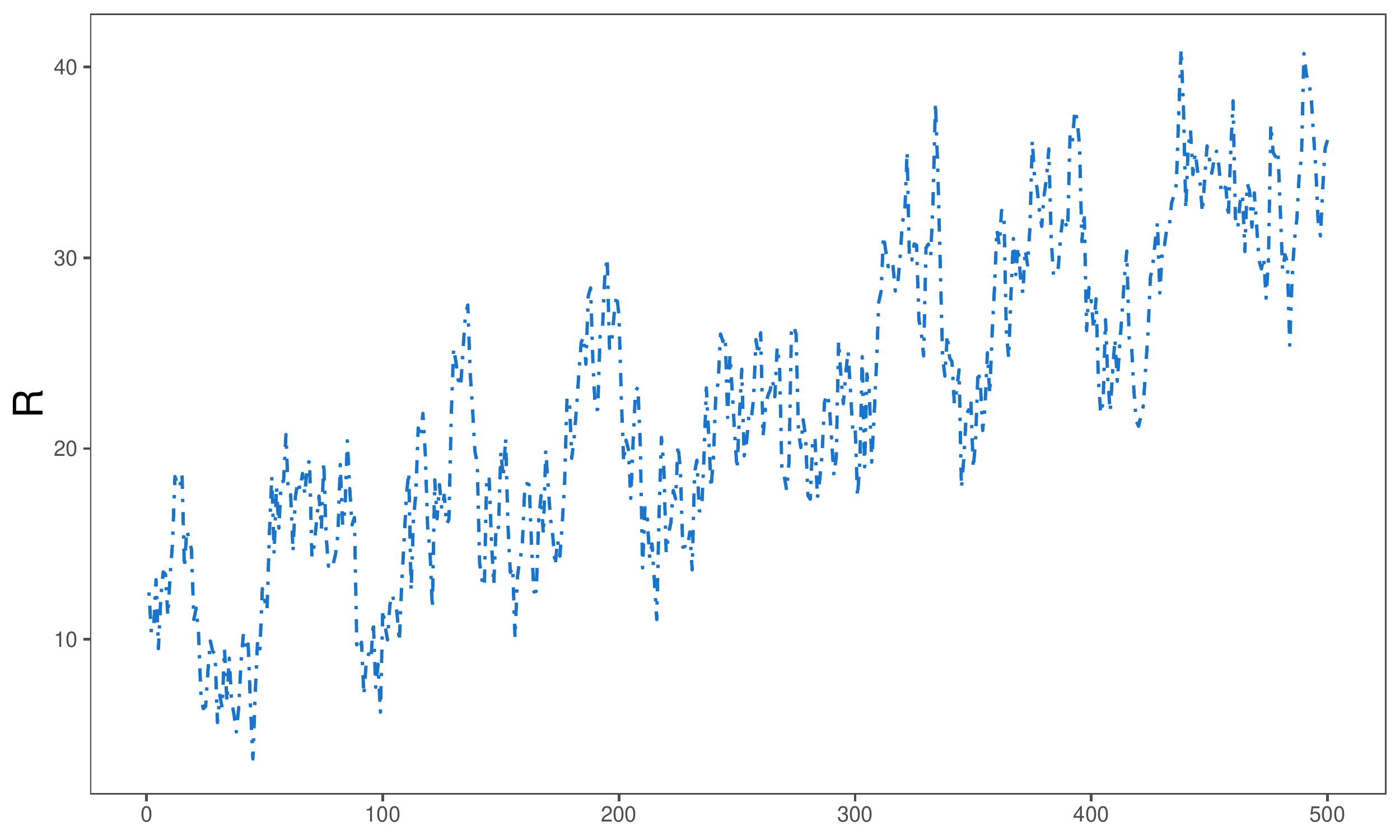}
\par\end{centering}
\caption{{\small{}Dynamic Entry and Exit - Exogenous Operating Profit\label{fig:DDCRt}}}
\end{figure}

For each experiment, we simulate $N=10,000$ firms for $\mathcal{T}=500$
periods. The first $T=250$ periods are used for training and the
last $\mathcal{T}-T=250$ periods are used to assess the out-of-domain
performance of our estimators. The training data thus consist of $\text{\ensuremath{\mathcal{D}}}=\left\{ \left\{ a_{it}\right\} _{i=1}^{N},R_{t}\right\} _{t=1}^{T}$.
We simulate $R_{t}$ to follow a rising time trend so that the environment
is non-stationary. Figure \ref{fig:DDCRt} shows a realized path of
$R_{t}$. The model parameters for each experiment are chosen so that
the entry and exit dynamics over the first $T$ periods are significantly
different from the last $\mathcal{T}-T$ periods, allowing us to better
distinguish the performance of the estimators. Appendix \hyperlink{APP}{A.2}
reports the parameter values we use as well as other details of the
simulation. 

\paragraph*{Statistical Estimation }

Our goal is to predict $n_{t}$ -- the number of firms operating
in the market in each period. The data we need for statistical modeling
are $\left(n_{t},R_{t}\right)$. We fit the following nonlinear ARX
model to the data:
\begin{equation}
n_{t}=\gamma_{0}+\sum_{j=1}^{p}\gamma_{j}R_{t}^{j}+\sum_{\ell=1}^{q}\rho_{\ell}n_{t-\ell}+e_{t}\label{eq:DDCStatModel}
\end{equation}
, where $\left(p,q\right)$ are again determined based on information
criteria. 

\paragraph*{Structural Estimation}

To estimate the DDC model, we use a strategy that builds on \citet{arcidiacono_conditional_2011}
and estimates an Euler-type equation constructed out of CCPs. Here
we sketch the strategy while presenting its details in Appendix \hyperlink{APP}{A.2}\footnote{See \citet{arcidiacono_practical_2011} for a review of related CCP
estimators. For empirical implementations, see, e.g. \citet{artuc_trade_2010,scott_dynamic_2014}.}. 

A key to our strategy is a rational expectations assumption: we assume
that because agents have rational expectations, their expected continuation
values do not deviate systematically from the realized values, i.e.
$\overline{V}_{t+1}^{j}=\mathbb{E}_{t}\left[\overline{V}_{t+1}^{j}\right]+\xi_{t}^{j}$,
where $\xi_{t}^{j}$ is a time-$t$ expectational error with $\mathbb{E}\left(\xi_{t}^{j}\right)=0$.
Given this assumption, and since our model has the finite dependence
property of \citet{arcidiacono_conditional_2011}, solution to \eqref{eq:DDCMValueFcn}
can be written in the form of the following Euler equation: 
\begin{align}
\ln\frac{p_{t}\left(k|j\right)}{p_{t}\left(j|j\right)} & =\left(\pi_{t}^{j,k}-\pi_{t}^{j,j}+\beta\left(\pi_{t+1}^{k,k}-\pi_{t+1}^{j,k}\right)\right)-\beta\ln\frac{p_{t+1}\left(k|k\right)}{p_{t+1}\left(k|j\right)}+\epsilon_{t}^{j,k}\label{eq:DDCMEuler}
\end{align}
, where $\epsilon_{t}^{j,k}=\beta\left(\xi_{t}^{k}-\xi_{t}^{j}\right)$. 

Replacing the CCPs with their sample analogues, i.e. let $\widehat{p}_{t}\left(k|j\right)=$
observed percentage of firms that are in state $j$ in $t-1$ and
state $k$ in time $t$, we obtain the following estimating equations:
for all $j\ne k$,
\begin{equation}
\ln\frac{\widehat{p}_{t}\left(k|j\right)}{\widehat{p}_{t}\left(j|j\right)}+\beta\ln\frac{\widehat{p}_{t+1}\left(k|k\right)}{\widehat{p}_{t+1}\left(k|j\right)}=\begin{cases}
\mu+\alpha R_{t}-\left(1-\beta\right)c+e_{t}^{01} & \left(j,k\right)=\left(0,1\right)\\
-\mu-\alpha R_{t}+e_{t}^{10} & \left(j,k\right)=\left(1,0\right)
\end{cases}\label{eq:DDCEstimate}
\end{equation}
, where $e_{t}=\left(e_{t}^{01},e_{t}^{10}\right)$ is an error term
that captures both the expectational errors in $\epsilon_{t}^{j,k}$
and the approximation errors in $\widehat{p}_{t}\left(k|j\right)$. 

We assume that the value of the discount factor $\beta$ is known.
Estimating \eqref{eq:DDCEstimate} gives us an estimate of the model
parameters $\left(\mu,\alpha,c\right)$. These estimates are consistent
for a model that assumes rational expectations. Therefore, the DDC
model estimated using this strategy is correctly specified for Experiment
1, but misspecified in Experiment 2 and 3. 

\paragraph*{Structural Regularization}

For structural regularization, we use the DDC model with rational
expectations as the benchmark and use \eqref{eq:DDCStatModel} with
$\left(p,q\right)=\left(2,4\right)$ as the specification for the
statistical model we regularize. Since the target variable $n_{t}$
is serially correlated, we use Algorithm~\ref{alg:sp} with a cross-validation
procedure based on a rolling-window design that is commonly used for
time series modeling. 

\paragraph*{Results}

\begin{figure}
\subfloat[\label{fig:ddc11}]{\includegraphics[width=1\columnwidth]{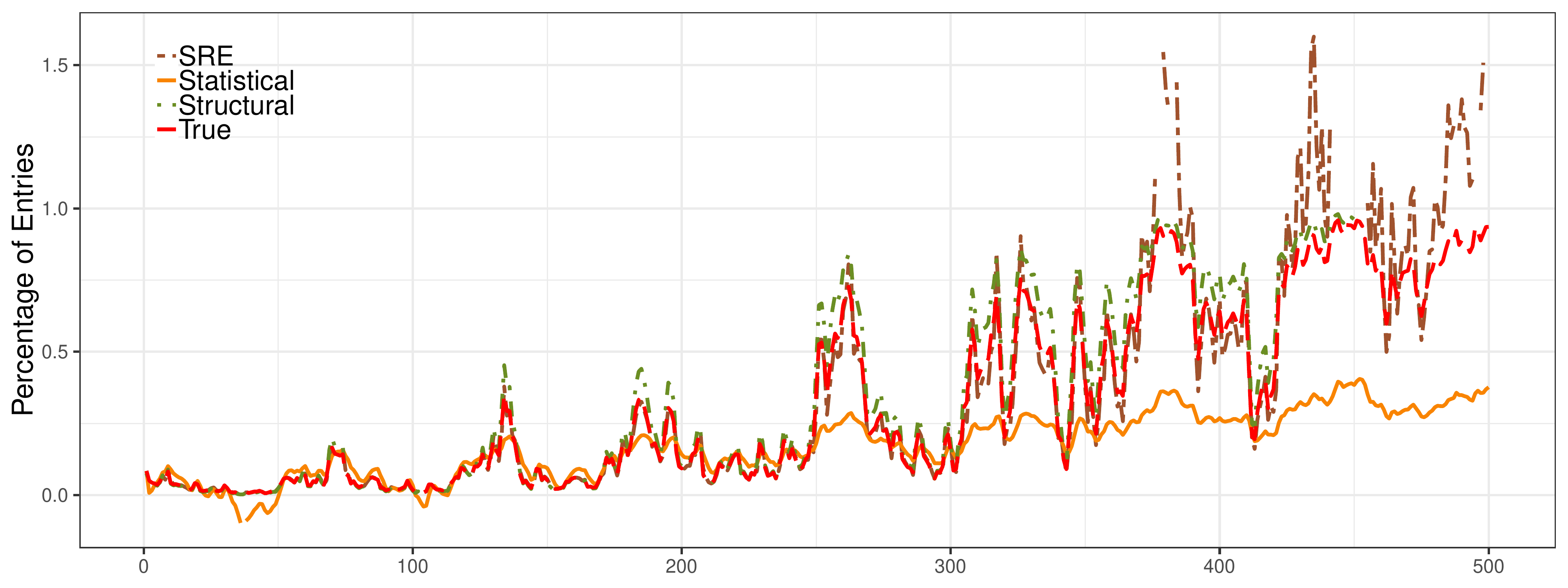}

}

\subfloat[\label{fig:ddc12}]{\includegraphics[width=0.5\columnwidth]{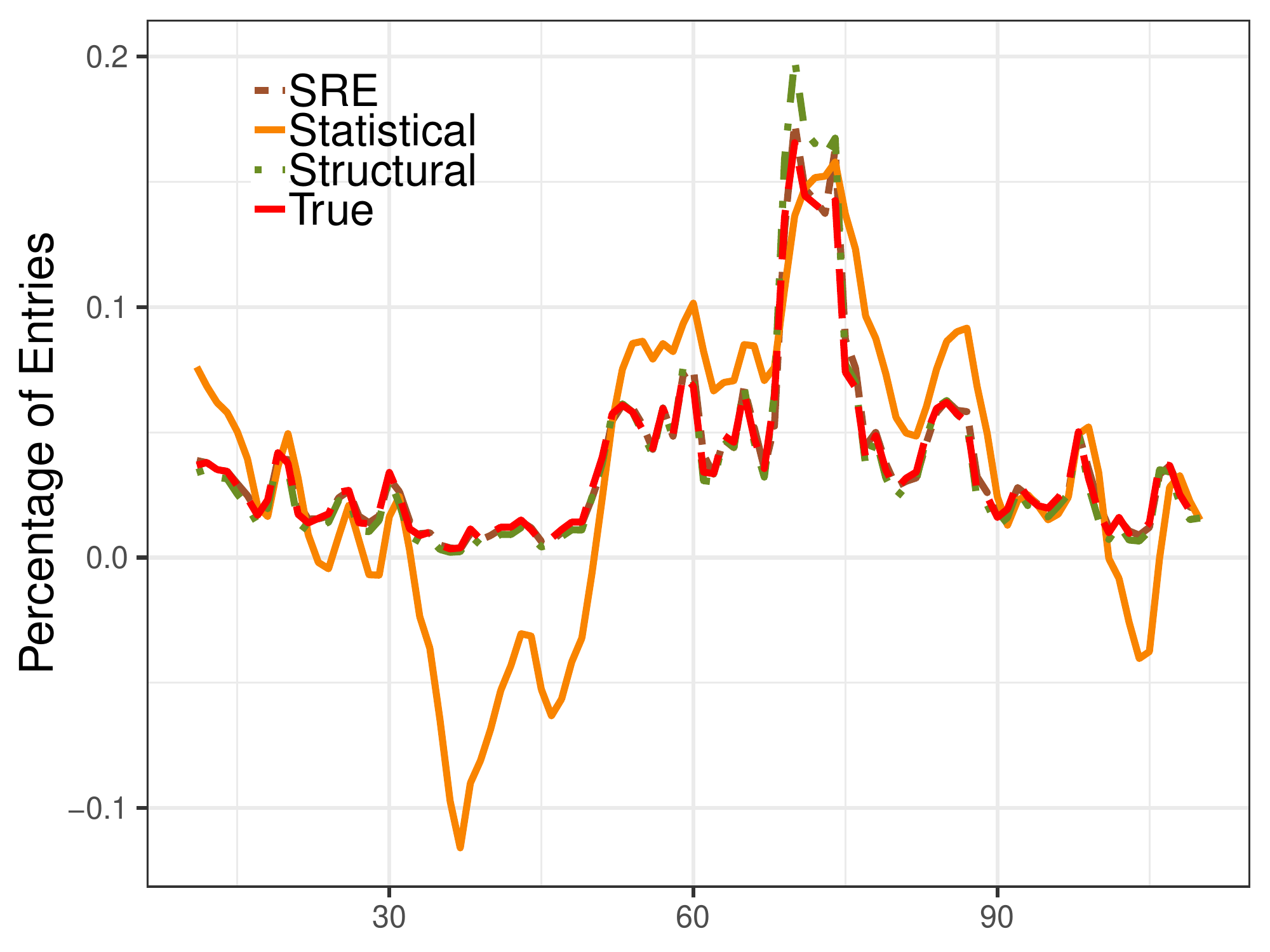}

}\subfloat[\label{fig:ddc13}]{\includegraphics[width=0.5\columnwidth]{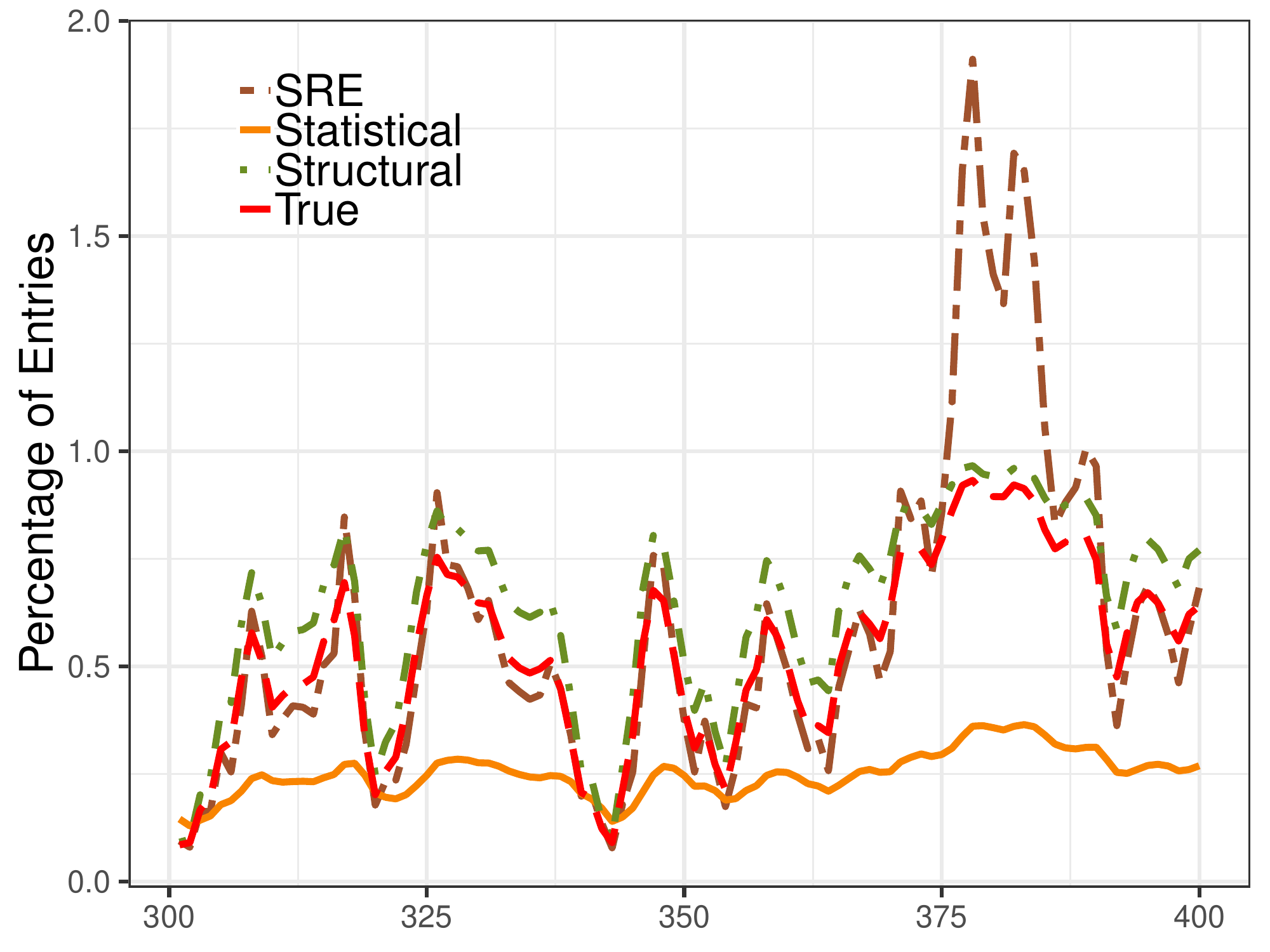}

}

\caption{{\small{}Dynamic Entry and Exit - Experiment 1. Plotted are the true
expected percentage of firms in the market (red) along with the predictions
by the three estimators. Training data are not plotted for clarity.
In (a), the entire periods of $t=1-500$ are plotted, which covers
both the in-domain periods of $t=1-250$ and the out-of-domain periods
of $t=251-500$. (b) and (c) plot respectively the in-domain periods
of $t=11-110$ and the out-of-domain periods of $t=301-400$ in order
to show a more detailed picture (the plot in (b) starts at $t=11$
due to $n_{t}$ during the initial periods being influenced by the
initial states, where we randomly assign half of the firms as incumbents
and the other half as potential entrants). \label{fig:DDCI}}}
\end{figure}

\begin{figure}
\subfloat[\label{fig:ddc21}]{\includegraphics[width=1\columnwidth]{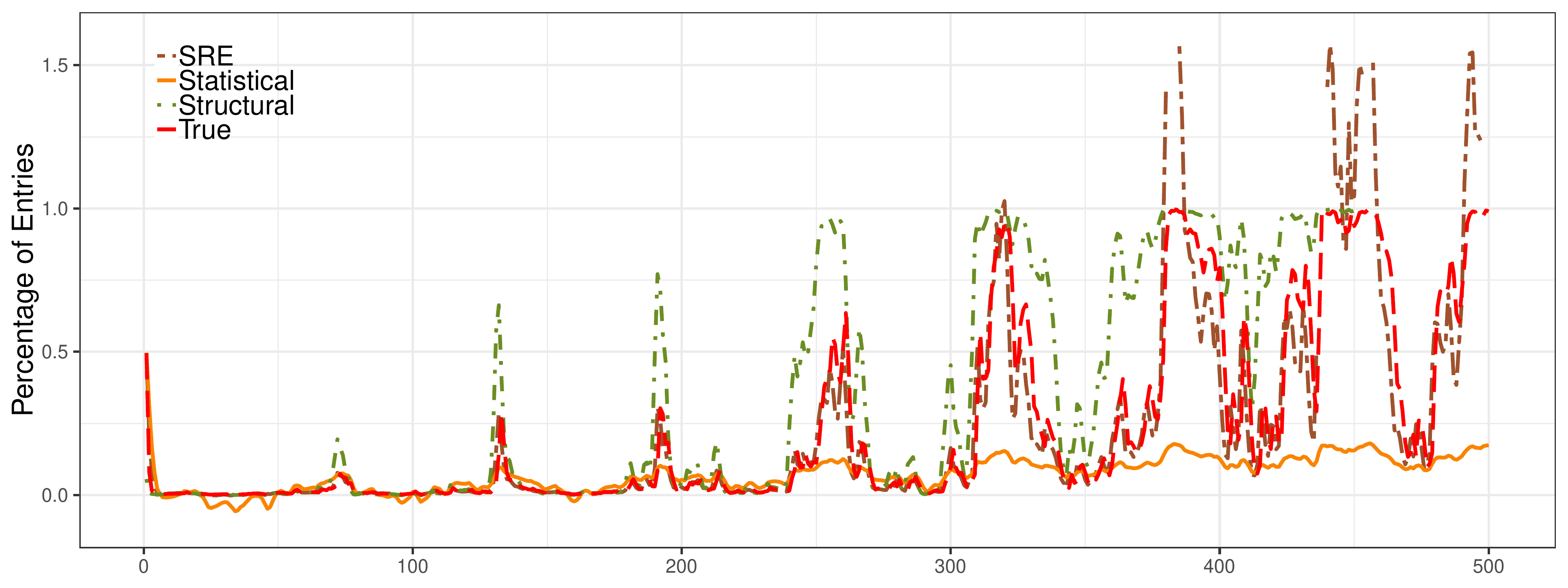}

}

\subfloat[\label{fig:ddc22}]{\includegraphics[width=0.5\columnwidth]{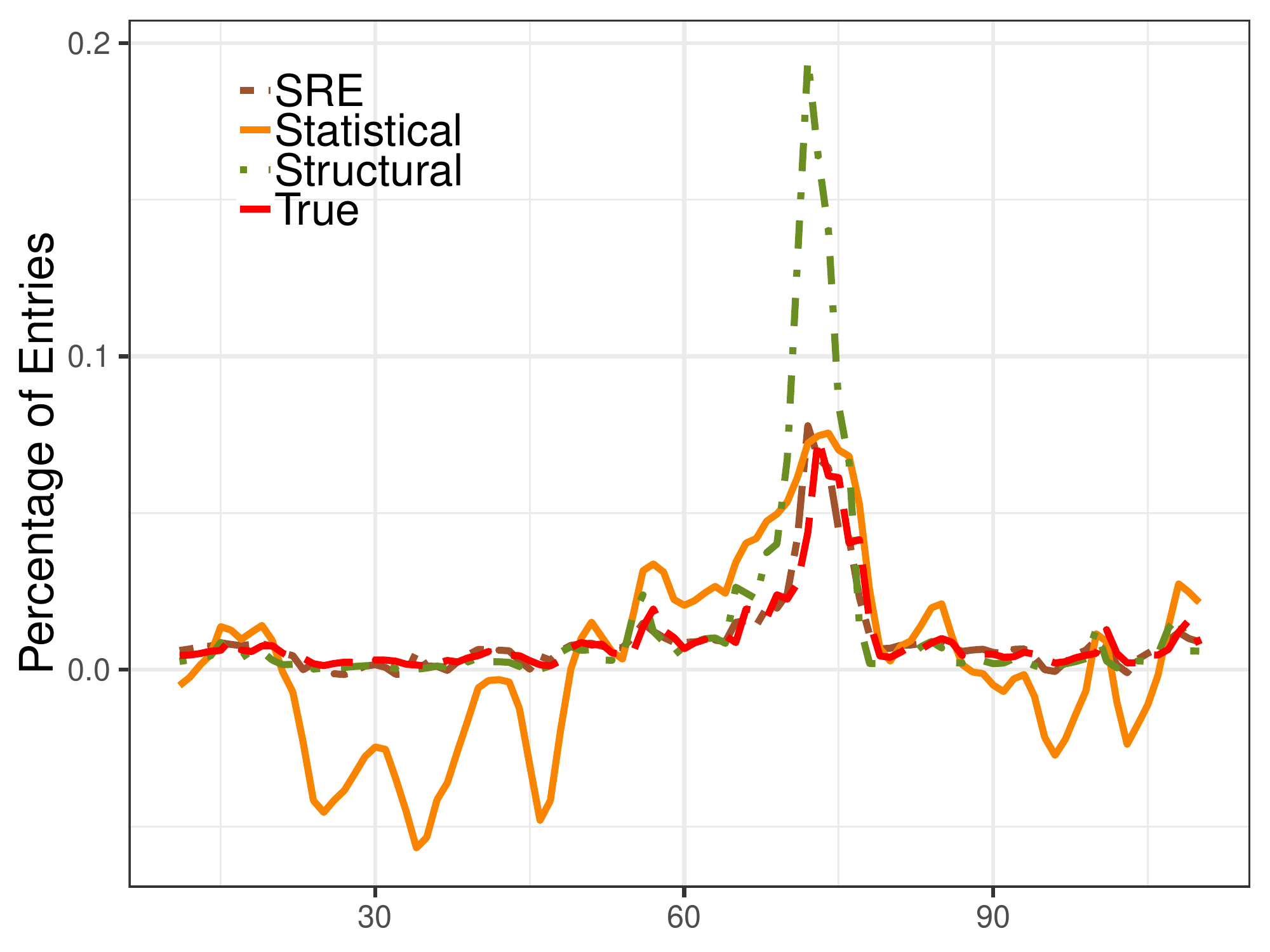}

}\subfloat[\label{fig:ddc23}]{\includegraphics[width=0.5\columnwidth]{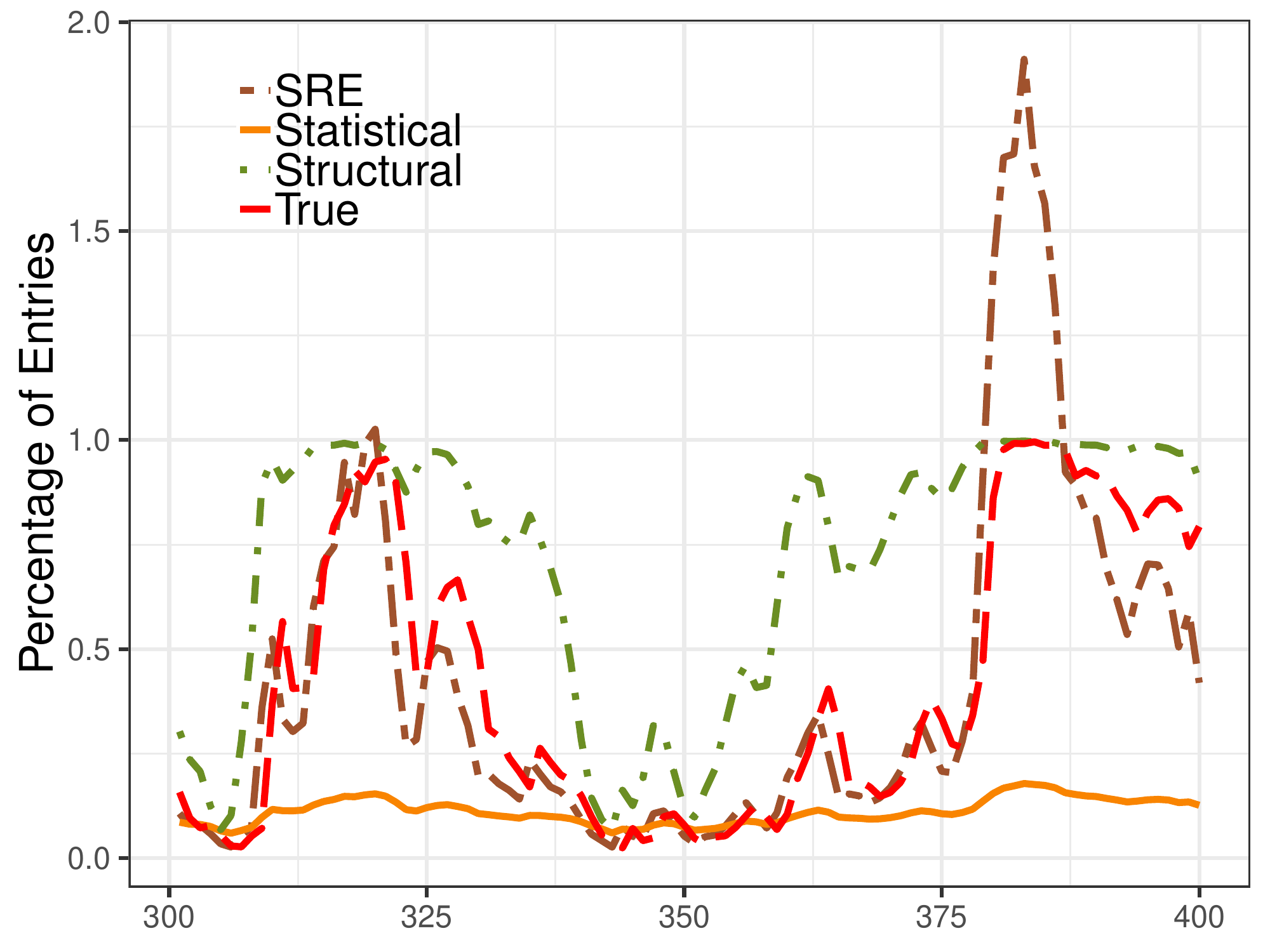}

}

\caption{{\small{}Dynamic Entry and Exit - Experiment 2. Plotted are the true
expected percentage of firms in the market (red) along with the predictions
by the three estimators. Training data are not plotted for clarity.
In (a), the entire periods of $t=1-500$ are plotted, which covers
both the in-domain periods of $t=1-250$ and the out-of-domain periods
of $t=251-500$. (b) and (c) plot respectively the in-domain periods
of $t=11-110$ and the out-of-domain periods of $t=301-400$ in order
to show a more detailed picture.\label{fig:DDCII}}}
\end{figure}

\begin{figure}
\subfloat[\label{fig:ddc31}]{\includegraphics[width=1\columnwidth]{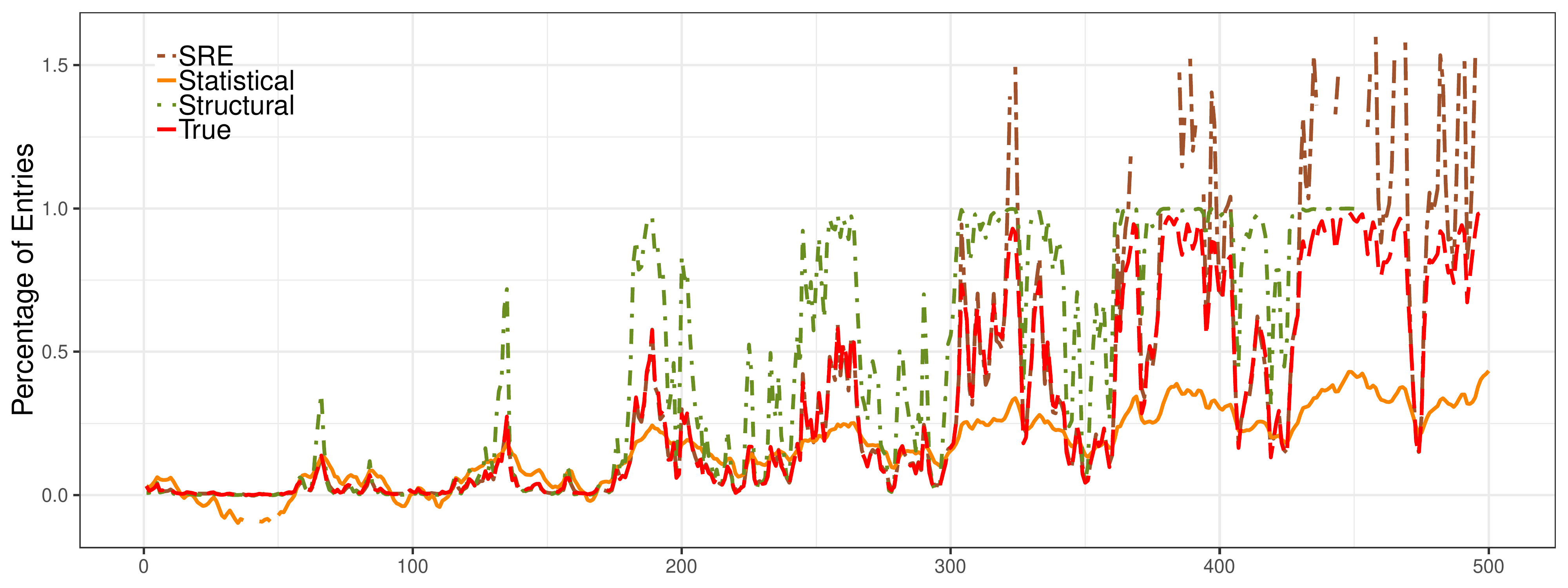}

}

\subfloat[\label{fig:ddc32}]{\includegraphics[width=0.5\columnwidth]{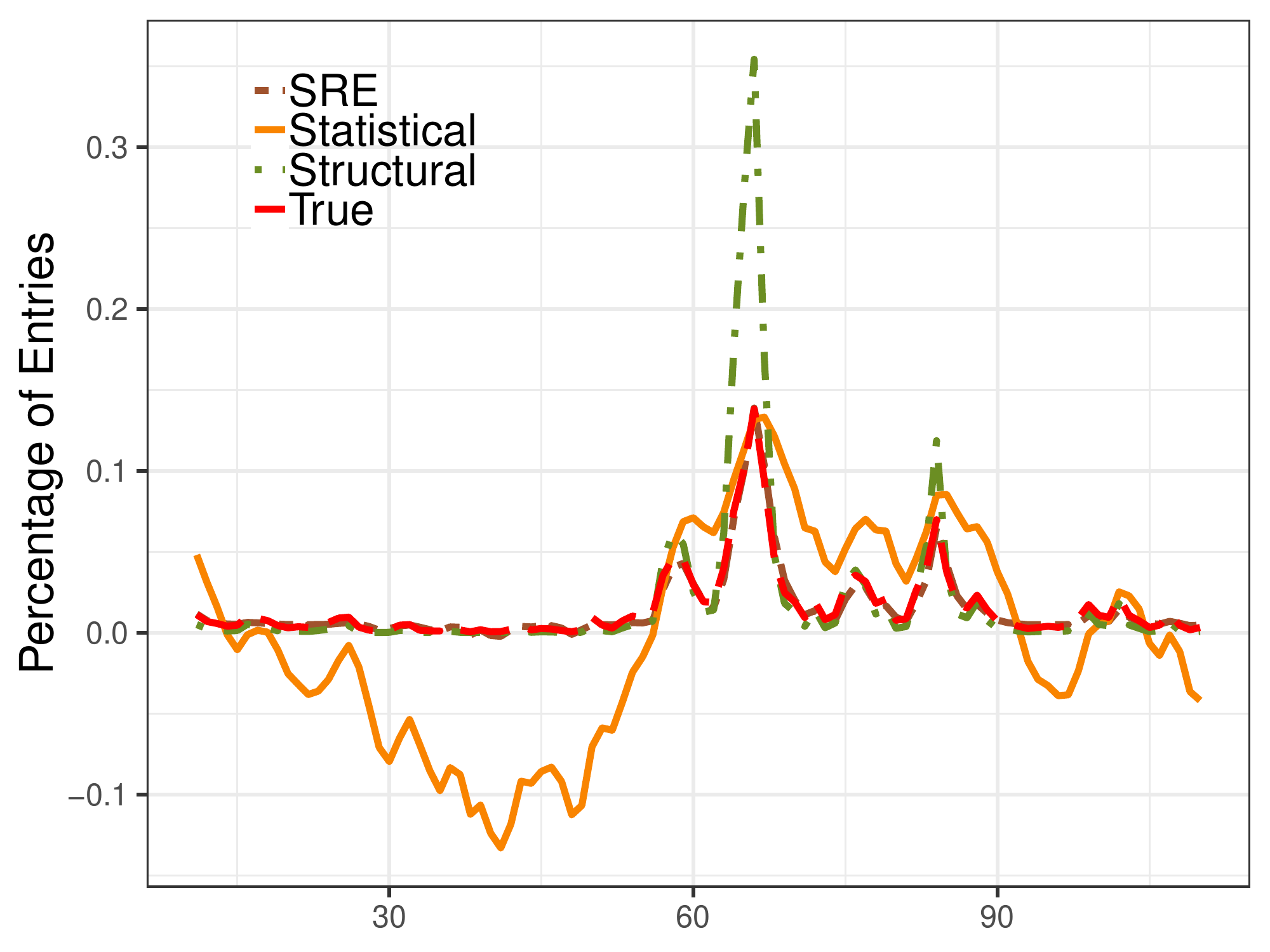}

}\subfloat[\label{fig:ddc33}]{\includegraphics[width=0.5\columnwidth]{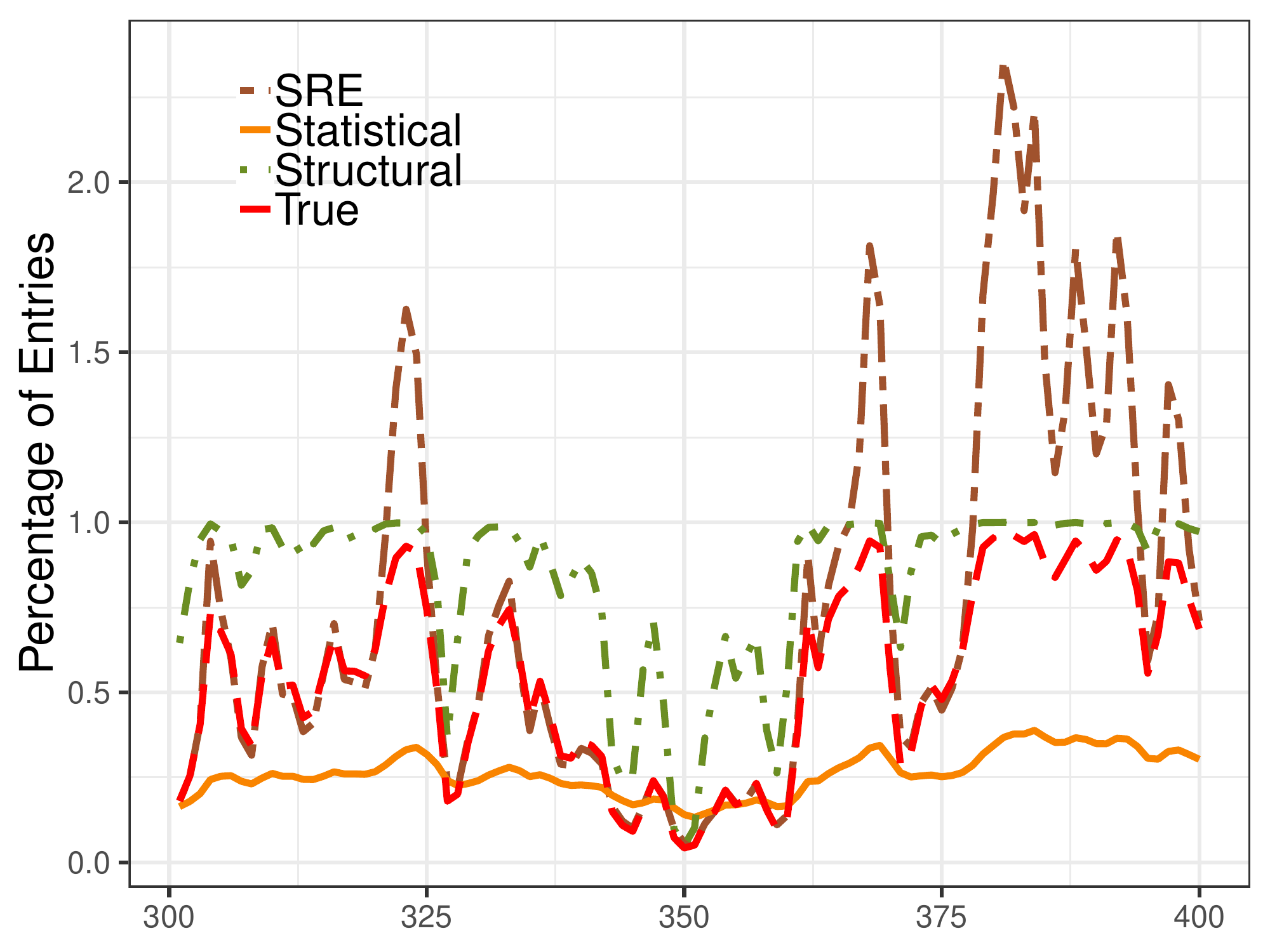}

}

\caption{{\small{}Dynamic Entry and Exit - Experiment 3. Plotted are the true
expected percentage of firms in the market (red) along with the predictions
by the three estimators. Training data are not plotted for clarity.
In (a), the entire periods of $t=1-500$ are plotted, which covers
both the in-domain periods of $t=1-250$ and the out-of-domain periods
of $t=251-500$. (b) and (c) plot respectively the in-domain periods
of $t=11-110$ and the out-of-domain periods of $t=301-400$ in order
to show a more detailed picture.\label{fig:DDCIII}}}
\end{figure}

\begin{table}
\caption{Dynamic Entry and Exit - Results\label{tab:DDCTable}}

\medskip{}

\begin{centering}
\begin{tabular}{clccccccc}
\toprule 
\toprule &  & \multicolumn{3}{c}{In-Domain} & \  & \multicolumn{3}{c}{Out-of-Domain}\tabularnewline
\cmidrule{3-5} \cmidrule{4-5} \cmidrule{5-5} \cmidrule{7-9} \cmidrule{8-9} \cmidrule{9-9} 
Experiment & Estimator & Bias & Var & MSE &  & Bias & Var & MSE\tabularnewline
\midrule 
\multirow{5}{*}{1} &  &  &  &  &  &  &  & \tabularnewline
 & Statistical & 0.0321 & 0.0018 & 0.0033 &  & 0.1347 & 0.0027 & 0.0400\tabularnewline
 & Structural & 0.0012 & 0.0009 & 0.0009 &  & 0.0180 & 0.0143 & 0.0148\tabularnewline
 & SRE & 0.0014 & 0.0009 & 0.0009 &  & 0.0353 & 0.0310 & 0.0392\tabularnewline
 &  &  &  &  &  &  &  & \tabularnewline
\midrule
\multirow{5}{*}{2} &  &  &  &  &  &  &  & \tabularnewline
 & Statistical & 0.0323 & 0.0046 & 0.0085 &  & 0.1258 & 0.6969 & 0.7371\tabularnewline
 & Structural & 0.0074 & 0.0002 & 0.0023 &  & 0.1621 & 0.0259 & 0.0731\tabularnewline
 & SRE & 0.0037 & 0.0001 & 0.0004 &  & 0.0472 & 0.0506 & 0.0668\tabularnewline
 &  &  &  &  &  &  &  & \tabularnewline
\midrule
\multirow{5}{*}{3} &  &  &  &  &  &  &  & \tabularnewline
 & Statistical & 0.0329 & 0.0015 & 0.0031 &  & 0.1558 & 0.0047 & 0.0583\tabularnewline
 & Structural & 0.0073 & 0.0025 & 0.0026 &  & 0.2111 & 0.0335 & 0.0986\tabularnewline
 & SRE & 0.0010 & 0.0004 & 0.0004 &  & 0.0736 & 0.0661 & 0.1091\tabularnewline
 &  &  &  &  &  &  &  & \tabularnewline
\bottomrule
\end{tabular}
\par\end{centering}
\bigskip{}

\emph{\small{}\hspace{27bp}Notes:}{\small{} results are based on
100 simulation trials. Reported are the mean bias, variance, }{\small\par}

\emph{\small{}\hspace{27bp}}{\small{}and MSE, averaged over time
$t$.}{\small\par}
\end{table}

Figure \ref{fig:DDCI} shows the results of the first experiment.
Figure \ref{fig:ddc11} plots the expected percentage of firms in
the market, $\mathbb{E}\left[\frac{n_{t}}{N}\right]$, for the entire
periods of $t=1-500$, covering both the in-domain periods of $t=1-250$
and the out-of-domain periods of $t=251-500$, together with the predictions
of the three estimators. The predictions are made using one-step ahead
forecasting\footnote{Given an estimated model, in each period $t$, we predict $n_{t}$
based on $\left\{ \left(n_{t-1},n_{t-2},\ldots\right),\left(R_{t},R_{t-1},\ldots\right)\right\} $.
To generate predictions for the structural model, we also assume agents
have perfect foresight regarding $\left(R_{t+1},R_{t+2},\ldots\right)$
.}. To display the results more clearly, Figure \ref{fig:ddc12} and
\ref{fig:ddc13} plot selected in-domain and out-of-domain periods
to offer a more detailed picture. All three estimators fit relatively
well in-domain. However, out-of-domain, the time series model is completely
unable to capture the rising market entries as $R_{t}$ increases.
This is partly by design: as we have discussed, we intentionally choose
parameter values so that out-of-domain dynamics differ markedly from
those in-domain. A statistical model that fits to the in-domain data
is apparently unable to extrapolate well in this case. On the other
hand, the structural model, which is correctly specified in this experiment,
extrapolates very well, as expected. The SRE performs as well as the
structural model in-domain. Out-of-domain, its predictions generally
match the true values closely, except when the true percentages are
close to $1$. In those cases the SRE fit tends to overshoot, which
is not surprising as the SRE model does not bind $n_{t}$ to be within
$\left[0,N\right]$. Nonetheless, it is apparent that the SRE is able
to capture the rising entries unlike the time series model. 

Figure \ref{fig:DDCII} shows the results of the second experiment.
In Experiment 2, agents have adaptive expectations in the sense that
they always assume $R_{t'}=R_{t}\ \forall t'>t$. Since in our simulations,
$R_{t}$ follows a rising trend, this means that agents systematically
underestimate future profits. The realized dynamics show that for
most of the in-domain periods, there is little entry into the market.
Entry increases significantly during the out-of-domain periods and
indeed, for multiple periods of time, almost all firms are in the
market. This marked difference between in-domain and out-of-domain
dynamics pose significant challenges. Looking at the model fits, the
time series model again fits relatively well in-domain but is completely
unable to extrapolate out-of-domain. The structural model, being misspecified,
is able to capture the rising entries, but tends to over-estimate
the percentages of firms in the market. In particular, its predicted
percentages tend to rise earlier and decline later than the real ones.
The model that fits the best is the SRE, which is able to match the
true dynamics closely both in-domain and out-of-domain, with the exception
of periods in which the true percentages are close to $1$, as the
SRE fit is unbounded. 

Figure \ref{fig:DDCIII} shows the results of the third experiment.
In this experiment, agents are myopic in that they only care about
current period returns when making entry and exit decisions. The data-generating
model is therefore static in nature. Looking at estimator performance,
the story is broadly similar to that of Experiment 2, with the time
series predictions biased toward $0$ out-of-domain, the structural
predictions biased toward $1$, and the SRE offering the most accurate
predictions both in-domain and out-of-domain.

Table \ref{tab:DDCTable} reports the mean bias, variance, and mean
squared error of the estimators with respect to the true $\mathbb{E}\left[\frac{n_{t}}{N}\right]$
over 100 trials. When correctly specified, the structural model performs
the best, as can be expected. When misspecified, the structural model
exhibits relatively large biases. The SRE consistently performs well
both in-domain and out-of-domain throughout the experiments. In particular,
it delivers significantly smaller biases, both in-domain and out-of--domain,
than the statistical and the structural model when the latter is misspecified.
Although it has a higher out-of-domain variance, presumably due to
its predictions not being bounded within $\left[0,1\right]$, its
overall performance is clearly superior to that of the misspecified
structural model in Experiment 2 and 3. 

\subsection{Demand Estimation}

\begin{table}
\caption{Demand Estimation - Setup\label{tab:DemandSetup}}

\medskip{}

\centering{}%
\begin{tabular}{clll}
\toprule 
\toprule Experiment & \multicolumn{1}{c}{True Mechanism} & Reduced-Form & Structural\tabularnewline
\midrule 
\multirow{2}{*}{1} & linear demand, optimal & \multirow{2}{*}{linear demand} & \tabularnewline
 & monopoly pricing &  & \tabularnewline
\multirow{2}{*}{2} & linear demand, non-optimal & \multirow{2}{*}{linear demand} & \tabularnewline
 & monopoly pricing &  & linear demand, optimal\tabularnewline
\multirow{2}{*}{3} & linear demand, optimal & \multirow{2}{*}{log-log demand} & monopoly pricing\tabularnewline
 & monopoly pricing &  & \tabularnewline
\multirow{2}{*}{4} & linear demand, non-optimal & \multirow{2}{*}{log-log demand} & \tabularnewline
 & monopoly pricing &  & \tabularnewline
\bottomrule
\end{tabular}
\end{table}

In our final application, we revisit the demand estimation problem
under a different setting. Suppose now that instead of observing consumer
demand under exogenously varying prices, the prices we observe are
set by a monopolist. In this case, changes in prices are endogenous
and the relationship between price and quantity sold is confounded.
As in the motivating example of section \ref{sec:Motivating-Example},
we are interested in learning the demand curve. To this end, if we
have access to a variable that shifts the cost of production for the
monopoly firm but does not affect demand directly, then it can be
used as an instrumental variable to help identify the true demand
curve. This is the reduced-form approach. Alternatively, we can estimate
a structural model that fully specifies monopoly pricing behavior.
This is the structural approach. Finally, we can combine the two using
the SRE. 

In this exercise, we conduct four experiments. In all four experiments,
we assume that we do have access to a valid instrument so that the
demand curve is nonparametrically identified. However, the functional
form of the reduced-form statistical model may still be misspecified.
On the other hand, using the structural approach, we estimate a model
that assumes the observed prices are optimally set by a profit-maximizing
monopoly firm. When this assumption is violated, as when for example
the firm's pricing is not optimal or it does not have monopoly power,
the structural model will also be misspecified. The four experiments
we conduct are thus arranged as follows: in the first experiment,
both the reduced-form and the structural models are correctly specified.
In Experiment 2 and 3, only one of the two is correctly specified.
In Experiment 4, both are misspecified. Table \ref{tab:DemandSetup}
summarizes this setup. 

This exercise differs from the previous two in two important aspects.
First, our first two applications focus on the misspecification of
structural models. The statistical models they fit are chosen using
a model selection procedure so as to produce the best out-of-sample
fit of the observed data. In practice, applied reduced-form research
in economics often specifies simple linear models, so misspecification
concerns are nontrivial. In this exercise, we highlight the functional
form misspecifications of the reduced-form model as well as the structural.
Second, this exercise focuses on comparisons of \emph{in-domain} performance.
We show that when either the reduced-form or the structural model
is misspecified, the SRE will have better in-domain performance --
more \emph{internal validity} -- than the misspecified model and
has the ability to outperform both when both are misspecified. 

\paragraph{Setup}

Consider $M$ geographical markets in which a product is sold. The
equilibrium price and quantity sold in market $m$ are $\left(p_{m},q_{m}\right)$.
Assume that all markets share the same aggregate demand function $Q^{d}\left(p\right)$:
\begin{equation}
q_{m}=Q^{d}\left(p_{m}\right)=\alpha-\beta\cdot p_{m}+\epsilon_{m}\label{eq:demandCurve}
\end{equation}

In Experiment 1 and 3, we assume the product is sold by a monopoly
firm who sets the prices in each market to maximize its profit. The
firm has different marginal costs $c_{m}$ for operating in different
markets. Hence it sets 
\begin{align}
p_{m} & =\underset{p>0}{\arg\max}\left\{ \left(p-c_{m}\right)Q^{d}\left(p\right)\right\} \label{eq:demandP1}\\
 & =c_{m}+\frac{1}{\beta}q_{m}\label{eq:demandP2}
\end{align}

Assume that we also observe a cost-shifter $z_{m}$, e.g. transportation
costs, such that 
\begin{equation}
c_{m}=a+b\cdot z_{m}\label{eq:demandZ}
\end{equation}
, then $z_{m}$ can serve as an instrument for $p_{m}$ for identifying
the demand curve. 

In Experiment 2 and 4, we assume the monopoly firm fails to set optimal
prices or does not have complete monopoly power. Its pricing decisions
are given by 
\begin{equation}
p_{m}=c_{m}+\frac{\lambda}{\beta}q_{m}\label{eq:demandNonOptP}
\end{equation}
, where $\lambda\in\left(0,1\right)$. The firm thus earns a lower
markup than an optimal price-setting monopoly.

\paragraph*{Simulation }

For each experiment, we simulate $M=1000$ markets and generate an
observed data set of $\text{\ensuremath{\mathcal{D}}}=\left\{ \left(p_{m},q_{m},z_{m}\right)\right\} _{m=1}^{M}$.
See Appendix \hyperlink{APP}{A.3} for the parameter values we use
in simulation. 

\paragraph*{Reduced-Form Estimation}

Because $p_{m}$ is now endogenous -- $p_{m}$ and $\epsilon_{m}$
are correlated through \eqref{eq:demandP2} -- the statistical relation
between $p_{m}$ and $q_{m}$ is confounded and no longer represents
the demand function. To estimate the demand curve using the reduced-form
approach, we avail of the instrumental variable $z_{m}$ and estimate
$Q^{d}\left(p\right)$ by two-stage least squares (2SLS). In Experiment
1 and 2, our reduced-form model is correctly specified, i.e. we fit
\eqref{eq:demandCurve} to the data by 2SLS. In Experiment 3 an 4,
however, we assume the demand function takes on a log-log form:
\begin{equation}
\log q_{m}=\alpha-\beta\cdot\log p_{m}+\epsilon_{m}\label{eq:DemandRFWRONG}
\end{equation}
, and is therefore misspecified in these two experiments. 

\paragraph*{Structural Estimation}

We fit a structural model featuring linear demand function \eqref{eq:demandCurve}
and price-setting function \eqref{eq:demandP2}. This structural model
is correctly specified for Experiment 1 and 3, but misspecified for
Experiment 2 and 4. The structural parameters are $\left(\alpha,\beta,a,b\right)$
and can be estimated as follows: from \eqref{eq:demandCurve} and
\eqref{eq:demandP2}, we obtain 
\begin{equation}
p_{m}=a+b\cdot z_{m}+\frac{1}{\beta}q_{m}\label{eq:demandSol1}
\end{equation}

If our model is correct, \eqref{eq:demandSol1} is a deterministic
linear equation system from which we can solve directly for $\left(\widehat{a},\widehat{b},\widehat{\beta}\right)$.
Substituting $\widehat{\beta}$ into \eqref{eq:demandCurve}, we then
obtain $\widehat{\alpha}=\frac{1}{M}\sum_{m=1}^{M}\left(q_{m}+\widehat{\beta}p_{m}\right)$. 

\paragraph*{Structural Regularization}

To estimate the demand curve using the SRE, we employ the structural
model described above as the benchmark model and specify a 2nd degree
polynomial $g\left(p;\theta\right)=\theta_{0}+\theta_{1}p+\theta_{2}p^{2}$
as the statistical model for $Q^{d}\left(p\right)$. As in reduced-form
estimation, we rely on the use of the instrumental variable $z_{m}$
and identify $\theta$ via the following moment conditions:
\begin{equation}
\mathbb{E}\left[\left.\left(q_{m}-g\left(p_{m};\theta\right)\right)\right|z_{m}\right]=0\label{eq:demandMC}
\end{equation}

The SRE proceeds in two stages. In the first stage, we estimate the
structural model and generate synthetic data $\left(p_{m}^{\mathcal{M}},Q^{\mathcal{M}}\left(p_{m}^{\mathcal{M}}\right)\right)$,
where $\text{\ensuremath{Q^{\mathcal{M}}\left(p\right)}}$ is the
model derived demand function, i.e. the structural estimate of $Q\text{\ensuremath{\left(p\right)}}$.
We obtain $\widehat{\theta}^{\mathcal{M}}$ by fitting $g$ directly
to the synthetic data. In the second stage, we minimize the objective
function \eqref{eq:GMM-main} with the following moment functions:
\begin{equation}
m_{m}\left(\theta\right)=\left(q_{m}-g\left(p_{m};\theta\right)\right)\phi\left(z_{m}\right)\label{eq:demandMM}
\end{equation}
, where we let $\phi\left(z\right)=\left(1,z,z^{2},\ldots,z^{5}\right)$. 

Analytical solution to \eqref{eq:GMM-main} for a given $\lambda$
is given by \eqref{eq:solGMM}. For the weight matrix $W$, we use
the 2SLS weight $W=\left(\phi\left(z\right)'\phi\left(z\right)\right)^{-1}$\footnote{See footnote \ref{fn:OptimalW}.}.
The regularization procedure follows Algorithm~\ref{alg:sp}. Since
no out-of-domain predictions are involved in this exercise, the standard
cross-validation procedure for \emph{i.i.d.} data is used to choose
the optimal $\lambda$. 

\paragraph*{Results}

\begin{figure}
\subfloat[\label{fig:demand_a}]{\includegraphics[width=0.5\columnwidth]{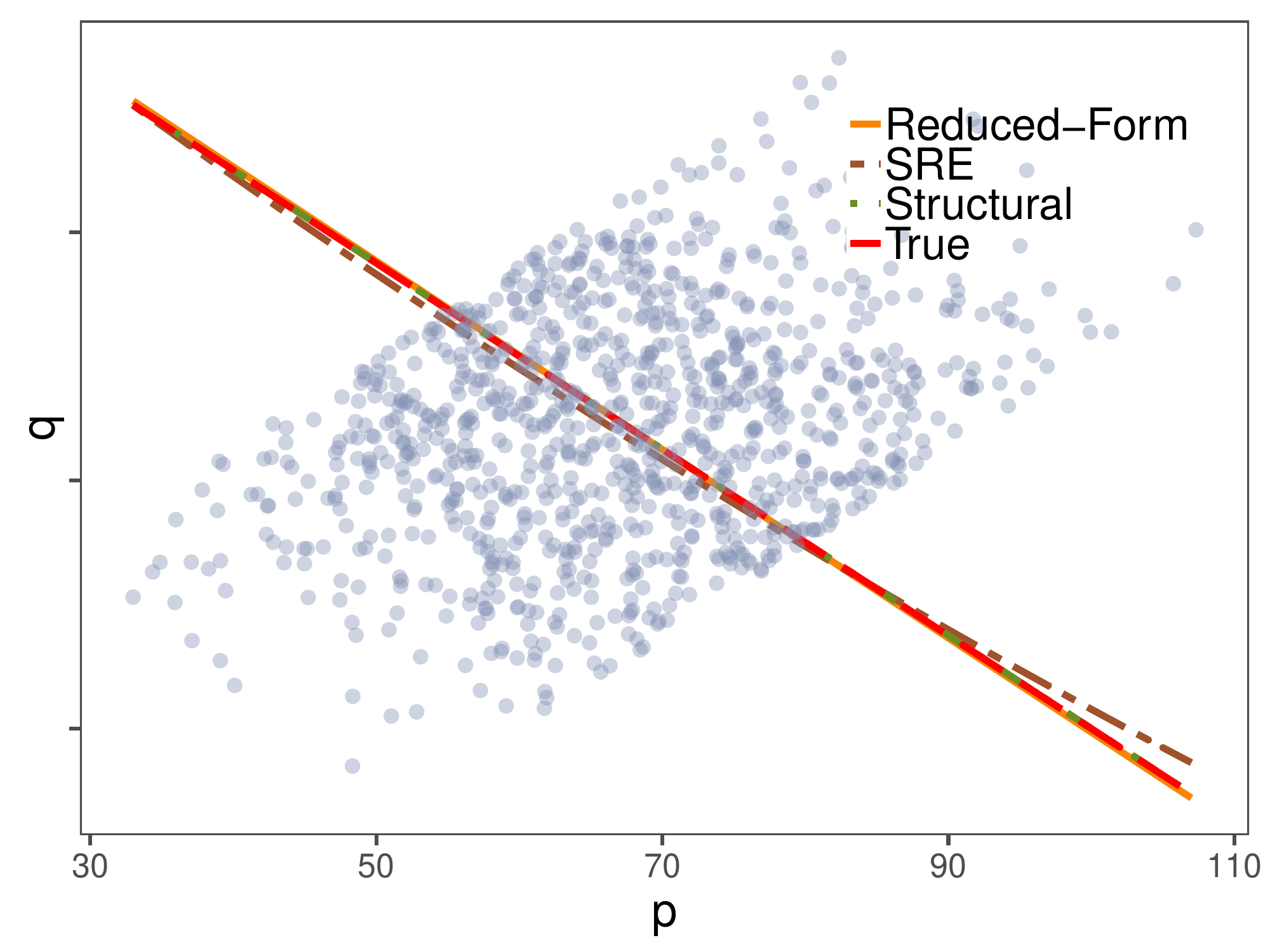}

}\subfloat[\label{fig:demand_b}]{\includegraphics[width=0.5\columnwidth]{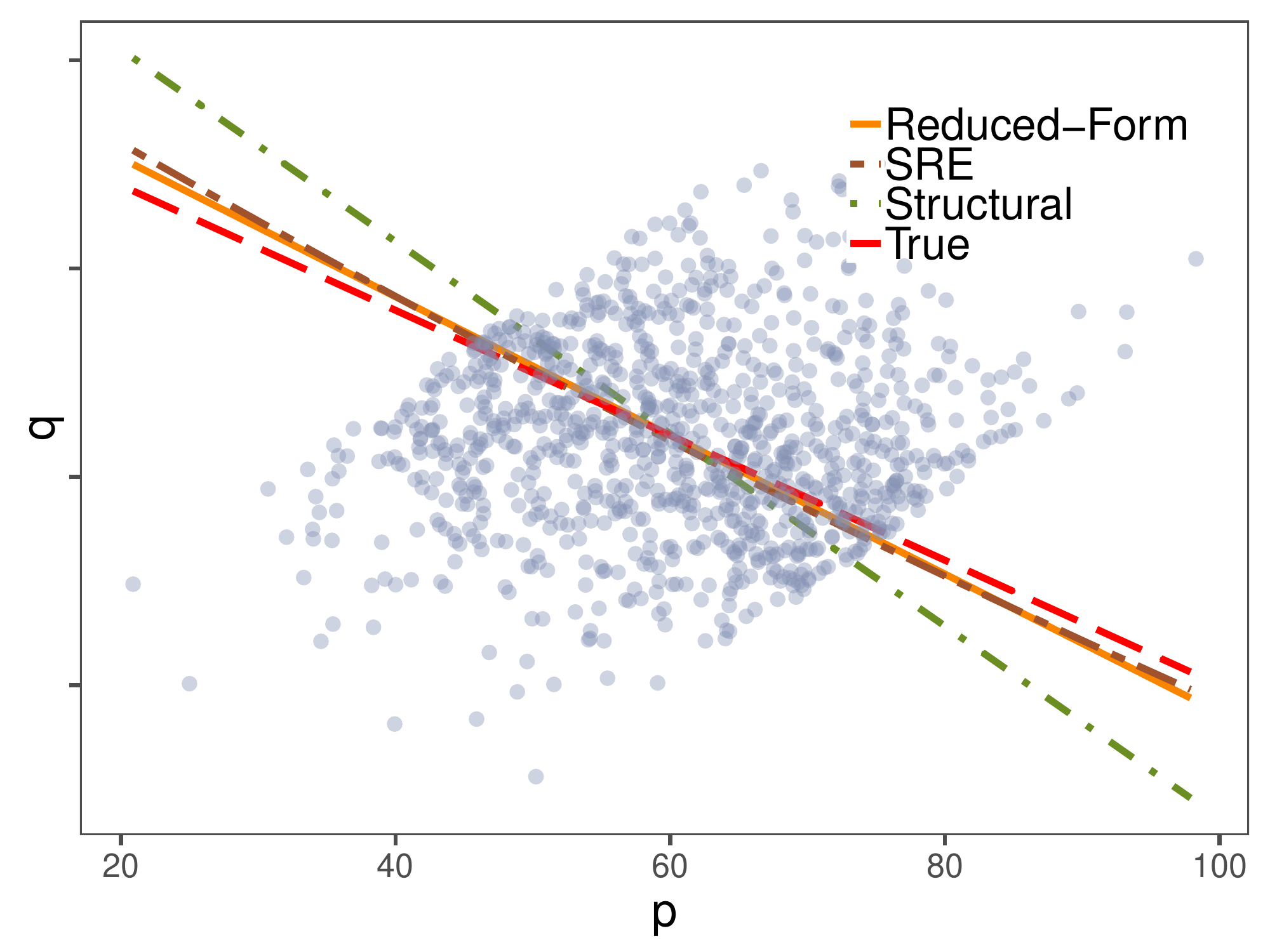}

}

\subfloat[\label{fig:demand_c}]{\includegraphics[width=0.5\columnwidth]{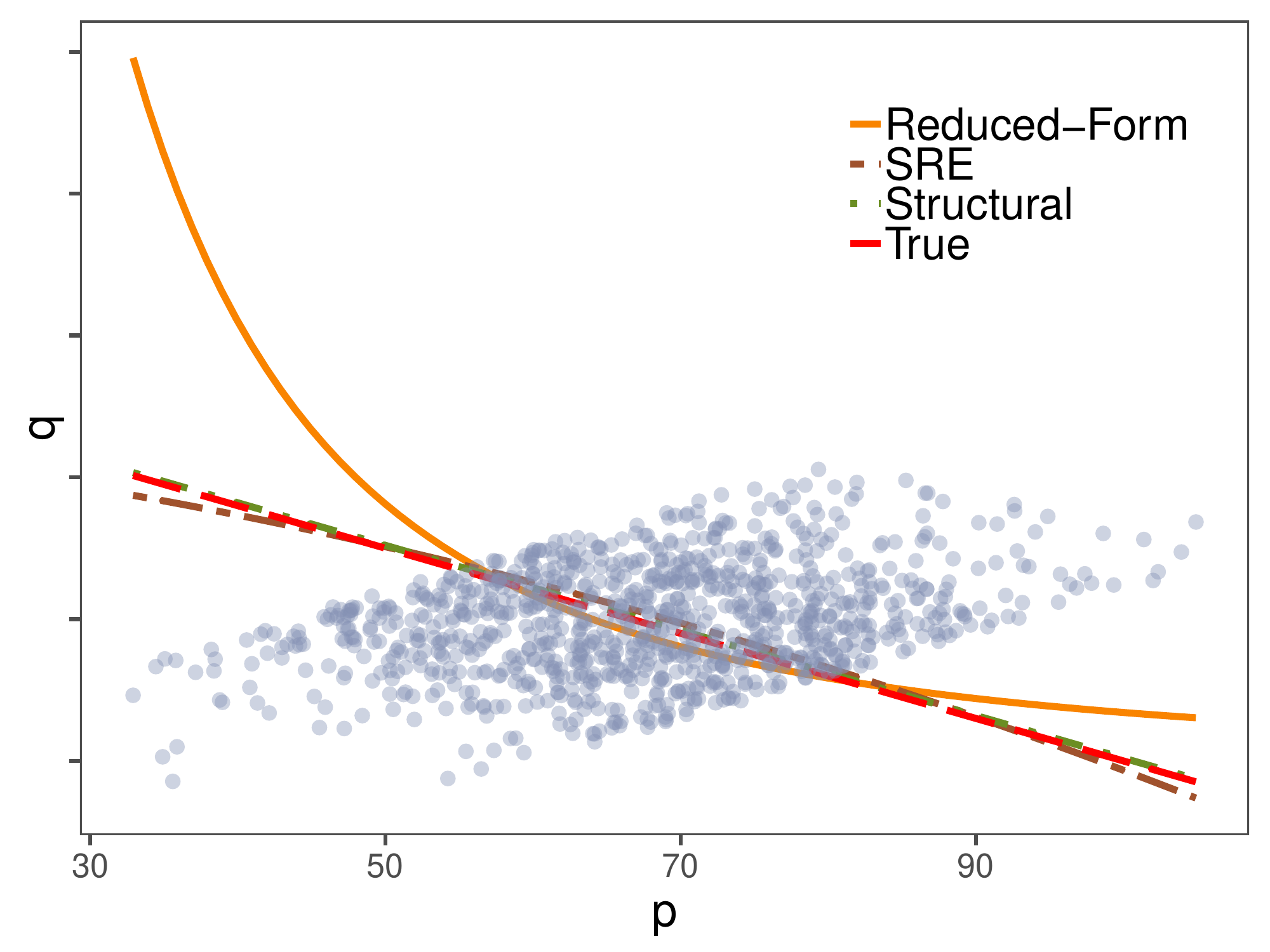}

}\subfloat[\label{fig:demand_d}]{\includegraphics[width=0.5\columnwidth]{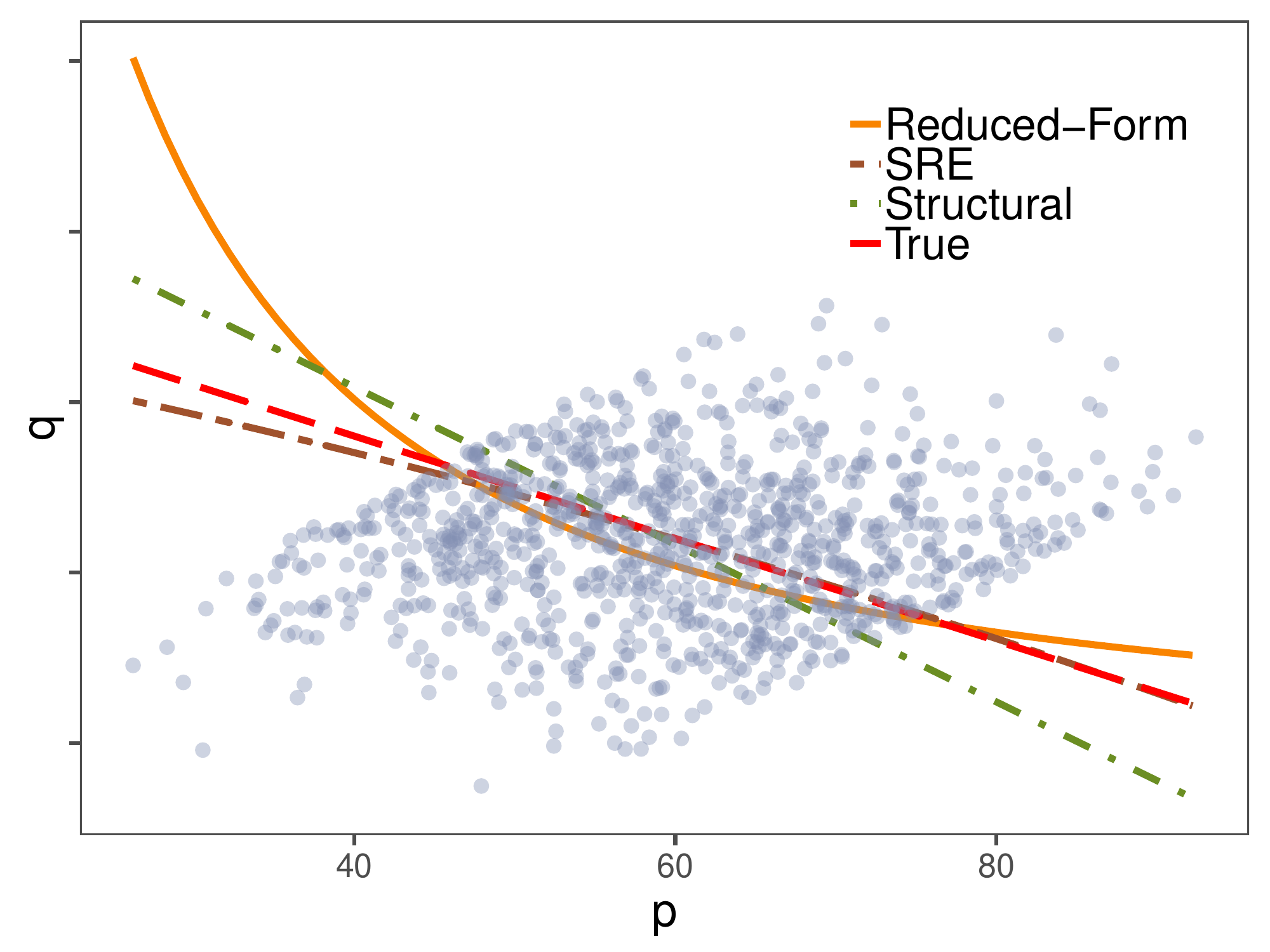}

}

\caption{{\small{}Demand Estimation. Dots represent training data. The true
demand curve is shown in red. (a) - (d) correspond respectively to
Experiment 1 - 4.}}
\end{figure}

\begin{table}
\caption{Demand Estimation - Results\label{tab:demandTable}}

\medskip{}

\begin{adjustwidth}{-0.52in}{-0.52in}
\begin{centering}
\begin{tabular}{cccccccccccc}
\toprule 
\toprule & \multicolumn{3}{c}{Reduced-Form} & \  & \multicolumn{3}{c}{Structural} & \  & \multicolumn{3}{c}{SRE}\tabularnewline
\cmidrule{2-4} \cmidrule{3-4} \cmidrule{4-4} \cmidrule{6-8} \cmidrule{7-8} \cmidrule{8-8} \cmidrule{10-12} \cmidrule{11-12} \cmidrule{12-12} 
Experiment & Bias & Var & MSE &  & Bias & Var & MSE &  & Bias & Var & MSE\tabularnewline
\midrule 
\multirow{1}{*}{1} & 0.2720 & 7.6780 & 7.7863 &  & 0.0770 & 0.9102 & 0.9161 &  & 0.9899 & 11.1375 & 13.3879\tabularnewline
\multirow{1}{*}{2} & 0.2884 & 5.1712 & 5.2821 &  & 12.3218 & 1.4233 & 203.8835 &  & 0.2783 & 11.4235 & 11.5223\tabularnewline
\multirow{1}{*}{3} & 25.9431 & 174.9081 & 2601.8750 &  & 0.1167 & 0.9648 & 0.9784 &  & 0.9703 & 13.0669 & 15.2066\tabularnewline
\multirow{1}{*}{4} & 11.8060 & 22.3152 & 423.3862 &  & 12.3212 & 1.4277 & 203.8519 &  & 0.3721 & 12.3088 & 12.5519\tabularnewline
\bottomrule
\end{tabular}
\par\end{centering}
\bigskip{}

\emph{\small{}Notes:}{\small{} results are based on 100 simulation
trials. Reported are the mean bias, variance, and MSE, averaged over
$p$.}{\small\par}

\end{adjustwidth}
\end{table}

Figure \ref{fig:demand_a} plots the results of the first experiment.
As the figure shows, the observed data $\left(p_{m},q_{m}\right)$
are significantly confounded -- fitting a least squares model to
the data would produce an upward-sloping curve. Despite the significant
confounding, reduced-form and structural estimation are both able
to identify the true demand curve. This is because both use correctly
specified models and $z$ is a valid instrument. In this case, the
SRE performs equally well. The three model fits and the true demand
curve almost coincide. 

Figure \ref{fig:demand_b} plots the results of the second experiment.
In this experiment, the reduced-form model is correctly specified,
while structural model is not. The structural fit therefore deviates
from the true demand curve, while the reduced-form model fits well.
Figure \ref{fig:demand_c} shows the other side of the coin. In Experiment
3, the structural model is correctly specified, but the reduced-form
model is not. In this case, even though the reduced-form fit manages
to capture the downward-sloping nature of the demand curve, it is
badly ``out of shape''. Finally, in Figure \ref{fig:demand_d},
we show the results of Experiment 4 in which both models are misspecified
and, as a result, produce fits that depart from the true relationship.
In all of these experiments, however, the SRE fits the true demand
curve well, regardless of which model -- the reduced-form or the
structural or even both -- is misspecified. 

Table \ref{tab:demandTable} reports the mean bias, variance, and
mean squared error of the estimators with respect to the true demand
curve over 100 trials. When they are correctly specified, reduced-form
and structural models exhibit low biases. The structural model, by
virtue of imposing more structure on the data, attains a lower variance.
When misspecified, both types of models exhibit large biases and MSEs.
The SRE, in comparison, consistently attains a low bias. Although
its variance is higher than that of structural estimation, its MSE
remains relatively low and is significantly lower than the other two
estimators when they are misspecified. 

\subsection{Discussion\label{subsec:Discussion}}

The tension between the goal of producing an accurate description
of the data and the goal of estimating externally valid structural
parameters that allow for counterfactual analysis and policy prediction
is a lasting legacy of Cowles Commission research program \citep{heckman_causal_2000}.
Structural estimation, in its effort to achieve the second goal, often
need to make strong and unrealistic assumptions, including both causal
assumptions such as rational expectations, and parametric assumptions
such as CES utility functions. Many efforts have been made to relax
these assumptions. In the context of dynamic structural models, for
example, these efforts include semiparametric estimation \citep{norets_semiparametric_2014},
robust estimation \citep{christensen_counterfactual_2019}, and alternative
specifications of expectations \citep{woodford_macroeconomic_2013}.
This paper offers an alternative: rather than seeking to minimize
assumptions and estimate partially identified models, or specify more
realistic models of behavior, which can be intractable and heterogeneous,
we show the feasibility of adopting a tractable structural model with
strong assumptions as an \emph{approximate} model and estimate the
data using structural regularization. A limitation with our approach
is that by doing so, the SRE estimator no longer permits a structural
interpretation and therefore cannot be used to conduct welfare analyses.
We leave addressing this limitation to future work. 

\section{Conclusion\label{sec:Conclusion}}

In this paper, we propose a general framework for incorporating theory
into statistical modeling for statistical prediction and causal inference.
We demonstrate the effectiveness of our method in a number of economic
applications including first-price auctions, dynamic models of entry
and exit, and demand estimation with instrumental variables. Many
more potential applications are possible, such as forecasting long-run
effects based on short-run observations or predicting effects at scale,
which we leave for future work. Our method has potential applications
not only in economics, but in other (social) scientific disciplines
whose theoretical models offer important insight but are subject to
significant misspecification concerns. \hypertarget{APP}{}

\bibliographystyle{apa}
\bibliography{mybib}

\begin{thebibliography}{}

\bibitem[\protect\astroncite{Aguirregabiria and
  Mira}{2010}]{aguirregabiria_dynamic_2010}
Aguirregabiria, V. and Mira, P. (2010).
\newblock Dynamic discrete choice structural models: {A} survey.
\newblock {\em Journal of Econometrics}, 156(1):38--67.
\newblock Publisher: Elsevier.

\bibitem[\protect\astroncite{Angrist and
  Krueger}{1995}]{angrist_split-sample_1995}
Angrist, J.~D. and Krueger, A.~B. (1995).
\newblock Split-{Sample} {Instrumental} {Variables} {Estimates} of the {Return}
  to {Schooling}.
\newblock {\em Journal of Business \& Economic Statistics}, 13(2):225--235.
\newblock Publisher: Taylor \& Francis.

\bibitem[\protect\astroncite{Angrist and
  Pischke}{2010}]{angrist_credibility_2010}
Angrist, J.~D. and Pischke, J.-S. (2010).
\newblock The credibility revolution in empirical economics: {How} better
  research design is taking the con out of econometrics.
\newblock {\em Journal of economic perspectives}, 24(2):3--30.

\bibitem[\protect\astroncite{Arcidiacono and
  Ellickson}{2011}]{arcidiacono_practical_2011}
Arcidiacono, P. and Ellickson, P.~B. (2011).
\newblock Practical {Methods} for {Estimation} of {Dynamic} {Discrete} {Choice}
  {Models}.
\newblock {\em Annual Review of Economics}, 3(1):363--394.
\newblock \_eprint: https://doi.org/10.1146/annurev-economics-111809-125038.

\bibitem[\protect\astroncite{Arcidiacono and
  Miller}{2011}]{arcidiacono_conditional_2011}
Arcidiacono, P. and Miller, R.~A. (2011).
\newblock Conditional choice probability estimation of dynamic discrete choice
  models with unobserved heterogeneity.
\newblock {\em Econometrica}, 79(6):1823--1867.
\newblock Publisher: Wiley Online Library.

\bibitem[\protect\astroncite{Artuc et~al.}{2010}]{artuc_trade_2010}
Artuc, E., Chaudhuri, S., and McLaren, J. (2010).
\newblock Trade {Shocks} and {Labor} {Adjustment}: {A} {Structural} {Empirical}
  {Approach}.
\newblock {\em American Economic Review}, 100(3):1008--1045.

\bibitem[\protect\astroncite{Athey and Haile}{2007}]{athey_nonparametric_2007}
Athey, S. and Haile, P.~A. (2007).
\newblock Nonparametric approaches to auctions.
\newblock {\em Handbook of econometrics}, 6:3847--3965.
\newblock Publisher: Elsevier.

\bibitem[\protect\astroncite{Bajari et~al.}{2013}]{bajari_game_2013}
Bajari, P., Hong, H., and Nekipelov, D. (2013).
\newblock Game theory and econometrics: {A} survey of some recent research.
\newblock In {\em Advances in economics and econometrics, 10th world congress},
  volume~3, pages 3--52.

\bibitem[\protect\astroncite{Bajari and Hortacsu}{2005}]{bajari_are_2005}
Bajari, P. and Hortacsu, A. (2005).
\newblock Are {Structural} {Estimates} of {Auction} {Models} {Reasonable}?
  {Evidence} from {Experimental} {Data}.
\newblock {\em Journal of Political Economy}, 113(4):703--741.
\newblock Publisher: The University of Chicago Press.

\bibitem[\protect\astroncite{Ben-David et~al.}{2010}]{ben-david_theory_2010}
Ben-David, S., Blitzer, J., Crammer, K., Kulesza, A., Pereira, F., and Vaughan,
  J.~W. (2010).
\newblock A theory of learning from different domains.
\newblock {\em Machine Learning}, 79(1):151--175.

\bibitem[\protect\astroncite{Bickel}{1982}]{bickel_adaptive_1982}
Bickel, P.~J. (1982).
\newblock On adaptive estimation.
\newblock {\em The Annals of Statistics}, pages 647--671.
\newblock Publisher: JSTOR.

\bibitem[\protect\astroncite{Bonhomme and
  Weidner}{2018}]{bonhomme_minimizing_2018}
Bonhomme, S. and Weidner, M. (2018).
\newblock Minimizing {Sensitivity} to {Model} {Misspecification}.
\newblock {\em arXiv:1807.02161 [econ, stat]}.
\newblock arXiv: 1807.02161.

\bibitem[\protect\astroncite{Chernozhukov
  et~al.}{2017}]{chernozhukov_doubledebiasedneyman_2017}
Chernozhukov, V., Chetverikov, D., Demirer, M., Duflo, E., Hansen, C., and
  Newey, W. (2017).
\newblock Double/debiased/neyman machine learning of treatment effects.
\newblock {\em American Economic Review}, 107(5):261--65.

\bibitem[\protect\astroncite{Chernozhukov
  et~al.}{2016}]{chernozhukov_double_2016}
Chernozhukov, V., Chetverikov, D., Demirer, M., Duflo, E., Hansen, C., and
  Newey, W.~K. (2016).
\newblock Double machine learning for treatment and causal parameters.
\newblock Technical report, cemmap working paper.

\bibitem[\protect\astroncite{Chetty}{2009}]{chetty_sufficient_2009}
Chetty, R. (2009).
\newblock Sufficient {Statistics} for {Welfare} {Analysis}: {A} {Bridge}
  {Between} {Structural} and {Reduced}-{Form} {Methods}.
\newblock {\em Annual Review of Economics}, 1(1):451--488.

\bibitem[\protect\astroncite{Chopra et~al.}{2013}]{chopra_dlid_2013}
Chopra, S., Balakrishnan, S., and Gopalan, R. (2013).
\newblock Dlid: {Deep} learning for domain adaptation by interpolating between
  domains.
\newblock In {\em {ICML} workshop on challenges in representation learning},
  volume~2.

\bibitem[\protect\astroncite{Christensen and
  Connault}{2019}]{christensen_counterfactual_2019}
Christensen, T. and Connault, B. (2019).
\newblock Counterfactual {Sensitivity} and {Robustness}.
\newblock {\em arXiv:1904.00989 [econ]}.
\newblock arXiv: 1904.00989.

\bibitem[\protect\astroncite{Dai et~al.}{2007}]{dai_boosting_2007}
Dai, W., Yang, Q., Xue, G.-R., and Yu, Y. (2007).
\newblock Boosting for transfer learning.
\newblock In {\em Proceedings of the 24th international conference on {Machine}
  learning}, {ICML} '07, pages 193--200, Corvalis, Oregon, USA. Association for
  Computing Machinery.

\bibitem[\protect\astroncite{Deaton}{2010}]{deaton_instruments_2010}
Deaton, A. (2010).
\newblock Instruments, randomization, and learning about development.
\newblock {\em Journal of economic literature}, 48(2):424--55.

\bibitem[\protect\astroncite{Donahue et~al.}{2014}]{donahue_decaf_2014}
Donahue, J., Jia, Y., Vinyals, O., Hoffman, J., Zhang, N., Tzeng, E., and
  Darrell, T. (2014).
\newblock Decaf: {A} deep convolutional activation feature for generic visual
  recognition.
\newblock In {\em International conference on machine learning}, pages
  647--655.

\bibitem[\protect\astroncite{Fessler and Kasy}{2019}]{fessler_how_2019}
Fessler, P. and Kasy, M. (2019).
\newblock How to {Use} {Economic} {Theory} to {Improve} {Estimators}:
  {Shrinking} {Toward} {Theoretical} {Restrictions}.
\newblock {\em The Review of Economics and Statistics}, 101(4):681--698.
\newblock Publisher: MIT Press.

\bibitem[\protect\astroncite{Ganin and
  Lempitsky}{2014}]{ganin_unsupervised_2014}
Ganin, Y. and Lempitsky, V. (2014).
\newblock Unsupervised domain adaptation by backpropagation.
\newblock {\em arXiv preprint arXiv:1409.7495}.

\bibitem[\protect\astroncite{Gao et~al.}{2008}]{gao_knowledge_2008}
Gao, J., Fan, W., Jiang, J., and Han, J. (2008).
\newblock Knowledge transfer via multiple model local structure mapping.
\newblock In {\em Proceedings of the 14th {ACM} {SIGKDD} international
  conference on {Knowledge} discovery and data mining}, pages 283--291.

\bibitem[\protect\astroncite{Giacomini
  et~al.}{2019}]{giacomini_estimation_2019}
Giacomini, R., Kitagawa, T., and Uhlig, H. (2019).
\newblock Estimation {Under} {Ambiguity}.
\newblock {\em Working Paper}.

\bibitem[\protect\astroncite{Glorot et~al.}{2011}]{glorot_domain_2011}
Glorot, X., Bordes, A., and Bengio, Y. (2011).
\newblock Domain adaptation for large-scale sentiment classification: {A} deep
  learning approach.

\bibitem[\protect\astroncite{Gopalan et~al.}{2011}]{gopalan_domain_2011}
Gopalan, R., Li, R., and Chellappa, R. (2011).
\newblock Domain adaptation for object recognition: {An} unsupervised approach.
\newblock In {\em 2011 international conference on computer vision}, pages
  999--1006. IEEE.

\bibitem[\protect\astroncite{Gourieroux
  et~al.}{1993}]{gourieroux_indirect_1993}
Gourieroux, C., Monfort, A., and Renault, E. (1993).
\newblock Indirect inference.
\newblock {\em Journal of applied econometrics}, 8(S1):S85--S118.
\newblock Publisher: Wiley Online Library.

\bibitem[\protect\astroncite{Guerre et~al.}{2000}]{guerre_optimal_2000}
Guerre, E., Perrigne, I., and Vuong, Q. (2000).
\newblock Optimal nonparametric estimation of first-price auctions.
\newblock {\em Econometrica}, 68(3):525--574.
\newblock Publisher: Wiley Online Library.

\bibitem[\protect\astroncite{Hansen and Sargent}{2001}]{hansen_robust_2001}
Hansen, L. and Sargent, T.~J. (2001).
\newblock Robust control and model uncertainty.
\newblock {\em American Economic Review}, 91(2):60--66.

\bibitem[\protect\astroncite{Hansen}{1982}]{hansen_large_1982}
Hansen, L.~P. (1982).
\newblock Large sample properties of generalized method of moments estimators.
\newblock {\em Econometrica: Journal of the Econometric Society}, pages
  1029--1054.
\newblock Publisher: JSTOR.

\bibitem[\protect\astroncite{Hansen and
  Marinacci}{2016}]{hansen_ambiguity_2016}
Hansen, L.~P. and Marinacci, M. (2016).
\newblock Ambiguity {Aversion} and {Model} {Misspecification}: {An} {Economic}
  {Perspective}.
\newblock {\em Statistical Science}, 31(4):511--515.
\newblock Publisher: Institute of Mathematical Statistics.

\bibitem[\protect\astroncite{Hansen and Sargent}{2010}]{hansen_wanting_2010}
Hansen, L.~P. and Sargent, T.~J. (2010).
\newblock Wanting robustness in macroeconomics.
\newblock In {\em Handbook of monetary economics}, volume~3, pages 1097--1157.
  Elsevier.

\bibitem[\protect\astroncite{Hansen and Sargent}{2020}]{hansen_structured_2020}
Hansen, L.~P. and Sargent, T.~J. (2020).
\newblock Structured {Uncertainty} and {Model} {Misspecification}.
\newblock {SSRN} {Scholarly} {Paper} ID 3280597, Social Science Research
  Network, Rochester, NY.

\bibitem[\protect\astroncite{Heckman}{2000}]{heckman_causal_2000}
Heckman, J.~J. (2000).
\newblock Causal parameters and policy analysis in economics: {A} twentieth
  century retrospective.
\newblock {\em The Quarterly Journal of Economics}, 115(1):45--97.

\bibitem[\protect\astroncite{Heckman}{2010}]{heckman_building_2010}
Heckman, J.~J. (2010).
\newblock Building bridges between structural and program evaluation approaches
  to evaluating policy.
\newblock {\em Journal of Economic literature}, 48(2):356--98.

\bibitem[\protect\astroncite{Heckman and
  Vytlacil}{2007}]{heckman_econometric_2007}
Heckman, J.~J. and Vytlacil, E.~J. (2007).
\newblock Econometric {Evaluation} of {Social} {Programs}, {Part} {I}: {Causal}
  {Models}, {Structural} {Models} and {Econometric} {Policy} {Evaluation}.
\newblock In Heckman, J.~J. and Leamer, E.~E., editors, {\em Handbook of
  {Econometrics}}, volume~6, pages 4779--4874. Elsevier.

\bibitem[\protect\astroncite{Hickman et~al.}{2012}]{hickman_structural_2012}
Hickman, B.~R., Hubbard, T.~P., and Saglam, Y. (2012).
\newblock Structural econometric methods in auctions: {A} guide to the
  literature.
\newblock {\em Journal of Econometric Methods}, 1(1):67--106.
\newblock Publisher: De Gruyter.

\bibitem[\protect\astroncite{Huang et~al.}{2007}]{huang_correcting_2007}
Huang, J., Gretton, A., Borgwardt, K., Scholkopf, B., and Smola, A.~J. (2007).
\newblock Correcting {Sample} {Selection} {Bias} by {Unlabeled} {Data}.
\newblock In Scholkopf, B., Platt, J.~C., and Hoffman, T., editors, {\em
  Advances in {Neural} {Information} {Processing} {Systems} 19}, pages
  601--608. MIT Press.

\bibitem[\protect\astroncite{James et~al.}{2013}]{james_introduction_2013}
James, G., Witten, D., Hastie, T., and Tibshirani, R. (2013).
\newblock {\em An introduction to statistical learning}, volume 112.
\newblock Springer.

\bibitem[\protect\astroncite{Jiang and Zhai}{2007}]{jiang_instance_2007}
Jiang, J. and Zhai, C. (2007).
\newblock Instance weighting for domain adaptation in {NLP}.
\newblock In {\em Proceedings of the 45th annual meeting of the association of
  computational linguistics}, pages 264--271.

\bibitem[\protect\astroncite{Keane}{2010a}]{keane_structural_2010}
Keane, M.~P. (2010a).
\newblock A structural perspective on the experimentalist school.
\newblock {\em Journal of Economic Perspectives}, 24(2):47--58.

\bibitem[\protect\astroncite{Keane}{2010b}]{keane_structural_2010-1}
Keane, M.~P. (2010b).
\newblock Structural vs. atheoretic approaches to econometrics.
\newblock {\em Journal of Econometrics}, 156(1):3--20.
\newblock Publisher: Elsevier.

\bibitem[\protect\astroncite{Kuang et~al.}{2020}]{kuang_stable_2020}
Kuang, K., Xiong, R., Cui, P., Athey, S., and Li, B. (2020).
\newblock Stable {Prediction} with {Model} {Misspecification} and {Agnostic}
  {Distribution} {Shift}.
\newblock {\em arXiv:2001.11713 [cs, stat]}.
\newblock arXiv: 2001.11713.

\bibitem[\protect\astroncite{Li and Goel}{2006}]{li_regularized_2006}
Li, B. and Goel, P.~K. (2006).
\newblock Regularized optimization in statistical learning: {A} {Bayesian}
  perspective.
\newblock {\em Statistica Sinica}, pages 411--424.

\bibitem[\protect\astroncite{Long et~al.}{2015}]{long_learning_2015}
Long, M., Cao, Y., Wang, J., and Jordan, M.~I. (2015).
\newblock Learning transferable features with deep adaptation networks.
\newblock {\em arXiv preprint arXiv:1502.02791}.

\bibitem[\protect\astroncite{Low and Meghir}{2017}]{low_use_2017}
Low, H. and Meghir, C. (2017).
\newblock The use of structural models in econometrics.
\newblock {\em Journal of Economic Perspectives}, 31(2):33--58.

\bibitem[\protect\astroncite{Mao and Xu}{2020}]{mao_ensemble_2020}
Mao, J. and Xu, J. (2020).
\newblock Ensemble {Learning} with {Statistical} and {Structural} {Models}.
\newblock {\em arXiv:2006.05308 [cs, econ]}.
\newblock arXiv: 2006.05308.

\bibitem[\protect\astroncite{Murphy}{2012}]{murphy_machine_2012}
Murphy, K.~P. (2012).
\newblock {\em Machine learning: a probabilistic perspective}.
\newblock MIT press.

\bibitem[\protect\astroncite{Muth}{1961}]{muth_rational_1961}
Muth, J.~F. (1961).
\newblock Rational expectations and the theory of price movements.
\newblock {\em Econometrica: Journal of the Econometric Society}, pages
  315--335.
\newblock Publisher: JSTOR.

\bibitem[\protect\astroncite{Nevo and Whinston}{2010}]{nevo_taking_2010}
Nevo, A. and Whinston, M.~D. (2010).
\newblock Taking the dogma out of econometrics: {Structural} modeling and
  credible inference.
\newblock {\em Journal of Economic Perspectives}, 24(2):69--82.

\bibitem[\protect\astroncite{Norets and
  Tang}{2014}]{norets_semiparametric_2014}
Norets, A. and Tang, X. (2014).
\newblock Semiparametric inference in dynamic binary choice models.
\newblock {\em Review of Economic Studies}, 81(3):1229--1262.

\bibitem[\protect\astroncite{Paarsch and
  Hong}{2006}]{paarsch_introduction_2006}
Paarsch, H.~J. and Hong, H. (2006).
\newblock An introduction to the structural econometrics of auction data.
\newblock {\em MIT Press Books}, 1.
\newblock Publisher: The MIT Press.

\bibitem[\protect\astroncite{Pan et~al.}{2010}]{pan_domain_2010}
Pan, S.~J., Tsang, I.~W., Kwok, J.~T., and Yang, Q. (2010).
\newblock Domain adaptation via transfer component analysis.
\newblock {\em IEEE Transactions on Neural Networks}, 22(2):199--210.

\bibitem[\protect\astroncite{Pan and Yang}{2010}]{pan_survey_2010}
Pan, S.~J. and Yang, Q. (2010).
\newblock A {Survey} on {Transfer} {Learning}.
\newblock {\em IEEE Transactions on Knowledge and Data Engineering},
  22(10):1345--1359.

\bibitem[\protect\astroncite{Pearl}{2009}]{pearl_causality_2009}
Pearl, J. (2009).
\newblock {\em Causality}.
\newblock Cambridge university press.

\bibitem[\protect\astroncite{Perrigne and
  Vuong}{2019}]{perrigne_econometrics_2019}
Perrigne, I. and Vuong, Q. (2019).
\newblock Econometrics of {Auctions} and {Nonlinear} {Pricing}.
\newblock {\em Annual Review of Economics}, 11(1):27--54.
\newblock \_eprint: https://doi.org/10.1146/annurev-economics-080218-025702.

\bibitem[\protect\astroncite{Reiss and Wolak}{2007}]{reiss_structural_2007}
Reiss, P.~C. and Wolak, F.~A. (2007).
\newblock Structural {Econometric} {Modeling}: {Rationales} and {Examples} from
  {Industrial} {Organization}.
\newblock In Heckman, J.~J. and Leamer, E.~E., editors, {\em Handbook of
  {Econometrics}}, volume~6, pages 4277--4415. Elsevier.

\bibitem[\protect\astroncite{Rojas-Carulla
  et~al.}{2018}]{rojas-carulla_invariant_2018}
Rojas-Carulla, M., Scholkopf, B., Turner, R., and Peters, J. (2018).
\newblock Invariant models for causal transfer learning.
\newblock {\em The Journal of Machine Learning Research}, 19(1):1309--1342.

\bibitem[\protect\astroncite{Rosenbaum and
  Rubin}{1983}]{rosenbaum_central_1983}
Rosenbaum, P.~R. and Rubin, D.~B. (1983).
\newblock The central role of the propensity score in observational studies for
  causal effects.
\newblock {\em Biometrika}, 70(1):41--55.
\newblock Publisher: Oxford University Press.

\bibitem[\protect\astroncite{Rosenzweig and
  Wolpin}{2000}]{rosenzweig_natural_2000}
Rosenzweig, M.~R. and Wolpin, K.~I. (2000).
\newblock Natural" natural experiments" in economics.
\newblock {\em Journal of Economic Literature}, 38(4):827--874.

\bibitem[\protect\astroncite{Rubin}{1974}]{rubin_estimating_1974}
Rubin, D.~B. (1974).
\newblock Estimating causal effects of treatments in randomized and
  nonrandomized studies.
\newblock {\em Journal of educational Psychology}, 66(5):688.
\newblock Publisher: American Psychological Association.

\bibitem[\protect\astroncite{Rust}{2014}]{rust_limits_2014}
Rust, J. (2014).
\newblock The {Limits} of {Inference} with {Theory}: {A} {Review} of {Wolpin}
  (2013).
\newblock {\em Journal of Economic Literature}, 52(3):820--850.

\bibitem[\protect\astroncite{Schwaighofer
  et~al.}{2005}]{schwaighofer_learning_2005}
Schwaighofer, A., Tresp, V., and Yu, K. (2005).
\newblock Learning {Gaussian} process kernels via hierarchical {Bayes}.
\newblock In {\em Advances in neural information processing systems}, pages
  1209--1216.

\bibitem[\protect\astroncite{Scott}{2014}]{scott_dynamic_2014}
Scott, P. (2014).
\newblock Dynamic discrete choice estimation of agricultural land use.
\newblock Publisher: TSE Working Paper.

\bibitem[\protect\astroncite{Sugiyama et~al.}{2008}]{sugiyama_direct_2008}
Sugiyama, M., Nakajima, S., Kashima, H., Buenau, P.~V., and Kawanabe, M.
  (2008).
\newblock Direct {Importance} {Estimation} with {Model} {Selection} and {Its}
  {Application} to {Covariate} {Shift} {Adaptation}.
\newblock In Platt, J.~C., Koller, D., Singer, Y., and Roweis, S.~T., editors,
  {\em Advances in {Neural} {Information} {Processing} {Systems} 20}, pages
  1433--1440. Curran Associates, Inc.

\bibitem[\protect\astroncite{Tibshirani}{1996}]{tibshirani_regression_1996}
Tibshirani, R. (1996).
\newblock Regression shrinkage and selection via the lasso.
\newblock {\em Journal of the Royal Statistical Society: Series B
  (Methodological)}, 58(1):267--288.

\bibitem[\protect\astroncite{Tzeng et~al.}{2014}]{tzeng_deep_2014}
Tzeng, E., Hoffman, J., Zhang, N., Saenko, K., and Darrell, T. (2014).
\newblock Deep domain confusion: {Maximizing} for domain invariance.
\newblock {\em arXiv preprint arXiv:1412.3474}.

\bibitem[\protect\astroncite{Wang and Deng}{2018}]{wang_deep_2018}
Wang, M. and Deng, W. (2018).
\newblock Deep visual domain adaptation: {A} survey.
\newblock {\em Neurocomputing}, 312:135--153.

\bibitem[\protect\astroncite{Wang and Schneider}{2014}]{wang_flexible_2014}
Wang, X. and Schneider, J. (2014).
\newblock Flexible transfer learning under support and model shift.
\newblock In {\em Advances in {Neural} {Information} {Processing} {Systems}},
  pages 1898--1906.

\bibitem[\protect\astroncite{Watson and Holmes}{2016}]{watson_approximate_2016}
Watson, J. and Holmes, C. (2016).
\newblock Approximate models and robust decisions.
\newblock {\em Statistical Science}, 31(4):465--489.
\newblock Publisher: Institute of Mathematical Statistics.

\bibitem[\protect\astroncite{Wolpin}{2013}]{wolpin_limits_2013}
Wolpin, K.~I. (2013).
\newblock {\em The {Limits} of {Inference} without {Theory}}.
\newblock MIT Press.
\newblock Google-Books-ID: ueXxCwAAQBAJ.

\bibitem[\protect\astroncite{Woodford}{2013}]{woodford_macroeconomic_2013}
Woodford, M. (2013).
\newblock Macroeconomic analysis without the rational expectations hypothesis.
\newblock {\em Annu. Rev. Econ.}, 5(1):303--346.
\newblock Publisher: Annual Reviews.

\bibitem[\protect\astroncite{Yosinski et~al.}{2014}]{yosinski_how_2014}
Yosinski, J., Clune, J., Bengio, Y., and Lipson, H. (2014).
\newblock How transferable are features in deep neural networks?
\newblock In {\em Advances in neural information processing systems}, pages
  3320--3328.

\bibitem[\protect\astroncite{Zadrozny}{2004}]{zadrozny_learning_2004}
Zadrozny, B. (2004).
\newblock Learning and evaluating classifiers under sample selection bias.
\newblock In {\em Proceedings of the twenty-first international conference on
  {Machine} learning}, {ICML} '04, page 114, Banff, Alberta, Canada.
  Association for Computing Machinery.

\end{thebibliography}

\end{document}